\documentclass[iop]{emulateapj}
\usepackage{natbib,apjfonts,graphicx,threeparttable,rotating}
\pdfoutput=1

\def\ts   {\thinspace}
\def\kms  {\ifmmode{{\rm \ts km\ts s}^{-1}}\else{\ts km\ts s$^{-1}$}\fi}
\def\kkms  {\ifmmode{{\rm \ts K\ts km\ts s}^{-1}}\else{\ts K\ts km\ts s$^{-1}$}\fi}
\def\lcou  {\ifmmode{{\rm \ts K\ts km\ts s}^{-1}\ts {\rm pc}^{2}}\else{\ts K\ts km\ts s$^{-1}$\ts pc$^{2}$}\fi}
\def\xcou  {\ifmmode{{\rm \ts cm^{-2}\ts (K\ts km\ts s^{-1})}^{-1}}\else{\ts cm$^{-2}$\ts (K\ts km\ts s$^{-1}$)$^{-1}$}\fi}
\def\cc  {\ifmmode{{\rm \ts cm}^{-2}}\else{\ts cm$^{-2}$}\fi}
\def\ccc  {\ifmmode{{\rm \ts cm}^{-3}}\else{\ts cm$^{-3}$}\fi}
\def\msol   {\ifmmode{{\rm M}_{\odot}}\else{M$_{\odot}$}\fi}
\def\mpcsq   {\ifmmode{{\rm M}_{\odot}\ts {\rm pc}^{-2}}\else{M$_{\odot}$}\ts pc$^{-2}$\fi}
\def\aco {\ifmmode{^{12}{\rm CO}(J=1\to0)}\else{$^{12}{\rm CO}(J=1\to0)$}\fi}
\def\bco {\ifmmode{^{12}{\rm CO}(J=2\to1)}\else{$^{12}{\rm CO}(J=2\to1)$}\fi}
\def\m  {\ifmmode{\mu {\rm m}}\else{$\mu$m}\fi}
\def\cco {\ifmmode{^{13}{\rm CO}(J=1\to0)}\else{$^{13}{\rm CO}(J=1\to0)$}\fi}
\def\hi  {\ifmmode{{\rm H}{\rm \scriptsize I}}\else{H\ts {\scriptsize I}}\fi}
\def\nh  {\ifmmode{N(\rm H)}\else{$N$(H)}\fi}
\def\hh   {\ifmmode{{\rm H}_2}\else{H$_2$}\fi}
\def\xco   {\ifmmode{X_{\rm CO}}\else{$X_{\rm CO}$}\fi}
\def\mh   {\ifmmode{{\rm H}_2}\else{H$_2$}\fi} 
\def\ha   {\ifmmode{{\rm H}{\alpha}}\else{H\ts {$\alpha$}}\fi}  
\def\hii  {\ifmmode{{\rm H}{\rm \small II}}\else{H\ts {\scriptsize II}}\fi}
\def\ico {\ifmmode{I(\rm CO)}\else{$I(\rm CO)$}\fi}
\def\D {\ifmmode{^{\circ}}\else{$^\circ$}\fi}
\def\nodata {\ifmmode{...}\else{$...$}\fi}
\def\punit  {\ifmmode{{\rm \ts cm}^{-3} {\rm \ts K}}\else{\ts cm$^{-3}$ \ts K}\fi}

\shorttitle{Properties of GMCs in M51, M33 and the LMC}
\slugcomment{\today\ draft}

\begin{document}

\title{A Comparative Study of Giant Molecular Clouds in M51, M33 and the Large Magellanic Cloud}

\author{
Annie Hughes\altaffilmark{1},
Sharon E. Meidt\altaffilmark{1},
Dario Colombo\altaffilmark{1},
Eva Schinnerer\altaffilmark{1},
Jer\^ome Pety\altaffilmark{2,3},
Adam K. Leroy\altaffilmark{4},
Clare L. Dobbs\altaffilmark{5},
Santiago Garc\'ia-Burillo\altaffilmark{6},
Todd A. Thompson\altaffilmark{7,8},
Ga\"elle Dumas\altaffilmark{2},
Karl F. Schuster\altaffilmark{2},
Carsten Kramer\altaffilmark{9}
\altaffiltext{1}{Max-Planck-Institut f\"ur Astronomie, K\"onigstuhl 17, D-69117, Heidelberg, Germany}
\altaffiltext{2}{Institut de Radioastronomie Millim\'etrique, 300 Rue de la Piscine, F-38406 Saint Martin d'H\`eres, France}
\altaffiltext{3}{Observatoire de Paris, 61 Avenue de l'Observatoire, F-75014 Paris, France}
\altaffiltext{4}{National Radio Astronomy Observatory, 520 Edgemont Road, Charlottesville, VA 22903, USA}
\altaffiltext{5}{School of Physics and Astronomy, University of Exeter, Stocker Road, Exeter EX4 4QL, UK}
\altaffiltext{6}{Observatorio Astron\'omico Nacional, Observatorio de Madrid, Alfonso XII, 3, 28014 Madrid, Spain}
\altaffiltext{7}{Department of Astronomy, The Ohio State University, 140 W. 18th Ave., Columbus, OH 43210, USA} 
\altaffiltext{8}{Center for Cosmology and AstroParticle Physics, The Ohio State University, 191 W. Woodruff Ave., Columbus, OH 43210, USA}
\altaffiltext{9}{Instituto Radioastronom\'ia Milim\'etrica, Av. Divina Pastora 7, Nucleo Central, 18012 Granada, Spain}
}

\begin{abstract}

%%%%%%%%%%%%%%%%%%%%%%%%%%%%%%%%
%%%%%%%%%%%%%%%%%%%%%%%%%%%%%%%%
%\section{Abstract}
%%%%%%%%%%%%%%%%%%%%%%%%%%%%%%%%
%%%%%%%%%%%%%%%%%%%%%%%%%%%%%%%%
\label{sect:abstract}

\noindent We compare the properties of giant molecular clouds
  (GMCs) in M51 identified by the Plateau de Bure Interferometer
  Whirlpool Arcsecond Survey (PAWS) with GMCs identified in
  wide-field, high resolution surveys of CO emission in M33 and the
  Large Magellanic Cloud (LMC). We find that GMCs in M51 are larger,
  brighter and have higher velocity dispersions relative to their size
  than equivalent structures in M33 and the LMC. These differences
  imply that there are genuine variations in the average mass surface
  density $\langle \Sigma_{\rm H_{2}} \rangle$ of the different GMC
  populations. To explain this, we propose that the pressure in the
  interstellar medium surrounding the GMCs plays a role in regulating
  their density and velocity dispersion. We find no evidence for a
  correlation between size and linewidth in any of M51, M33 or the LMC
  when the CO emission is decomposed into GMCs, although moderately
  robust correlations are apparent when regions of contiguous CO
  emission (with no size limitation) are used. Our work demonstrates
  that observational bias remains an important obstacle to the
  identification and study of extragalactic GMC populations using CO
  emission, especially in molecule-rich galactic environments.

\end{abstract}

\setlength{\parskip}{8pt}

%%%%%%%%%%%%%%%%%%%%%%%%%%%%%%%%%%%
%%%%%%%%%%%%%%%%%%%%%%%%%%%%%%%%%%%
\section{Introduction}
%%%%%%%%%%%%%%%%%%%%%%%%%%%%%%%%%%%
%%%%%%%%%%%%%%%%%%%%%%%%%%%%%%%%%%%
\label{sect:intro}

\noindent Among the different phases of the interstellar medium (ISM),
the dense molecular hydrogen gas is especially deserving of study. It
is the primary component by mass of the ISM in the central regions of
spiral galaxies, and the principal -- perhaps only -- site of star
formation \citep[e.g.][]{youngscoville91}. In regions with high
pressure and high extinction, the molecular gas may be extensive and
diffuse \citep{elmegreen93}, but under more typical interstellar
conditions a significant fraction \citep[$\sim50$\%,][]{sawadaetal12}
of the molecular gas is organized into discrete cloud complexes with
masses of $\sim10^{4}$ to $10^{6}$\,\msol\ and sizes of $\sim20$ to
50\,pc \citep{blitz93}. The study of these giant molecular clouds
(GMCs) is of great importance, since their properties determine
whether, where and how stars form. \\

\noindent GMCs in the Milky Way and other nearby galaxies are observed
to follow correlations between their size, line width, and CO
luminosity. These scaling relations have become a standard metric for
comparing molecular cloud populations. As originally formulated by
\citet{larson81}, GMCs exhibit: i) a power-law relationship between
their size and velocity dispersion, with a slope of $\sim0.5$; ii) a
nearly linear correlation between their virial mass and mass estimates
based on other tracers of \hh\ column density, which would seem to
imply that the clouds are self-gravitating and in approximate virial
balance; and iii) an inverse relationship between their size and
volume-averaged density. \citet[][henceforth S87]{solomonetal87} were
subsequently able to measure the coefficients and exponents of these
correlations for 273 GMCs in the inner Milky Way, establishing the
empirical expressions for ``Larson's Laws'' that have become the
yardstick for studies of GMCs in other galaxies and in different
interstellar environments \citep[e.g.][henceforth
  B08]{bolattoetal08}.\\

\noindent While resolved studies of extragalactic GMC populations will
become routine with the Atacama Large Millimeter Array (ALMA), the
twin requirements of high resolution and high sensitivity mean that
obtaining extragalactic datasets comparable to the S87 catalogue has
thus far only been feasible for a few nearby galaxies. Using either
\aco\ or \bco\ to trace the molecular gas distribution, wide-field
surveys covering a significant fraction of a galactic disk with a
linear resolution of $\sim50$\,pc or better have recently been
completed for M31, M33, IC10, M64, the Magellanic Clouds, IC342,
NGC~6822 and NGC~6946
\citep{rosolowskyetal07,engargiolaetal03,gardanetal07,gratieretal12,leroyetal06,rosolowskyblitz05,fukuietal08,mizunoetal01,
  wongetal11,mulleretal10,gratieretal10,hirotaetal11,donovanmeyeretal12,
  rebolledoetal12}. These surveys have found some evidence that the
properties of molecular clouds vary with environment and their level
of star formation activity. In IC342, the LMC and M33, GMCs with signs
of ongoing massive star formation are found to exhibit higher peak CO
brightness temperatures than non-star-forming clouds
\citep{hirotaetal11,hughesetal10,gratieretal12}. Other examples
include larger linewidths for molecular structures without high-mass
star formation \citep[IC342 and M83,][]{hirotaetal11,muraokaetal09}
and in the central regions of galaxies \citep[the Galactic Centre and
  NGC6946,][]{okaetal01,donovanmeyeretal12}; a decrease in CO
brightness at large galactocentric radii \citep[the Milky Way and
  M33,][]{heyeretal01,gratieretal12}; higher mass surface densities in
high pressure environments \citep[e.g. M64,][]{rosolowskyblitz05}; and
a lower CO surface brightness and narrower linewidths for GMCs in
dwarf galaxies
\citep[e.g. B08,][]{rubioetal93,mulleretal10,hughesetal10,gratieretal10}. Yet
much of the apparent galaxy-to-galaxy variation in GMC properties
could be due to the disparate sensitivity and resolution of the
observations and/or methodological differences \citep[as noted by
  e.g.][]{shethetal08}. Using a consistent method to identify and
measure the properties of $\sim100$ resolved GMCs in a sample of
twelve galaxies, B08 concluded that GMCs in fact demonstrate nearly
uniform properties across the Local Group.\\

\noindent In this paper, we compare the properties of GMCs identified
using high angular resolution CO surveys of three galaxies: M51, M33
and the LMC. Technically, the main difference between our work and
previous comparative studies is that each of our datasets covers a
significant fraction of the underlying galactic disk and therefore
provides a statistically significant sample of clouds for each galaxy
(from $\sim100$ for M33, to more than $\sim1500$ for M51, although the
precise number depends on the decomposition method). All three
datasets have sufficient resolution to resolve individual GMCs, but
were obtained either with a combination of single-dish and
interferometric observations, or with a single-dish telescope
alone. Spatial filtering of large-scale emission should therefore not
be of concern. We use a consistent methodology to identify significant
emission and decompose it into cloud-like structures, and we
explicitly test whether differences in the sensitivity, resolution and
gridding scheme of the CO data influence the derived GMC properties. A
second important difference is physical: the galaxies targeted by
previous GMC studies did not include a massive, grand design spiral
galaxy like M51 where the ISM is \hh-dominated over a significant
fraction of the galactic disk \citep[e.g.][]{schusteretal07}. Some of
the observed uniformity of extragalactic GMC populations may be due to
the limited range of interstellar environments where high resolution
CO surveys have been conducted to date. In this sense, a comparison
between the GMCs in M51, M33 and the LMC is of particular interest,
since galactic properties such as the metallicity, strength of the
spiral potential and the average interstellar pressure vary
significantly between these three galaxies (see also
Table~\ref{tbl:galcmp}). \\

\noindent This paper is structured as follows. In
Section~\ref{sect:data}, we briefly describe the origin and
characteristics of the CO datasets that we have
used. Section~\ref{sect:cloudidentification} describes the approach
that we have used to identify GMCs and to determine their physical
properties. Our comparative analysis of GMC properties and Larson-type
scaling relations is presented in Section~\ref{sect:results}. Our
primary result is that GMCs in the inner disk of M51 have different
physical properties to the GMCs in M33 and the LMC. In
Section~\ref{sect:discussion}, we consider possible physical origins
for the differences that we observe, and suggest reasons why our
conclusion differs from previous comparative studies of GMC
populations (e.g. B08). As part of this discussion, we describe
several observational effects that should be considered when
intepreting empirical correlations between GMC properties. We
summarize the key results of our analysis in
Section~\ref{sect:conclusions}. \\

%%%%%%%%%%%%%%%%%%%%%%%%%%%%%%
%%%%%%%%%%%%%%%%%%%%%%%%%%%%%%
\section{Molecular Gas Data}
%%%%%%%%%%%%%%%%%%%%%%%%%%%%%%
%%%%%%%%%%%%%%%%%%%%%%%%%%%%%%
\label{sect:data}

\subsection{M51}
\label{sect:datam51}

\noindent The CO data for M51 were obtained by the Plateau de Bure
Arcsecond Whirlpool Survey
\citep[PAWS][]{schinnereretal13,petyetal13}. PAWS observations mapped
a total field-of-view of approximately 270\arcsec\ $\times$
170\arcsec\ in the inner disk of M51 in the ABCD configurations of the
Plateau de Bure Interferometers (PdBI) between August 2009 and March
2010. Since an interferometer filters out low spatial frequencies, the
PdBI data were combined with observations of CO emission in M51
obtained using the IRAM 30\,m single-dish telescope in May 2010. The
effective angular resolution of the final combined PAWS data cube is
1\farcs16 $\times$ 0\farcs97, corresponding to a spatial resolution of
$\sim40$\,pc at our assumed distance to M51
\citep[7.6\,Mpc,][]{ciardulloetal02}. The data cube covers the LSR
velocity range 173 to 769\,\kms\ and the width of each velocity
channel is 5\,\kms.  The mean RMS of the noise fluctuations across the
survey is $\sim0.4$\,K in a 5.0\,\kms\ channel. The PAWS observing
strategy, data reduction and combination procedures, and flux
calibration are described by \citet{petyetal13}. Here we focus on the
properties of M51 clouds relative to the GMC populations of the other
low-mass galaxies; for some of our analysis, we also distinguish
between GMCs located in the spiral arms and central region of M51, and
GMCs in M51's interarm region. The methods that were used to define
these different zones (i.e. arm, interarm and central regions) are
described by Colombo et al. (submitted), where we also present the M51
GMC catalogue and conduct a detailed investigation of GMC properties
in different environments within M51. A CO integrated intensity image
of M51 by PAWS is shown in Figure~\ref{fig:maps}[a]. The total
  CO luminosity within the PAWS data cube is $9.2 \times
  10^{8}$\,\lcou\ \citep{petyetal13}. Over the same field-of-view,
  this agrees with the total CO flux obtained by the BIMA
  \citep{helferetal03} and CARMA \citep{kodaetal11} surveys of M51 to
  within $10$\% \citep{petyetal13}.

\subsection{M33}
\label{sect:datam33}

\noindent For M33, we use the CO data published by
\citet{rosolowskyetal07}, which combines observations by the
Berkeley-Illinois-Maryland Association (BIMA) array
\citep{engargiolaetal03} and the Five College Radio Astronomy
Observatory (FCRAO) 14\,m single-dish telescope
\citep{heyeretal04}. The common field-of-view of the single-dish and
interferometer surveys is 0.25 square degrees, covering most of M33's
optical disk. The angular resolution of the combined cube is
$13\farcs2 \times 12\farcs9$, corresponding to a spatial resolution of
53\,pc for our assumed distance to M33 of 840\,kpc
\citep[e.g.][]{galletietal04}. The data covers the LSR velocity range
$[-400,40]$\,\kms, and the velocity channel width is 2.0\,\kms. The
RMS noise per channel is 0.24\,K. A CO integrated intensity image
constructed from the M33 data is shown in
Figure~\ref{fig:maps}[b]. By summing the emission in the
  BIMA+FCRAO M33 data cube, we estimate that the total CO luminosity
  of M33 is $3.2 \times 10^{7}$\,\lcou. This agrees with other recent
  observational estimates for M33's total CO luminosity to within
  $\sim30$\% \citep[see
    e.g.][]{gratieretal10,rosolowskyetal07,heyeretal04}, but is a
  factor of $\sim2.5$ higher than the total luminosity obtained by
  summing the emission within the NRO M33 All-Disk Survey map of CO
  integrated intensity \citep{tosakietal11}.

\subsection{The Large Magellanic Cloud}
\label{sect:datalmc}

\begin{figure*}
\begin{center}
\hspace{-0.5cm}
\includegraphics[width=52mm,angle=270]{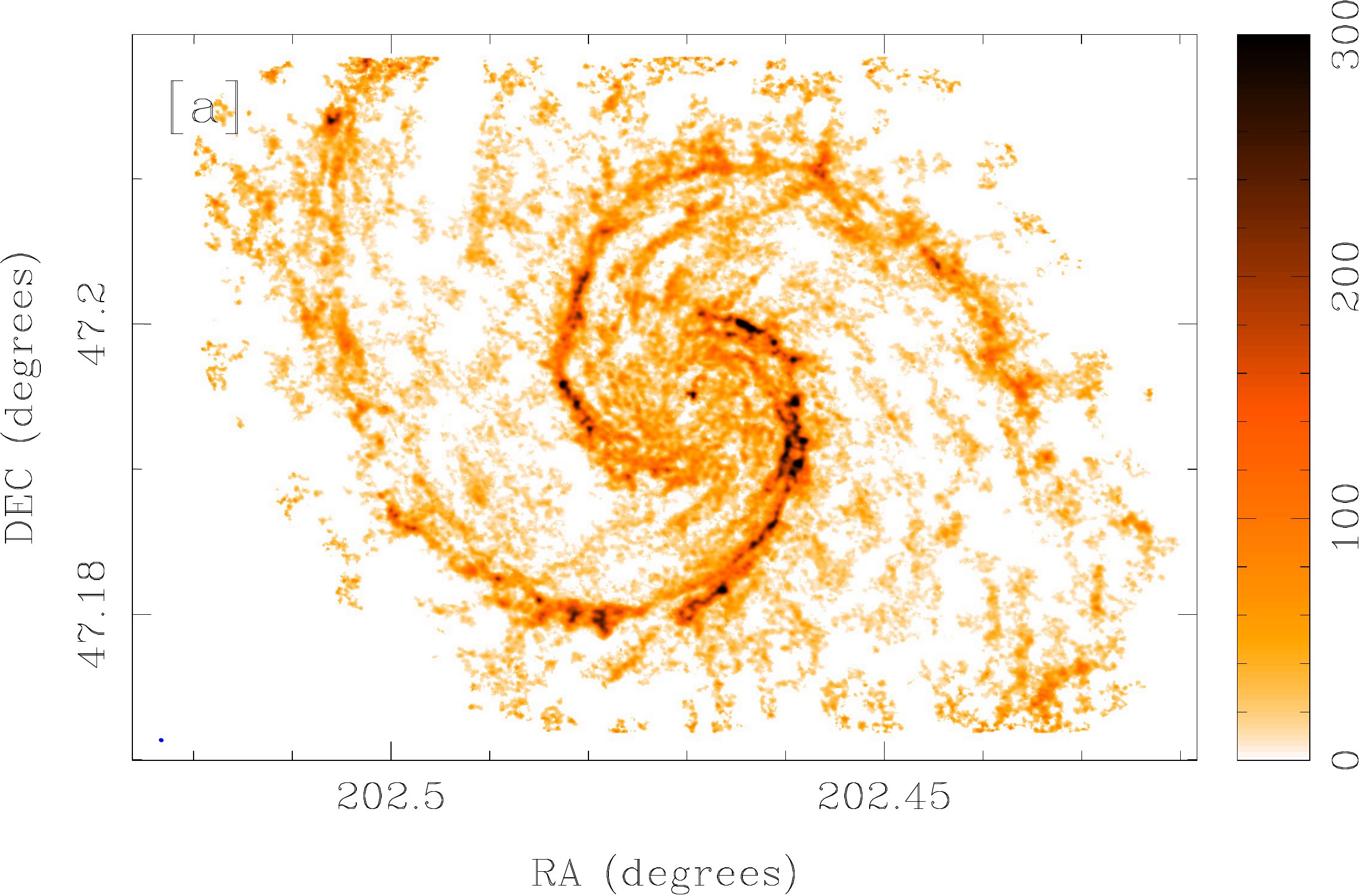}
\hspace{0.5cm}
\includegraphics[width=52mm,angle=270]{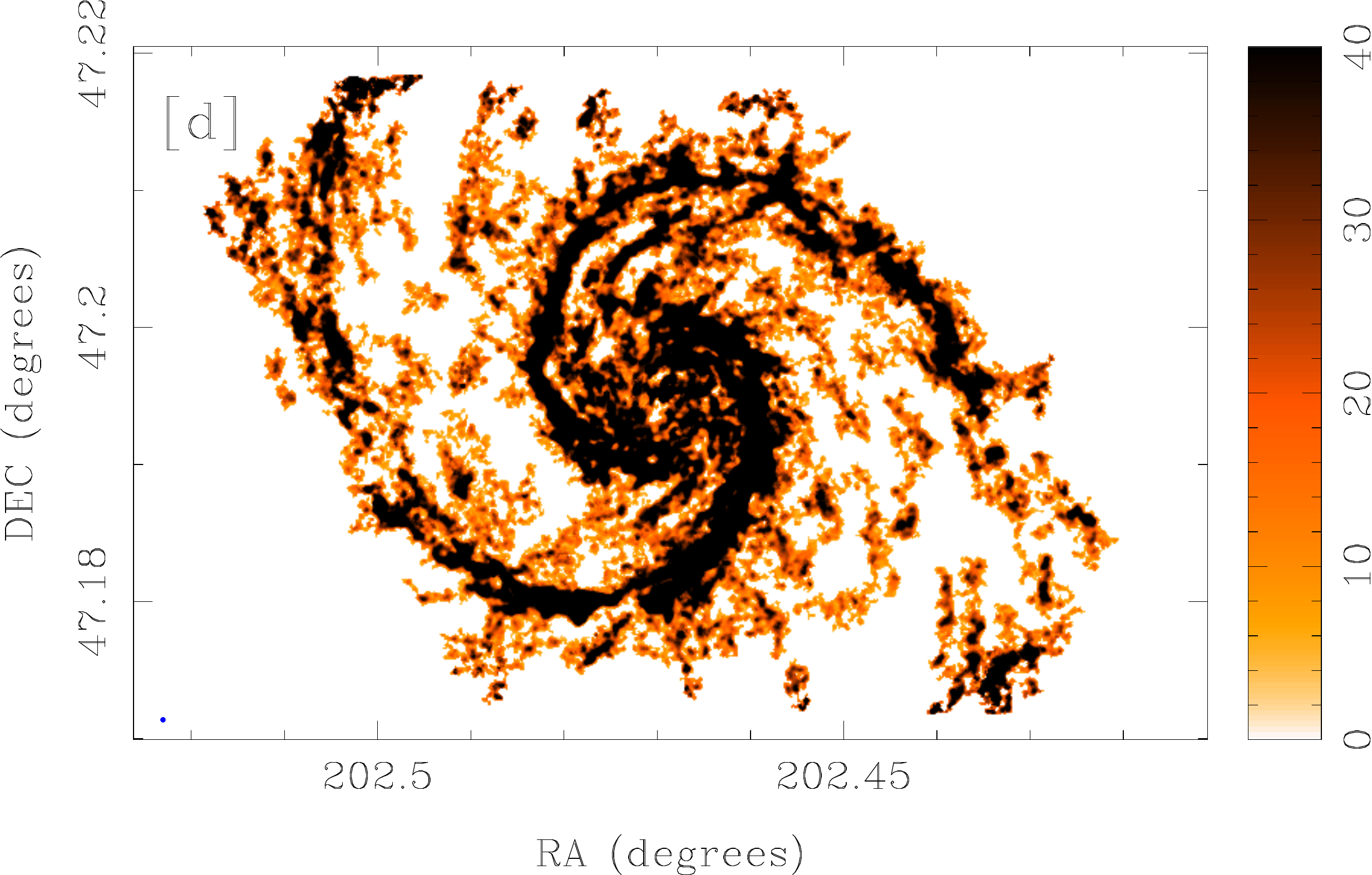}
\par \addvspace{0.5cm}
\hspace{-0.1cm}
\includegraphics[width=68mm,angle=270]{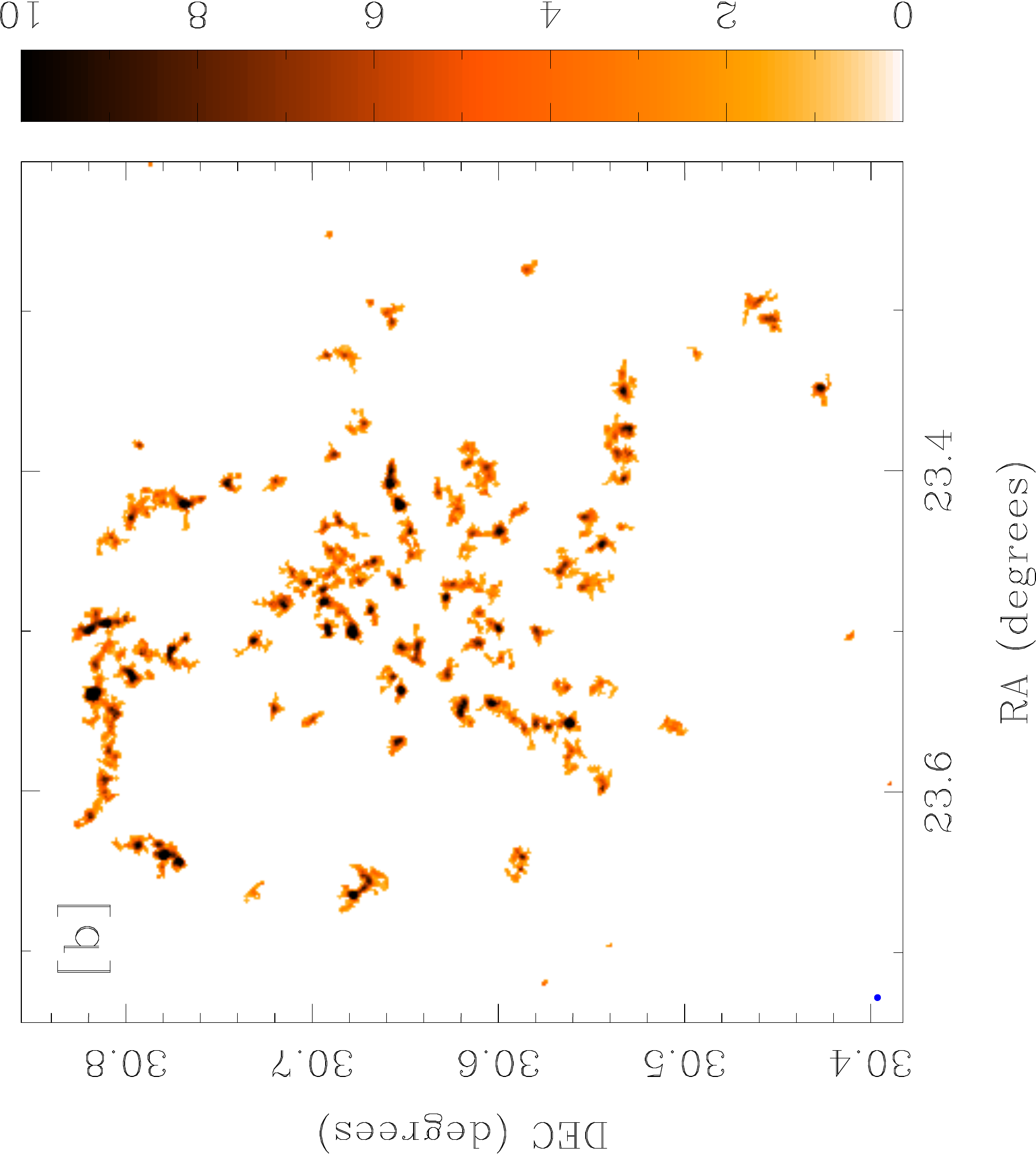}
\hspace{1.1cm}
\includegraphics[width=68mm,angle=270]{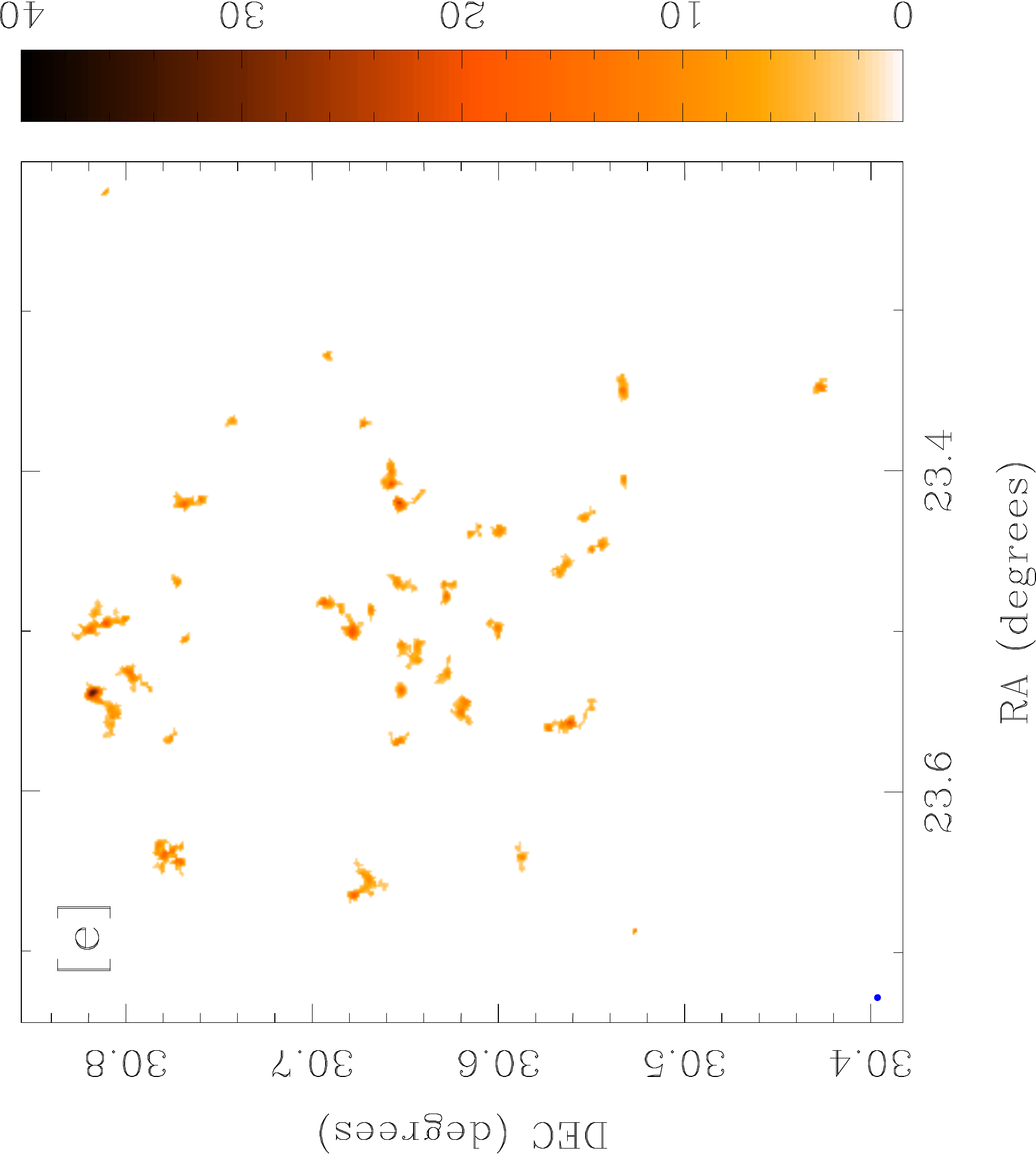}
\par \addvspace{0.5cm}
\hspace{-0.1cm}
\includegraphics[width=60mm,angle=270]{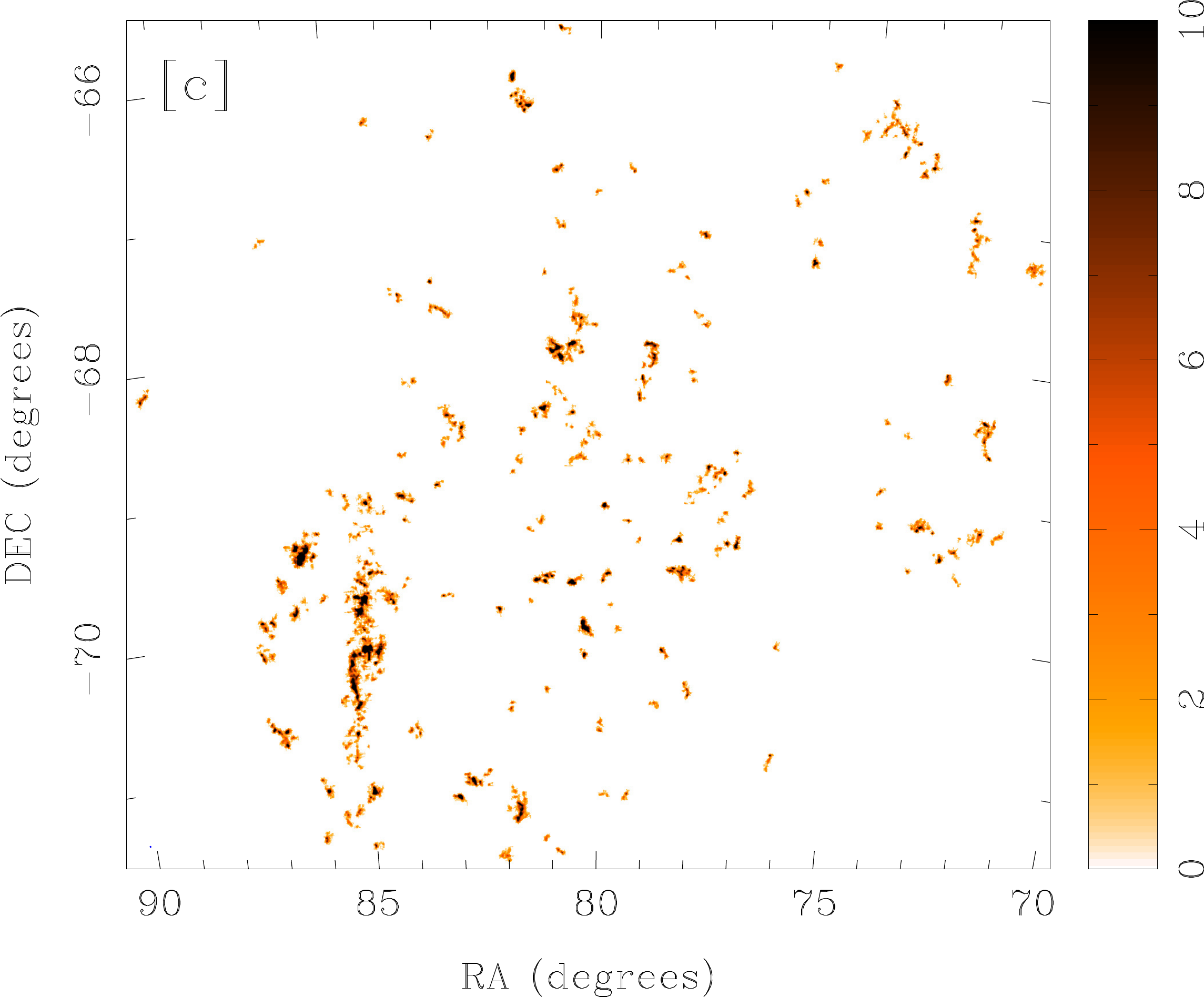}
\hspace{1.1cm}
\includegraphics[width=62mm,angle=270]{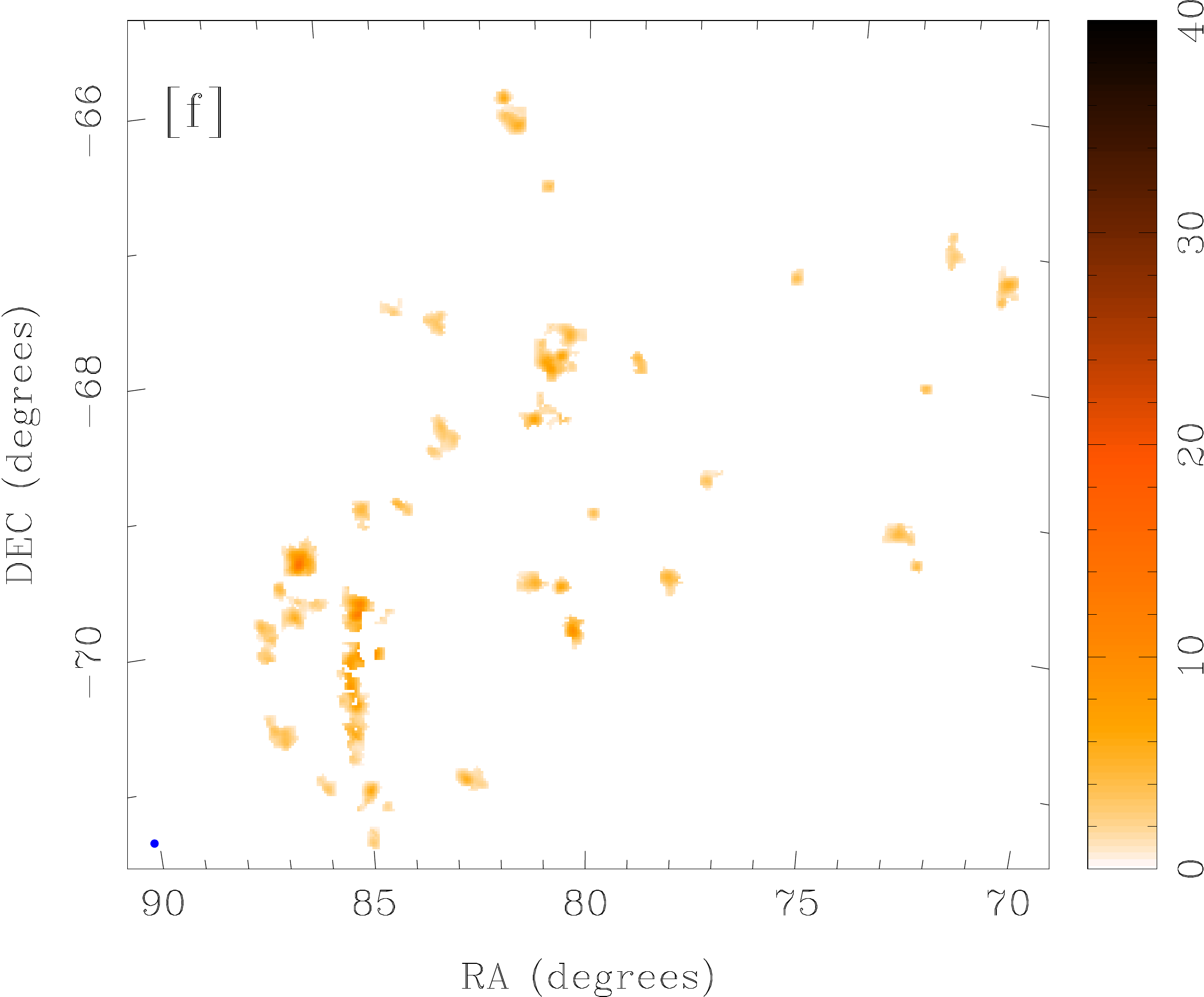}

\caption{\small Maps of CO integrated intensity in [a] M51
  \citep{schinnereretal13,petyetal13}, [b] M33
  \citep{rosolowskyetal07}, and [c] the LMC \citep{wongetal11}. Panels
        [d] to [f] present the corresponding CO integrated intensity
        maps after matching the spatial and spectral resolution of the
        datacubes and interpolating them onto a pixel grid with the
        same physical dimensions (see text). For all panels, the
        telescope beam is shown as a blue circle in the bottom left
        corner. All maps are presented using a square-root intensity
        scale in \kkms\ units. The limits of the colour stretch in
        panels [a] to [c] are chosen to provide an optimal overview of
        the spatial distribution of CO emission within each galaxy. In
        panels [d] to [f], we use the same limits for the intensity
        scale for each galaxy to highlight the difference in CO
        brightness between the three galaxies. }
\label{fig:maps}
\end{center}
\end{figure*}

\noindent The CO data for the LMC were obtained by the Magellanic
Mopra Assessement (MAGMA). The MAGMA survey design, data acquisition,
reduction procedures and calibration are described in detail by
\citet{wongetal11}. MAGMA mapped CO cloud complexes that had been
identified at lower resolution by NANTEN \citep{fukuietal08},
targeting 114 NANTEN GMCs with CO luminosities greater than
$7000$\,\lcou, and peak integrated intensities greater than
$1$\,\kkms. The combined field-of-view of the MAGMA survey is
$\sim3.6$ square degrees. Although the clouds targeted for mapping
represent only $\sim50$\% of the clouds in the NANTEN catalogue, the
region surveyed by MAGMA contributes $\sim80$\% of the total CO flux
measured by NANTEN. The MAGMA LMC data cube has an effective
resolution of 45\arcsec, corresponding to a linear resolution of
$\sim11$\,pc at the distance of the LMC
\citep[$50.1$\,kpc,][]{alves04}. The velocity channel width is
0.53\,\kms, and the total LSR velocity range of the cube is 200 to
305\,\kms. The average RMS noise per channel across the MAGMA survey
is $0.3$\,K. A CO integrated intensity image constructed from the
MAGMA LMC data is shown in Figure~\ref{fig:maps}[c]. The total
  CO luminosity within the MAGMA data cube is $5.3 \times
  10^{6}$\,\lcou\ \citep{wongetal11}. This is $\sim30$\% larger than
  the total CO flux obtained by the NANTEN survey of the LMC
  \citep{fukuietal08} over the same field-of-view. As noted by
  \citet{wongetal11}, some of this discrepany is due to systematic
  errors in the spectral baselines of the MAGMA cube, which accumulate
  when summing large numbers of noise channels. Using a smoothed (to
  3\farcm0) 3$\sigma$ contour mask to identify regions of significant
  emission in the MAGMA cube yields a total CO flux of $3.2 \times
  10^{6}$\,\lcou, which agrees with the NANTEN measurement to within
  15\%.

%\begin{sidewaystable}
\begin{table*}
\centering
\caption{Survey parameters and global properties of M51, M33 and the LMC}
\label{tbl:galcmp}
\par \addvspace{0.2cm}
\begin{threeparttable}
{\small
\begin{tabular}{@{}lcccccccc}
\hline 
Galaxy & Resolution & Vel. Resolution & Distance & Sensitivity\tnote{a} & Inclination\tnote{b} & Morphology\tnote{c} & Metallicity\tnote{d}     & M$_{\rm B}$\tnote{c} \\
       &   [pc]     & [\kms]          &   [Mpc]    &     & [degrees] (Ref)  &                     & [12 + log(O/H)] (Ref) & [mag]     \\
\hline
LMC                        &  11    &  0.53      &   0.05  & 0.3\,\kkms & 35 (1) &  SB(s)m      &  8.26 (1)   & -18.0  \\
M33                        &  53    &  2.0       &   0.84  & 3.5\,\kkms & 56 (2) &  SA(s)cd     &  8.36 (2)   & -18.9  \\
M51                        &  40    &  5.0       &   7.6   & 0.8\,\kkms & 22 (3) &  SA(s)bc pec &  8.55 (3)   & -20.6  \\
\hline
\end{tabular}}
{\small
\begin{tablenotes}
\item[a]{RMS integrated intensity, assuming a linewidth
    corresponding to three spectral channels. For the corresponding
    mass surface density, these numbers should be multiplied by $4.4$,
    assuming $\xco = 2 \times 10^{20}$\,\xcou\ and a helium
    contribution of 1.36 by mass.}
\item[b]{References for galaxy inclination: (1)
  \citet{vandermarelcioni01} (2) \citet{patureletal03} (\textsc{HyperLeda}) (3) Colombo et al., submitted.}
\item[c]{The reference for galaxy types and magnitudes is
  \citet{devaucouleursetal91}.}
\item[d]{References for characteristic metallicity: (1)
  \citet{marbleetal10} (2) \citet{rosolowskysimon08} (3)
  \citet{moustakasetal10}.}
\end{tablenotes}}
\end{threeparttable}
%\end{sidewaystable}
\end{table*}

%\begin{sidewaystable}
\begin{table*}
\centering
\caption{\small CO Emission Structures Identified in Data Cubes}
\label{tbl:numbercmp}
\par \addvspace{0.2cm}
\begin{threeparttable}
{\small
\begin{tabular}{@{}lc|cc|cc}
\hline 
Galaxy              & Cube   & \multicolumn{2}{c|}{Clouds, $N$ ($L_{\rm CO} [\times 10^{7]}$\,\lcou) } & \multicolumn{2}{c}{Islands, $N$ ($L_{\rm CO} [\times 10^{7]}$\,\lcou)} \\
                    & ($L_{\rm CO} [\times 10^{7]}$\,\lcou)\tnote{a}   & All\tnote{b}      & Resolved   & All & Resolved  \\
\hline
M51                 & 91.8   & 1507 (48.65) & 971 (43.10) & 512 (90.02) & 247 (88.05)     \\
M51 arm+central\tnote{c}     & 68.0   & 1100 (40.69) & 735 (36.60) &   235 (82.10) & 122 (81.44) \\
M51 interarm        & 21.9   &  407 (7.96)  & 236 (6.49)    &  277 (7.92)  & 125 (6.60)   \\
M33                 & 3.2    & 114  (0.86)  & 75  (0.70)   &  88 (0.88)   & 66 (0.78)    \\ 
LMC                 & 0.53\tnote{d}   & 481  (0.24)  & 436 (0.23)  &  285 (0.31)  & 267 (0.31)     \\ 
\hline			
M51                 & 90.7  & 879 (52.46) & 676 (48.37)  & 144 (90.23) & 98 (89.74)  \\
M51 arm+central\tnote{c}     & 66.3  & 519 (40.01) & 417 (37.20) & 45 (83.11)  & 32 (83.00) \\
M51 interarm        & 24.4  & 360 (12.45) & 259 (11.16)  & 99 (7.12)   & 66 (6.74)   \\
M33                 & 3.3   & 58  (0.24)  &  33 (0.20)  & 38 (0.39)   & 15 (0.25)   \\ 
LMC                 & 0.46\tnote{d}  & 41  (0.38)  &  16 (0.24)  & 47 (0.27) & 32 (0.24)  \\ 
\hline
\end{tabular}}
{\small
\begin{tablenotes}
\item[a] CO flux obtained by summing all the emission within the
  spectral line cube.  Values in the upper half of the table are for
  the intrinsic resolution cubes; values in the lower half of the
  table are for the matched resolution cubes.
\item[b] The first value in each column is the number of objects (see
  Section~\ref{sect:cloudidentification}); the value in parentheses is
  the total CO flux that is assigned to the objects. The first column
  (`All') lists all identified objects. The second column (`Resolved')
  lists objects where the size and linewidth measurements can be
  successfully deconvolved. 
\item[c] For both the intrinsic and matched resolution cubes, an
  island decomposition identifies structures that are located across
  the boundary between the arm and interarm region. We classify all
  such islands as belonging to the arm+central environment.
\item[d] Direct summation may not produce a reliable estimate, see
  Section~\ref{sect:datalmc}.
\end{tablenotes}}
\end{threeparttable}
%\end{sidewaystable}
\end{table*}

% reminder about how these LCO values are calculated:
% M33 is cube value, R07 = 1.2e8 Msol, Gratier=1.8e8Msol, Us: 1.4e8Msol for XCO=2.0
% LMC is direct integration of DR1 cube (5.26e6) 
% NANTEN value in MAGMA area quoted by Tony is 3.7e6, matches 80% of NANTEN cube as expected
% M51 is value in cube (not Dario's mask)

%%%%%%%%%%%%%%%%%%%%%%%%%%%%%%
%%%%%%%%%%%%%%%%%%%%%%%%%%%%%%
\section{Cloud Identification}
%%%%%%%%%%%%%%%%%%%%%%%%%%%%%%
%%%%%%%%%%%%%%%%%%%%%%%%%%%%%%
\label{sect:cloudidentification}

\noindent For the identification of significant emission and
decomposition of cloud structures within our CO data cubes, we use the
algorithm presented by \citet[][henceforth RL06]{rosolowskyleroy06},
implemented in \textsc{IDL} as part of the \textsc{CPROPS}
package. \textsc{CPROPS} uses a dilated mask technique to isolate
regions of significant emission within spectral line cubes, and a
modified watershed algorithm to assign the emission into individual
clouds. Moments of the emission along the spatial and spectral axes
are used to determine the size, linewidth and flux of the clouds, and
corrections for the finite sensitivity and instrumental resolution are
applied to the measured cloud properties. Each step of the
\textsc{CPROPS} method is described in detail by RL06.\\

\noindent We adopt the default \textsc{CPROPS} definitions of GMC
properties. The cloud radius is defined as $R = 1.91 \sigma_{\rm R}$\,pc,
where $\sigma_{\rm R}$ is the geometric mean of the second moments of the
emission along the cloud's major and minor axes. The velocity
dispersion $\sigma_{\rm v}$ is the second moment of the emission
distribution along the velocity axis, which for a Gaussian line
profile is related to the FWHM linewidth, $\Delta v$, by $\Delta v =
\sqrt{8 \ln 2}\sigma_{\rm v}$. The CO luminosity of the cloud $L_{\rm
  CO}$ is the emission inside the cloud integrated over position and
velocity, i.e.
\begin{equation} 
L_{\rm CO} \; [\lcou] = D^{2} \left( \frac{\pi}{180 \times 3600} \right)^{2} \Sigma T \delta v \delta x \delta y\; ,
\label{eqn:lcodef}
\end{equation}
\noindent where $D$ is the distance to the galaxy in parsecs, $\delta x$
and $\delta y$ are the spatial dimensions of a pixel in arcseconds,
and $\delta v$ is the width of one channel in \kms. The mass of
molecular gas estimated from the GMC's CO luminosity $M_{\rm CO}$ is
calculated as
\begin{equation}
M_{\rm CO} \; [\msol] \equiv 4.4 \frac{\xco}{2 \times 10^{20} [\xcou]} L_{\rm CO}\; ,
\label{eqn:mcodef}
\end{equation}
\noindent where \xco\ is the assumed CO-to-\mh\ conversion factor, and
a factor of 1.36 is applied to account for the mass contribution of
helium. The fiducial value of \xco\ used by \textsc{CPROPS} is $\xco =
2.0 \times 10^{20}$\,\xcou. The virial mass is estimated as
\begin{equation}
M_{\rm vir} \; [\msol] = 1040 \sigma_{\rm v}^{2}R\; , 
\end{equation}
\noindent which assumes that molecular clouds are spherical with
truncated $\rho \propto r^{-1}$ density profiles
\citep{maclarenetal88}. \textsc{CPROPS} estimates the error
associated with a cloud property measurement using a bootstrapping
method, which is described in section~2.5 of RL06.  \\

\noindent Since molecular clouds exhibit hierarchical structure, it is
difficult to identify a scale that uniquely represents their intrinsic
physical properties. Recent analyses of extragalactic CO datasets have
tended to adopt the recommended \textsc{CPROPS} decomposition
parameters for identifying structures with similar properties as
Galactic GMCs \citep[i.e. spatial sizes greater than $\sim10$\,pc,
  linewidths of several \kms, and brightness temperatures less than
  $\sim10$\,K, e.g. B08,][]{hughesetal10}, but \textsc{CPROPS} offers
several tunable parameters that allow the user to modify the kinds of
emission structures that are identified by the algorithm. For the
comparisons in this paper, we decompose the CO data cubes using two
different approaches:
\begin{enumerate}
\item{ {\it Islands:} \textsc{CPROPS} identifies all contiguous
  regions of significant emission within the cube. Significant
  emission is initially identified by finding pixels with CO
  brightness $T_{\rm mb}$ above a $4\sigma_{RMS}$ threshold across two
  adjacent velocity channels, where the RMS noise $\sigma_{RMS}$ is
  estimated from the median absolute deviation (MAD) of each
  spectrum. The mask is then expanded to include all connected pixels
  with $T_{\rm mb} > 1.5\sigma_{RMS}$. Islands smaller than a
  telescope beam are rejected from the catalogue. }
\item{{\it Clouds:} Islands are further decomposed into emission
  structures that can be uniquely assigned to local maxima that are
  identified within a moving box with dimensions 150\,pc $\times$
  150\,pc $\times$ 15\,\kms. The dimensions of this box are 
  arbitrary: by default, \textsc{CPROPS} uses an $l \times l \times k$
  box, where $l$ and $k$ are defined to be three times the beam and
  channel width respectively. We prefer to adopt a box defined in
  physical space and apply it uniformly to all three datasets. The
  emission associated with a local maximum is required to lie at least
  $2\sigma_{RMS}$ above the merge level with any other maxima, and be
  larger than the telescope beam. We categorize all such emission
  regions as distinct clouds. Contrary to the default parameter
  values, we set the parameter $\textsc{SIGDISCONT} = 0$ so that the
  algorithm makes no attempt to merge the emission associated with
  pairs of local maxima into a single object.}
\end{enumerate}
\noindent For both decomposition approaches, the size, linewidth, and
flux measurements of each object include extrapolation to a
zero-intensity boundary and corrections for the finite spatial and
spectral resolution by deconvolving the spatial beam and channel width
from the measured cloud size and linewidth respectively. For M51, the
cloud decomposition is identical to the method used to construct the
PAWS GMC catalogue, which is presented in Colombo et
al. (submitted). \\

%%%%%%%%%%%%%%%%%%%%%%%%%%%%%%
%%%%%%%%%%%%%%%%%%%%%%%%%%%%%%
\section{Results}
%%%%%%%%%%%%%%%%%%%%%%%%%%%%%%
%%%%%%%%%%%%%%%%%%%%%%%%%%%%%%
\label{sect:results}

\begin{figure*}
\begin{center}
\hspace{-0.5cm}
\includegraphics[width=80mm,angle=270]{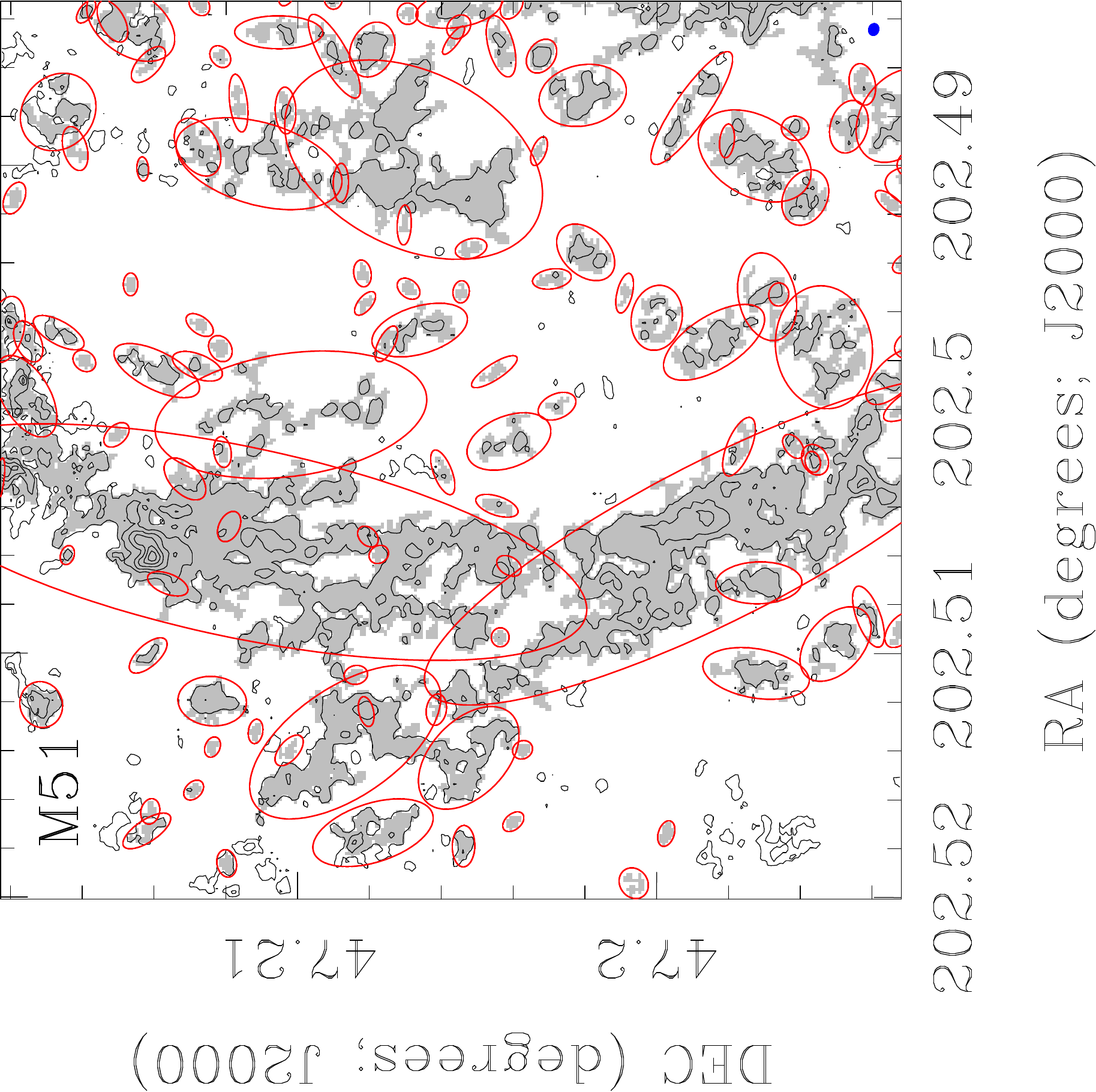}
\hspace{0.5cm}
\includegraphics[width=80mm,angle=270]{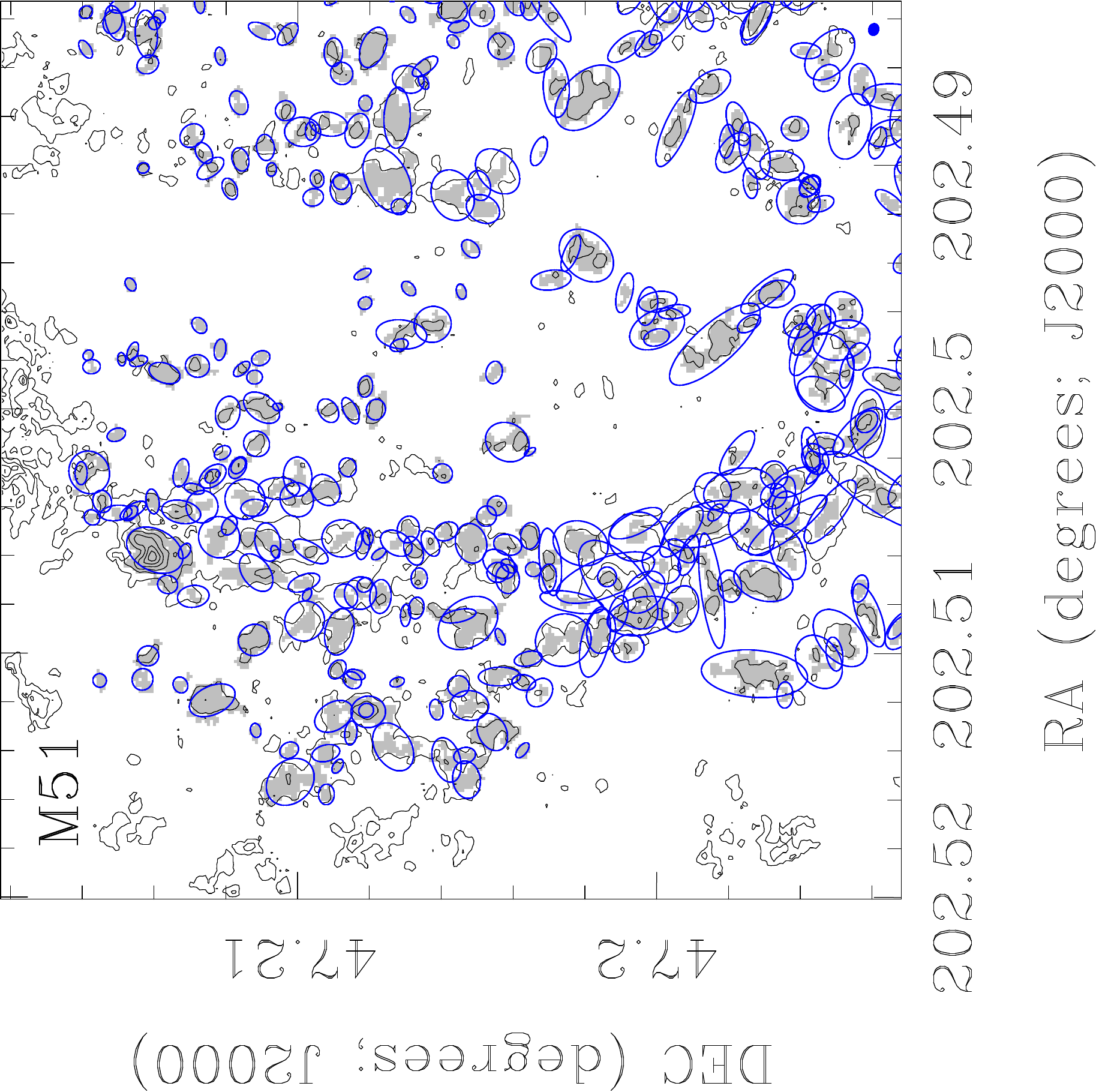}
\caption{\small A map of CO integrated intensity (black contours)
  for a subregion of the PAWS field, with the results of the
  islands (left) and cloud (right) decompositions overlaid. The
  greyscale image indicates emission that has been assigned to a
  catalogued structure. The FWHM major and minor axes and the orientation
  of each structure, as parameterised by \textsc{CPROPS}, are
  indicated by red and blue ellipses. The black contours represent
  steps of 50\,\kkms, with the lowest contour at 20\,\kkms.}
\label{fig:decompresultsM51}
\end{center}
\end{figure*}

\begin{figure*}
\begin{center}
\hspace{-0.5cm}
\includegraphics[width=80mm,angle=270]{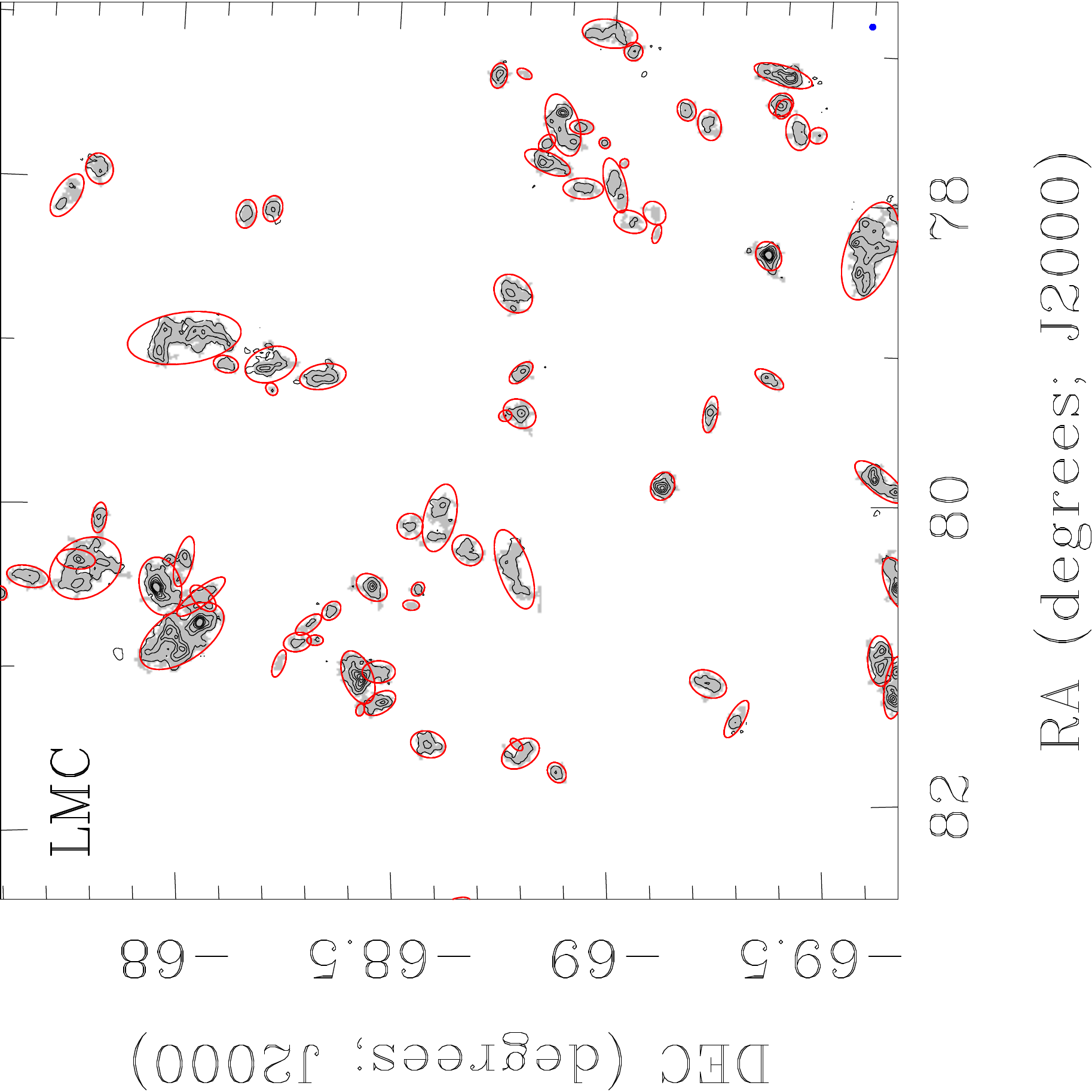}
\hspace{0.5cm}
\includegraphics[width=80mm,angle=270]{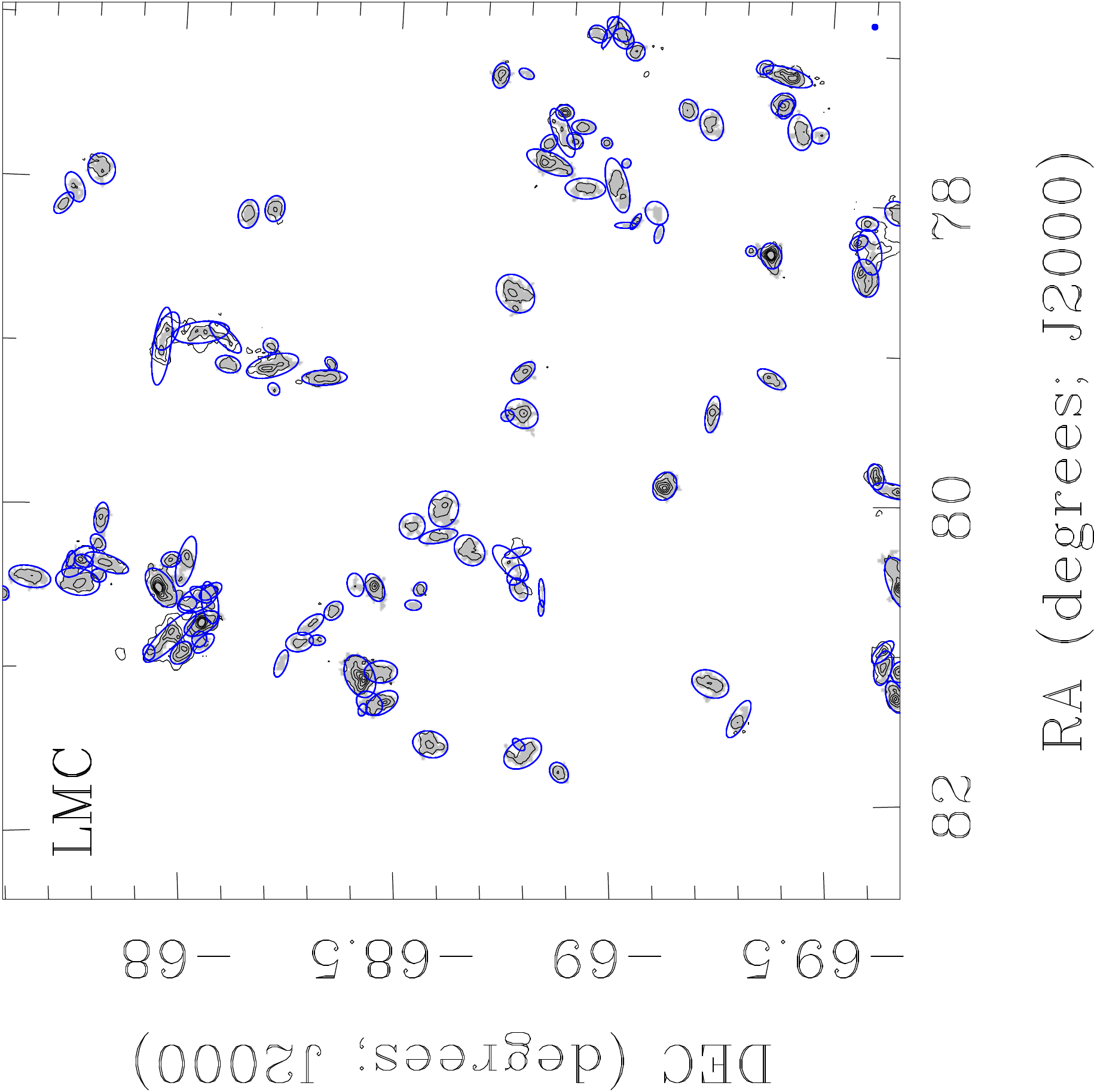}
\caption{\small Same as Figure~\ref{fig:decompresultsM51}, but for the
  LMC. The black contours represent CO integrated intensity in steps
  of 4\,\kkms, with the lowest contour at 1.5\,\kkms.}
\label{fig:decompresultsLMC}
\end{center}
\end{figure*}

\begin{figure*}
\begin{center}
\hspace{-0.5cm}
\includegraphics[width=75mm,angle=270]{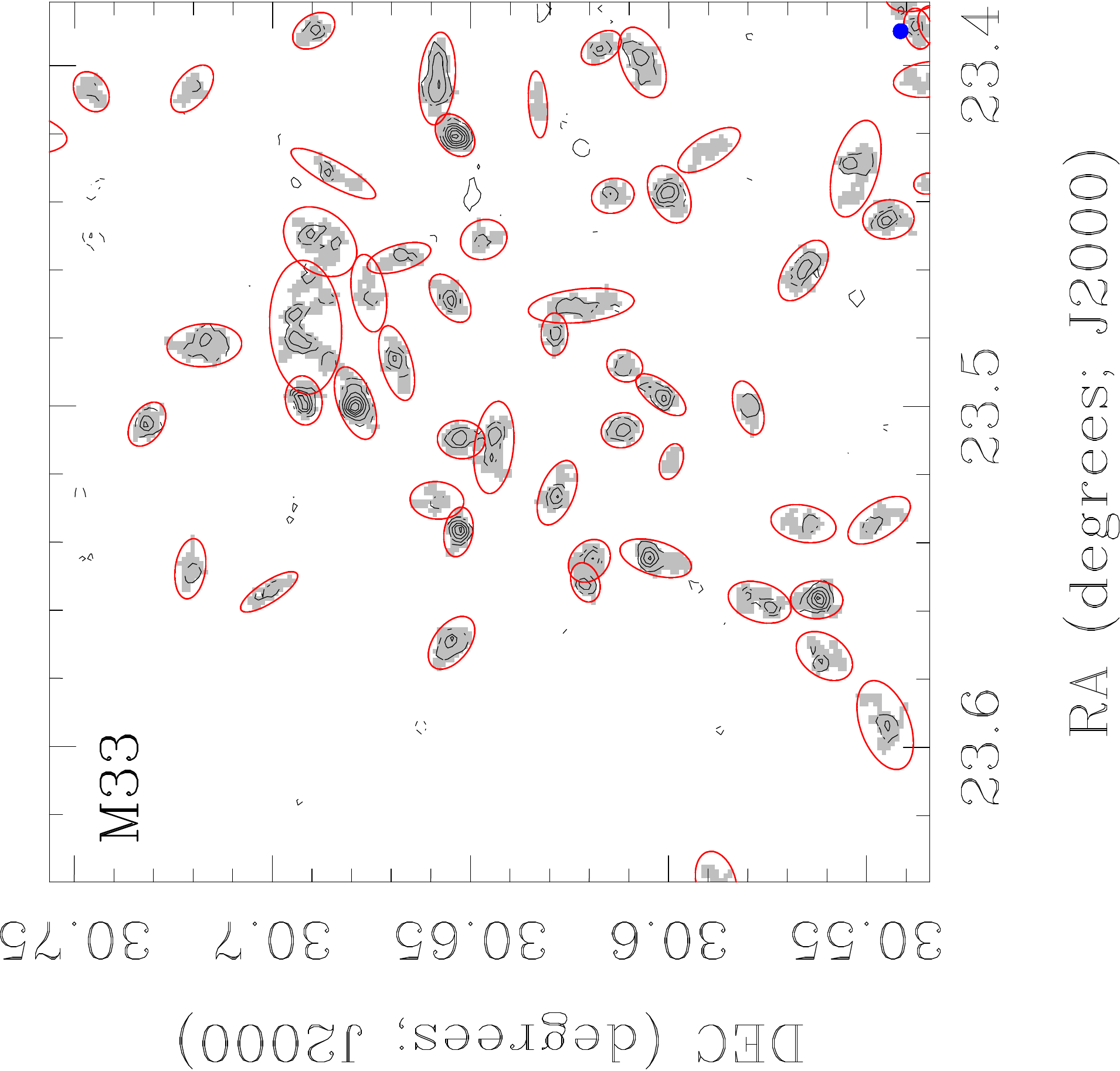}
\hspace{0.5cm}
\includegraphics[width=75mm,angle=270]{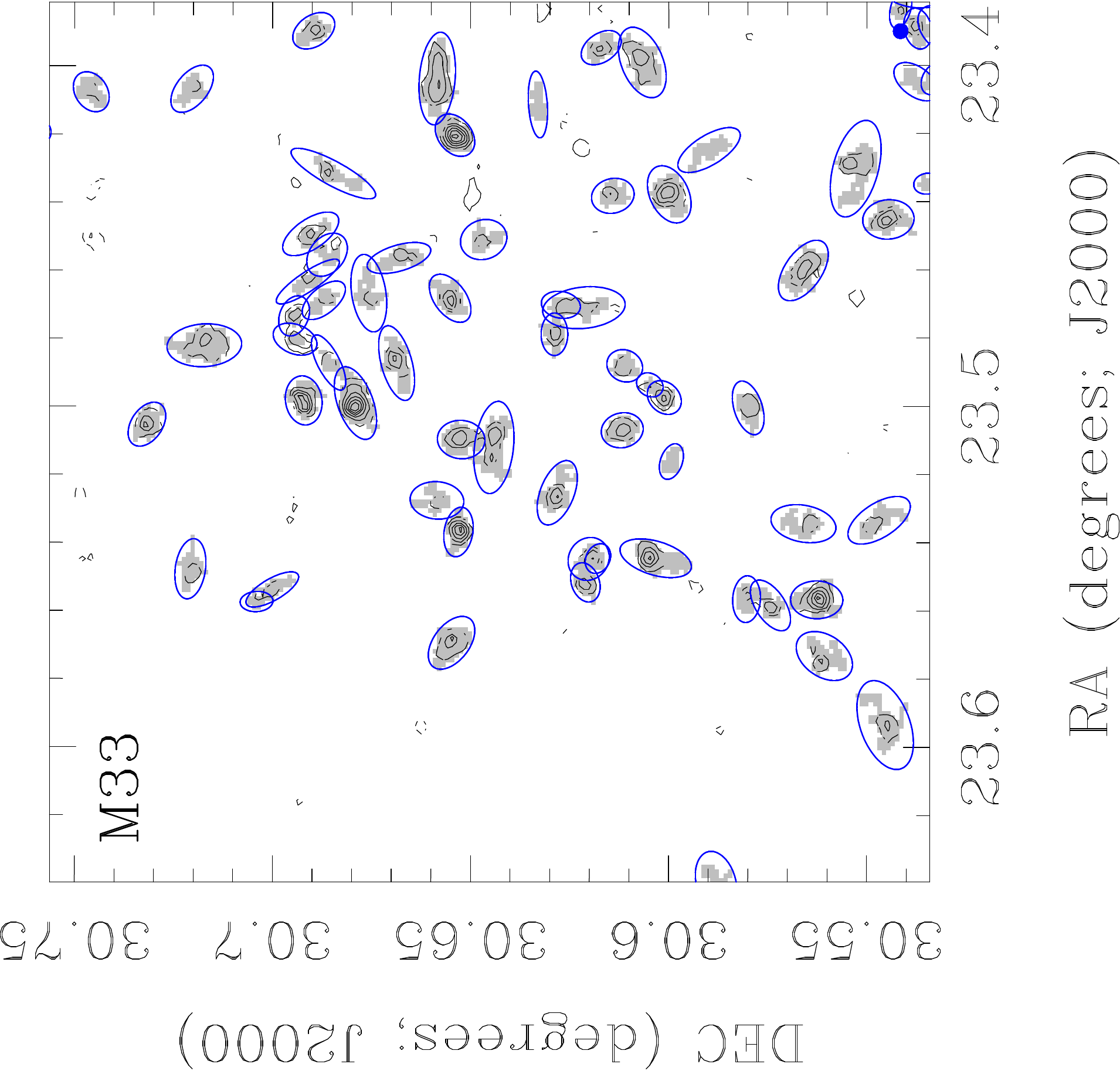}
\caption{\small Same as Figure~\ref{fig:decompresultsM51}, but for
  M33. The black contours represent CO integrated intensity in steps
  of 3\,\kkms, with the lowest contour at 2\,\kkms.}
\label{fig:decompresultsM33}
\end{center}
\end{figure*}

%%%%%%%%%%%%%%%%%%%%%%%%%%%%%%
\subsection{Physical Properties of GMCs}
%%%%%%%%%%%%%%%%%%%%%%%%%%%%%%
\label{sect:gmcproperties}

\noindent In this section, we compare the basic physical properties
(e.g. size, line width, and luminosity) of the cloud structures
identified in M51, M33 and the LMC. One advantage of \textsc{CPROPS}
over other GMC identification algorithms is that it attempts to
correct the cloud property measurements for the finite sensitivity and
resolution of the input dataset, and hence reduce some of the
observational bias that affects comparisons between heterogeneous
datasets. However, resolution and sensitivity should still have
considerable impact on whether emission is detected and considered
significant, so there may still be some residual bias in the \textsc{
  CPROPS} results (see RL06 for further discussion). We therefore
conduct our analysis on two sets of data cubes, an `intrinsic
resolution' set (as described in Section~\ref{sect:data}) and a
`matched resolution' set. In both cases, we work with cloud
  property measurements that have been extrapolated to the limit of
  perfect sensitivity, and deconvolved from the instrumental profile
  (i.e. spatial beam and velocity channel width).\\

\noindent To construct the `matched resolution' dataset, we degraded
the M51 and LMC CO data cubes to the same linear resolution as the M33
cube ($\sim53$\,pc), and folded the M33 and LMC datacubes along the
velocity axis to the same channel width as the M51 cube
($5\,\kms$). We also interpolated the matched cubes onto an ($x,y$)
grid with the same pixel dimensions in physical space ($15 \times
15$\,pc). The matched M33 and M51 cubes have a similar sensitivity
($0.2$\,K per 5\,\kms\ channel), but the LMC cubes are almost an order
of magnitude more sensitive. We tried to match the sensitivity of all
three matched datasets by adding Gaussian noise at the beam scale to
the LMC data, but the emission in the LMC is so faint that
\textsc{CPROPS} did not identify any clouds after the noise was
increased to this level. \\

\noindent In total, 1507 cloud structures are identified in
  the intrinsic resolution M51 data cube; 971 clouds have size and
  linewidth measurements that are deconvolved successfully. These
  clouds have radii between 5 and 150\,pc, velocity dispersions
  between 0.9 and 31\,\kms, and peak brightnesses between 1.2 and
  16.5\,K. For M33 and the LMC, the resolved cloud samples identified
  in the intrinsic resolution datasets contain 75 and 436 objects
  respectively. The M33 clouds have radii between 10 and 100\,pc,
  velocity dispersions between 1.2 and 9\,\kms, and peak brightnesses
  between 0.7 and 2.8\,K, while the LMC clouds have radii between 4 and
  40\,pc, velocity dispersions between 0.4 and 7\,\kms, and peak
  brightnesses between 0.7 and 8.1\,K. The median and median absolute
  deviation (MAD) of the basic physical properties of the clouds
  identified in the intrinsic resolution datasets for all three
  galaxies are listed in the upper half of
  Table~\ref{tbl:cldprops}. We note that the average values of the
  cloud size and velocity dispersion for each galaxy peak around the
  spatial and spectral resolution of the data cubes. This is a
  well-known bias \citep[e.g.][]{verschuur93} that reflects the
  hierarchical structure of the ISM from parsec to kiloparsec
  scales. A peak in the frequency distribution at the resolution limit
  occurs because structures close to the instrumental resolution are
  incompletely sampled, whereas larger structures tend to be resolved
  into smaller objects. As identified in the intrinsic resolution
  cubes, the trend for clouds in the low mass galaxies to be fainter
  and have narrower linewidths than clouds in M51 could therefore be
  mostly due to observational bias. The smaller size and narrower
  linewidth of the LMC clouds, for example, likely reflects the
  superior spatial and spectral resolution of the MAGMA survey, while
  the lower peak brightness of M33 clouds relative to LMC clouds
  probably arises because the CO emission in M33 suffers more strongly
  from dilution within the telescope beam (53\,pc versus 11\,pc, for
  our adopted distances to M33 and the LMC respectively). Due to
  resolution bias, it is thus very difficult to determine whether
  there are significant differences in the cloud populations of the
  three galaxies using the intrinsic resolution cubes. \\

\noindent The rationale for constructing the matched resolution
datacubes is that they allow us to assess whether differences in the
M51, M33 and LMC GMC populations exist, even after suppressing
resolution bias. It is worth noting, however, that the primary
consequence of degrading the M33 and the LMC cubes to a common
resolution is to greatly decrease the number of clouds that are
identified in the low-mass galaxies. In total, 879, 41 and 58 clouds
are identified in the matched resolution cubes for M51, M33 and the
LMC, while the corresponding resolved cloud populations (i.e. where
the size and linewidth can be successfully deconvolved) contain 676,
16, and 33 objects. The `loss' of resolved clouds from the matched
resolution cubes relative to the intrinsic resolution cubes indicates
that most of the CO emission in M33 and the LMC exists in structures
that are spatially compact and/or have narrow linewidths, and hence
diluted in the spatial and/or spectral domain below our detection
threshold. Only the largest and brightest CO clouds in M33 and the LMC
remain detectable in the matched resolution datasets. \\

\begin{figure*}
\begin{center}
\includegraphics[width=110mm,angle=270]{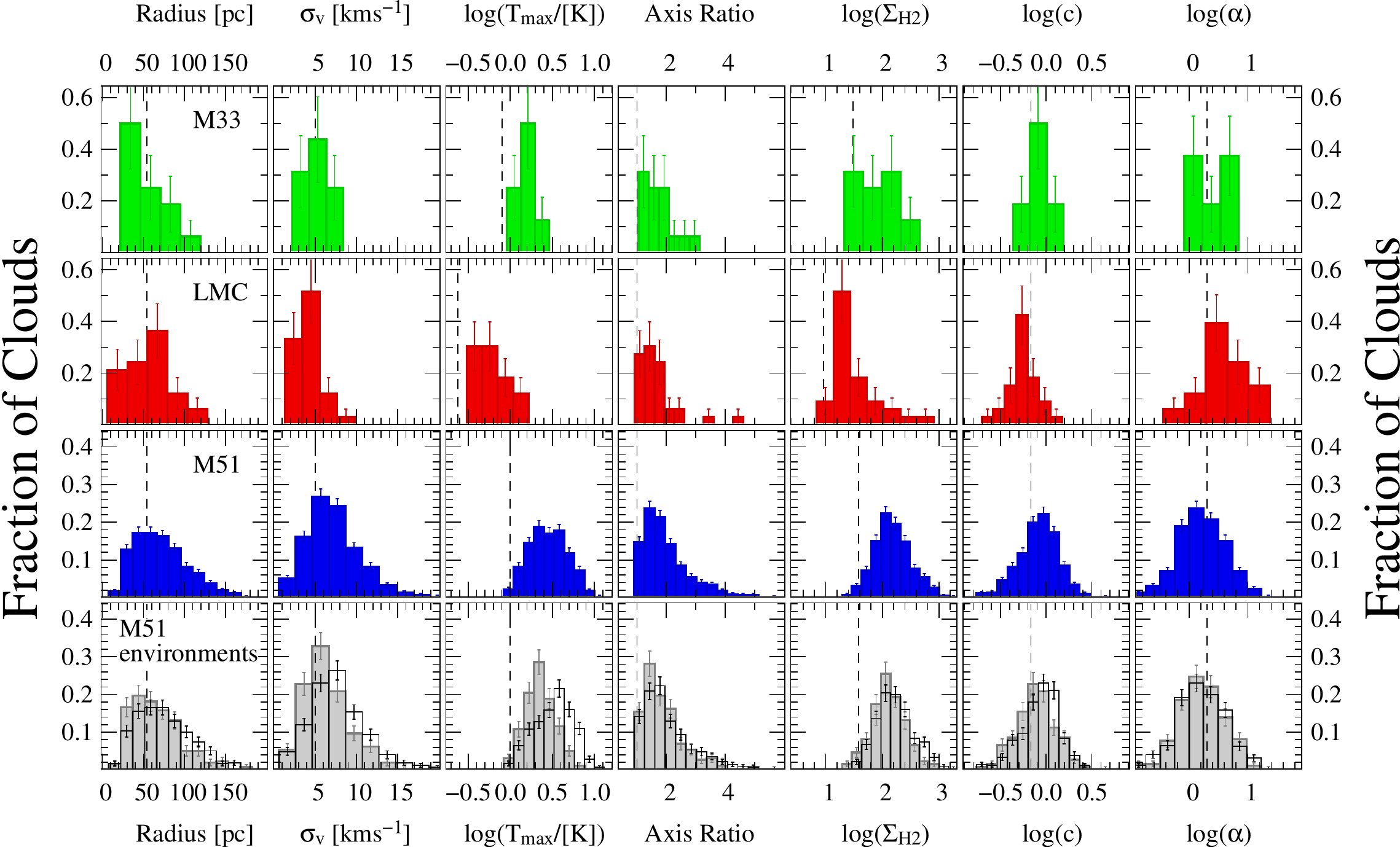}
\caption{\footnotesize Histograms of radius, velocity dispersion, peak
  CO brightness, axis ratio, \hh\ mass surface density $\Sigma_{\rm
    H_{2}}$, the scaling coefficient $c$, and the virial parameter
  $\alpha$ for clouds in the matched resolution datacubes of LMC, M33
  and M51. In the bottom row, we show the distributions for the spiral
  arm and central regions (black) and interarm region (grey) of M51
  separately. The error bars indicate simple counting ($\sqrt{N}$)
  errors. The dashed vertical lines in the first two columns
  correspond to resolution limits. For the third and fifth columns,
  the dashed vertical lines correspond to $4\sigma$ sensitivity
  limits; for the $\Sigma_{\rm H_{2}}$ sensitivity limit, we further
  assume that a cloud spans a minimum of three velocity channels and
  that $\xco = 2 \times 10^{20}$\,\xcou. The dashed vertical lines in
  the fourth column indicates an axis ratio of 1, i.e. a perfectly
  round cloud. The dashed vertical lines in the sixth column indicate
  the value of $c$ that would be expected if clouds have sizes and
  linewidths corresponding to the spatial and spectral resolution. The
  dashed lines in the final column indicate $\alpha \equiv
  5\sigma_{\rm v}^{2}R/GM_{\rm CO} = 2$, which is expected if the
  kinetic and gravitational energies of the cloud are
  balanced. For the LMC and M33 distributions, the width of
    the histogram bins is determined using the Freedman-Diaconis rule,
    i.e. $h = 2{\rm IQR}/n^{1/3}$, where IQR is the interquartile
    range, $n$ is the number of clouds in the sample, and $h$ is the
    bin width \citep{freedmandiaconis81}. We use the same rule to
    obtain a first estimate of the bin width for the M51
    distributions, but we increase the bin width if the increment is
    finer than our measurement uncertainties would allow us to
    determine robustly.}
\label{fig:histos_match_gmcs}
\end{center}
\end{figure*}

\noindent The clouds identified in the M51 matched resolution cube
have radii between 9 and 190\,pc, velocity dispersions between 0.7 and
28\,\kms, and peak brightnesses between 0.8 and 13.4\,K. In the M33
matched resolution cube, the clouds have radii between 25 and 108\,pc,
velocity dispersions between 2.5 and 7.6\,\kms, and peak brightnesses
between 0.9 and 2.6\,K, while in the LMC matched resolution cube, the
clouds have radii between 7 and 116\,pc, velocity dispersions between
1.5 and 8.9\,\kms, and peak brightnesses between 0.3 and 1.6\,K. In
Figure~\ref{fig:histos_match_gmcs}, we plot the distributions of
radius, velocity dispersion, peak CO brightness, mass surface density
(derived from the CO luminosity assuming a Galactic
CO-to-\hh\ conversion factor, $\xco = 2.0 \times 10^{20}$\,\xcou), the
virial parameter $\alpha \equiv 5\sigma_{\rm v}^{2}R/GM_{\rm CO}$, the
scaling coefficient $c \equiv \sigma_{\rm v}/\sqrt{R}$, and the axis
ratio for the cloud populations of each galaxy, derived using the
matched resolution cubes. The median and MAD of each of the cloud
property distributions are listed in the lower half of
Table~\ref{tbl:cldprops}. \\

\noindent We test whether the cloud property distributions are
  similar using a modified version of the two-sided Kolmogorov-Smirnov
  (KS) test that attempts to account for the uncertainties in the
  cloud property measurements. In practice, this involves repeating each KS test 500
  times, sampling the cloud property measurements within their $3\sigma$
  uncertainties using uniform random sampling, rather than only using
  the measurement reported by \textsc{CPROPS}. The results of these
tests are listed in Table~\ref{tbl:kstest}. We tabulate the median $p$
value, which indicates the probability that measurements in two
samples are drawn from the same parent population. We regard median
$p$ values of $\langle p \rangle \leq 0.05$ to indicate that there is
a statistically significant difference between two distributions. \\

%\begin{sidewaystable}
\begin{table*}
\centering
\caption{\small Average properties of the resolved GMC populations in M51, M33 and the LMC.}
\label{tbl:cldprops}
\par \addvspace{0.2cm}
\begin{threeparttable}
{\small
\begin{tabular}{@{}cccccc}
\hline 
Cloud Property\tnote{a}    &  \multicolumn{5}{c}{Galaxy/Region\tnote{b}}   \\ 
                  & M33    &  LMC   & M51 & M51-arm+central  & M51-interarm    \\
\hline
$R [{\rm pc}]$                                   & $51\pm13$  & $16\pm5$  & $48\pm14$  & $49\pm14$  & $45\pm14$    \\
$\sigma_{\rm v} [\kms]$                           & $3.8\pm0.7$  & $1.6\pm0.4$  & $6.4\pm1.8$ & $6.6\pm1.9$  & $5.5\pm1.5$   \\
$T_{\rm peak} [{\rm K}]$                          & $1.3\pm0.2$  & $2.0\pm0.5$  & $3.5\pm1.2$  & $3.9\pm1.4$  & $3.0\pm0.7$  \\
$c \equiv \sigma_{\rm v}/\sqrt{R} [\kms {\rm pc}^{-0.5}]$ & $0.5\pm0.1$  & $0.4\pm0.1$ & $0.9\pm0.3$ & $1.0\pm0.3$  & $0.8\pm0.2$  \\
$\Sigma_{\rm H_{2}} [\mpcsq]$                     & $46\pm20$  & $21\pm9$  & $180\pm82$  & $196\pm90$ & $141\pm53$    \\
$\alpha \equiv \sigma_{\rm v}^{2}R/5GM$          & $2.1\pm0.9$ & $2.8\pm1.2$  & $1.6\pm0.9$ & $1.6\pm0.9$ & $1.6\pm0.9$ \\
Axis Ratio                                     & $1.7\pm0.3$  & $1.6\pm0.3$ & $1.7\pm0.4$ & $1.7\pm0.4$ & $1.6\pm0.4$ \\
\hline
$R [{\rm pc}]$                                    & $46\pm16$  & $57\pm19$ & $67\pm22$  & $71\pm24$  & $60\pm19$ \\
$\sigma_{\rm v} [\kms]$                            & $5.4\pm1.1$  & $4.0\pm0.9$  & $6.8\pm2.0$  & $7.4\pm1.9$  & $5.8\pm1.5$  \\
$T_{\rm peak} [{\rm K}]$                           & $1.7\pm0.3$  & $0.5\pm0.2$ & $2.8\pm1.0$  & $3.5\pm1.3$  & $2.3\pm0.6$  \\
$c \equiv \sigma_{\rm v}/\sqrt{R} [\kms {\rm pc}^{-0.5}]$ & $0.7\pm0.2$  & $0.5\pm0.2$  & $0.9\pm0.3$  & $0.9\pm0.3$  & $0.8\pm0.2$   \\
$\Sigma_{\rm H_{2}} [\mpcsq]$\tnote{c}                     & $86\pm44$   & $22\pm7$  & $145\pm66$  & $167\pm77$  & $122\pm49$   \\
$\Sigma_{\rm H_{2}} [\mpcsq]$\tnote{d}                     & $124\pm64$ & $34\pm11$  & $116\pm63$  & $134\pm62$  & $98\pm39$\\
$\Sigma_{\rm H_{2}} [\mpcsq]$\tnote{e}                     & $62\pm9$    & $30\pm5$  & $98\pm41$  & $110\pm50$  & $79\pm27$ \\
$\alpha \equiv \sigma_{\rm v}^{2}R/5GM$          & $2.9\pm1.3$  & $3.1\pm2.0$ & $1.6\pm0.9$  & $1.6\pm0.9$  & $1.6\pm0.9$   \\
Axis Ratio                                     &  $1.6\pm0.3$ & $1.5\pm0.3$ & $1.8\pm0.4$ & $1.8\pm0.5$ & $1.7\pm0.4$   \\
\hline
\end{tabular}}
{\small
\begin{tablenotes}

\item[a] Properties were obtained using a cloud-based decomposition
  (see Section~\ref{sect:cloudidentification}). The upper half of the
  table refers to properties derived from the data cubes at their
  intrinsic resolution; the results in the lower section refer to the
  matched cubes.
\item[b] We list the median and median absolute deviation (MAD) of the
  cloud properties for each region. The tabulated values are for resolved
  clouds. 
\item[c] Assuming $\xco = 2 \times 10^{20}$\,\xcou\ for each population.
\item[d] Assuming a galaxy-dependent \xco\ value, such that the median
  virial parameter for each galaxy is $\langle \alpha \rangle = 2$.
\item[e] Assuming a galaxy-dependent \xco\ value, but for large clouds ($R>50$\,pc) only.
\end{tablenotes}}
\end{threeparttable}
%\end{sidewaystable}
\end{table*}

%\begin{sidewaystable}
\begin{table*}
\centering
\caption{\small Results for Kolomogorov-Smirnov tests.}
\label{tbl:kstest}
\par \addvspace{0.2cm}
\begin{threeparttable}
{\small
\begin{tabular}{@{}lccccccccc}
\hline 
Galaxy/Region\tnote{a} & $R$ & $\sigma_{\rm v}$ & $T_{\rm peak}$ & $c \equiv \sigma_{\rm v}/\sqrt{R}$ & $\Sigma_{\rm H_{2}}$\tnote{b}  & $\Sigma_{\rm H_{2}}$\tnote{c} & $\Sigma_{\rm H_{2}}$\tnote{d} & $\alpha \equiv \sigma_{\rm v}^{2}R/5GM$ & Axis Ratio \\
\hline
LMC - M33                      & 0.73       & 0.07     & $<0.01$  & 0.10       & $<0.01$     & $<0.01$  & $<0.01$     & 0.15     & 0.61  \\
LMC - M51                      & 0.02       & $<0.01$  & $<0.01$  & $<0.01$    & $<0.01$     & $<0.01$  & $<0.01$    & $<0.01$  & 0.07  \\
LMC - M51-arm+central          & $<0.01$    & $<0.01$  & $<0.01$  & $<0.01$    & $<0.01$     & $<0.01$  & $<0.01$    & $<0.01$  & 0.04  \\
LMC - M51-interarm             & 0.14       & $<0.01$  & $<0.01$  & $<0.01$    & $<0.01$     & $<0.01$  & $<0.01$    & $<0.01$  & 0.19  \\
M33 - M51                      & 0.06       & 0.02     & $<0.01$  & 0.39        & 0.10        & 0.48        & 0.21     & 0.15       & 0.25  \\
M33 - M51-arm+central          & 0.03       & $<0.01$  & $<0.01$  & 0.24        & 0.05        & 0.42        & 0.13     & 0.03        & 0.16  \\
M33 - M51-interarm             & 0.16       & 0.15     & $<0.01$  & 0.53        & 0.20        & 0.43        & 0.44     & 0.09        & 0.39  \\
M51-arm+central - M51-interarm & $<0.01$    & $<0.01$  & $<0.01$  & 0.03       &$<0.01$       &$<0.01$      & $<0.01$  & 0.19       & 0.10  \\
\hline
 \end{tabular}}
{\small
\begin{tablenotes}
 \item[a] Properties were obtained using a cloud-based decomposition
   of the matched resolution cubes (see
   Section~\ref{sect:cloudidentification}). Only resolved clouds
   (i.e. where the size and linewidth measurements can be successfully
   deconvolved) are included in the comparison. We tabulate the median
   $p$-value from 500 repeats of the KS test, where we uniformly
   sample the cloud property measurements within their $3\sigma$
   uncertainties.
\item[b] Assuming $\xco = 2 \times 10^{20}$\,\xcou\ for each population.
\item[c] Assuming a galaxy-dependent \xco\ value, such that the median
  virial parameter for each galaxy is $\langle \alpha \rangle = 2$.
\item[d] Assuming a galaxy-dependent \xco\ value, but for large
  $R>50$\,pc clouds only.
\end{tablenotes}}
\end{threeparttable}
\end{table*}
%\end{sidewaystable}

\noindent The KS tests indicate differences in the size and linewidth
for clouds in the low-mass galaxies compared to the spiral arm and
central regions of M51: on average, clouds in the spiral arms and
central region of M51 are larger and have higher velocity dispersions
than clouds in M33 and the LMC. This is not just a resolution effect,
since these differences are detected in the matched resolution cubes
(and, as noted above, the bulk of the cloud population in M33 and the
LMC have such small sizes and narrow linewidths that they are not
detected in the matched resolution cubes). In the case of the LMC, the
tendency for the size and linewidth distributions to extend to low
values may be partially due to the MAGMA survey's higher sensitivity,
which allows us to recover a greater proportion of clouds that are
small and/or have narrow linewidths. The matched cubes for M33 and M51
have almost identical sensitivity, however, so the differences in the
size and linewidth distributions for these galaxies are likely to be
physical. \\

\noindent Cloud properties related to CO brightness, such as $T_{\rm
  peak}$ and $\Sigma_{\rm H_{2}}$, also vary between the three
galaxies. The average peak CO brightness of clouds in M51 is
significantly higher than in the other galaxies. Assuming that the
CO-to-\mh\ conversion factor \xco\ does not vary significantly between
the different galaxies, this implies that the mass surface density of
a typical cloud in M51 is higher than in M33 and the LMC by a factor
of a few. We discuss the assumption of a constant \xco\ factor in
Section~\ref{sect:prev_results}. Once again, the lower average values
of $T_{\rm peak}$ and $\Sigma_{\rm H_{2}}$ for the LMC clouds are
partly due to the higher sensitivity of the LMC data, which allows us
to detect a greater fraction of faint clouds. However, there are no
well-resolved ($R>50$\,pc) clouds in the matched LMC cubes with
$\Sigma_{\rm H_{2}} > 50$\,\mpcsq, even though such clouds would have
easily been detected by MAGMA if present. By contrast, 90\% of clouds
with $R>50$\,pc in M51 have $\Sigma_{\rm H_{2}} > 50$\,\mpcsq. This
suggests that differences in the $T_{\rm peak}$ and $\Sigma_{\rm
  H_{2}}$ distributions for the LMC and M51 cloud populations would
remain even if we had a more sensitive CO survey of M51. Finally,
there appear to be genuine differences between the properties of
clouds in different M51 environments: clouds in the interarm region
tend to be fainter, and have narrower velocity dispersions than clouds
in the spiral arms and central region. A detailed comparison between
the properties of clouds within different M51 environments is
presented elsewhere (Colombo et al., submitted). \\

\noindent In summary, our analysis suggests that the properties of
GMCs are not the same across different galactic environments. More
precisely, clouds in the spiral arm and central region of M51 tend to
be larger, brighter and have larger velocity dispersions than the
clouds in M33 and the LMC. Clouds in the interarm region of M51 tend
to be more similar to clouds in the low-mass galaxies.  These
conclusions hold even after matching the spatial and spectral
resolution of the input datacubes and using the same methods to
identify and decompose the CO emission into cloud structures.

%%%%%%%%%%%%%%%%%%%%%%%%%%%%%%
\subsection{Scaling Relations}
%%%%%%%%%%%%%%%%%%%%%%%%%%%%%%
\label{sect:llaws}

\begin{figure*}
\begin{center}
\hspace{-0.5cm}
\includegraphics[width=70mm,angle=0]{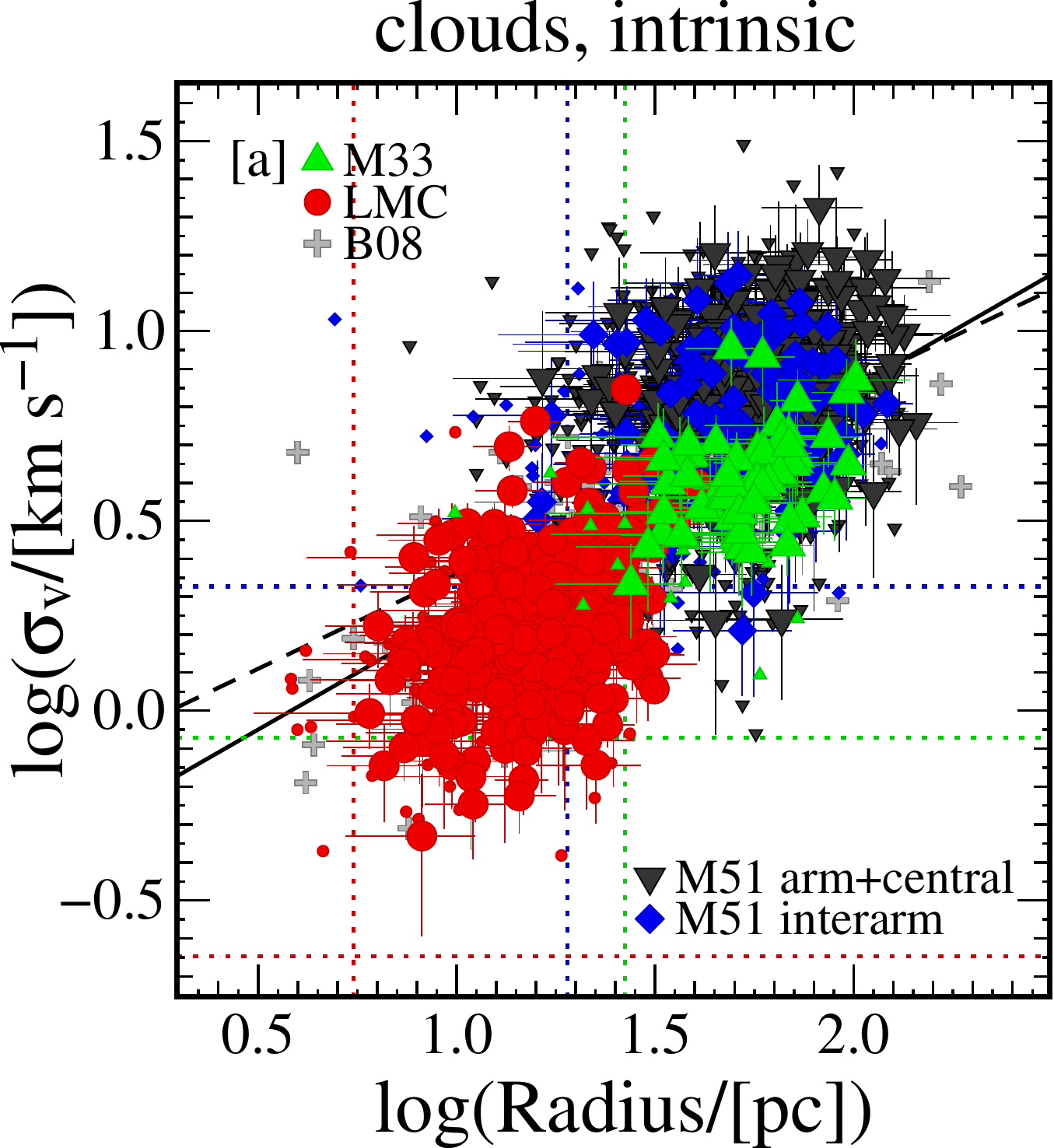}
\hspace{0.5cm}
\includegraphics[width=70mm,angle=0]{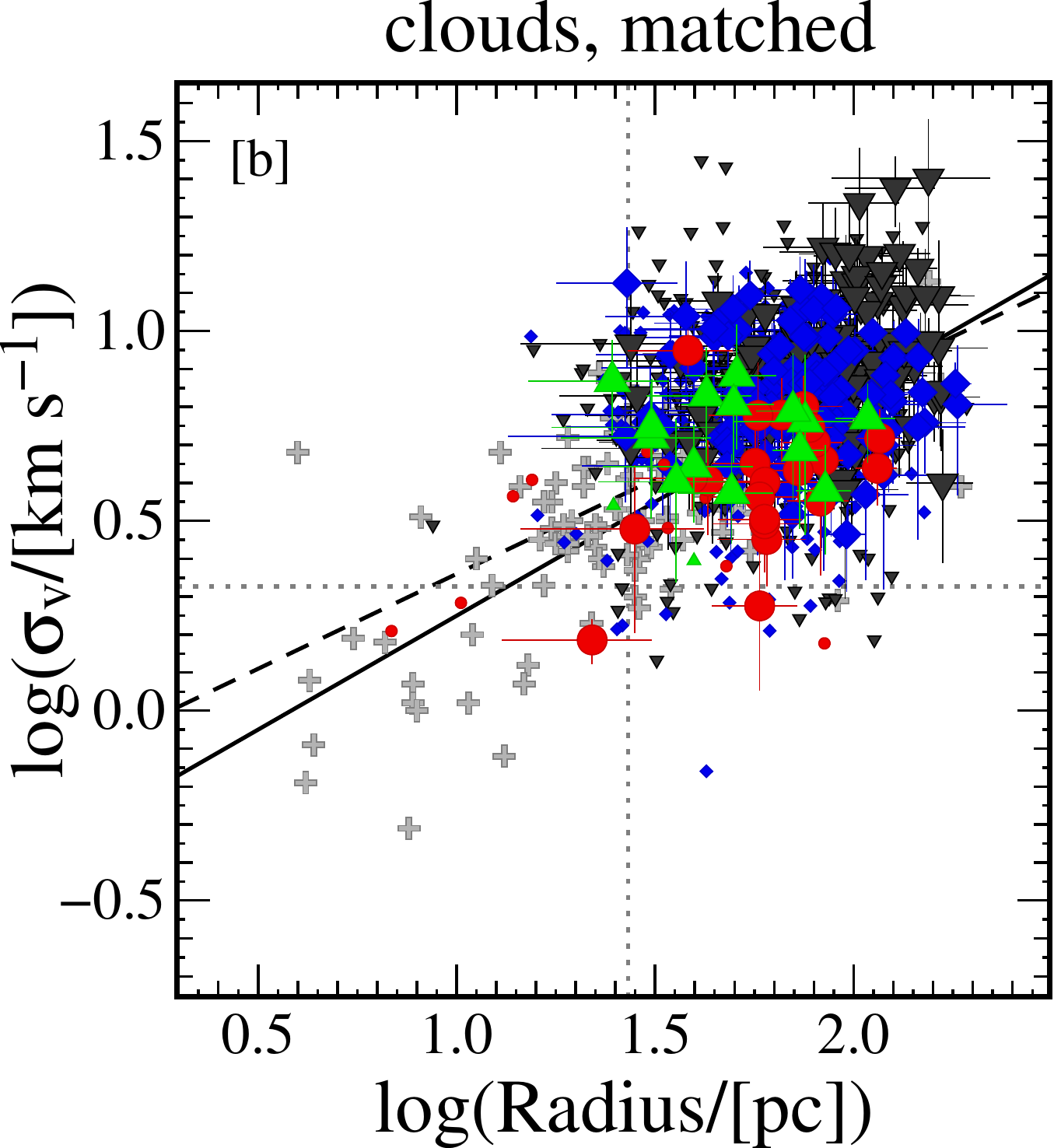}
\par \addvspace{0.5cm}
\hspace{-0.5cm}
\includegraphics[width=70mm,angle=0]{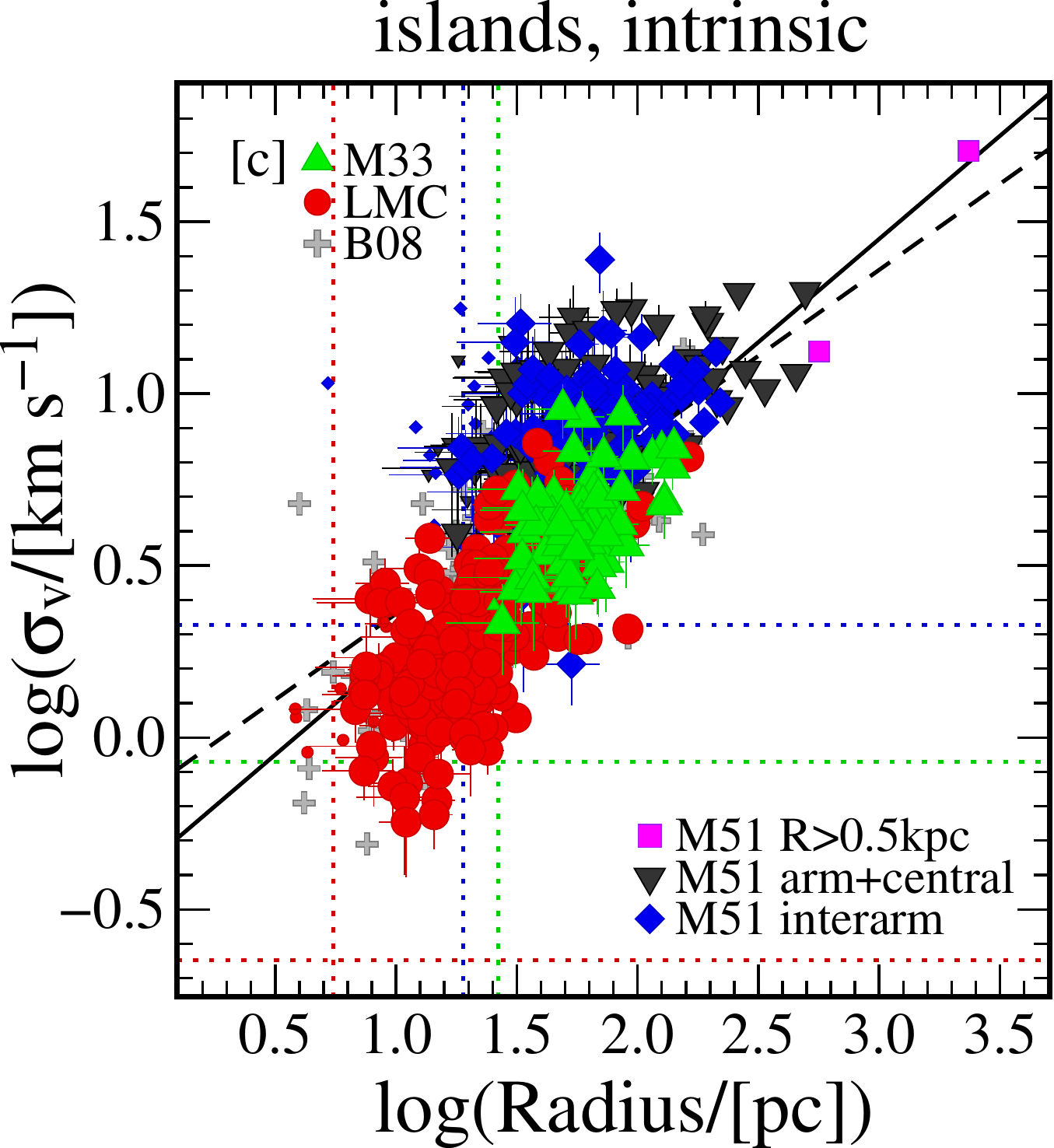}
\hspace{0.5cm}
\includegraphics[width=70mm,angle=0]{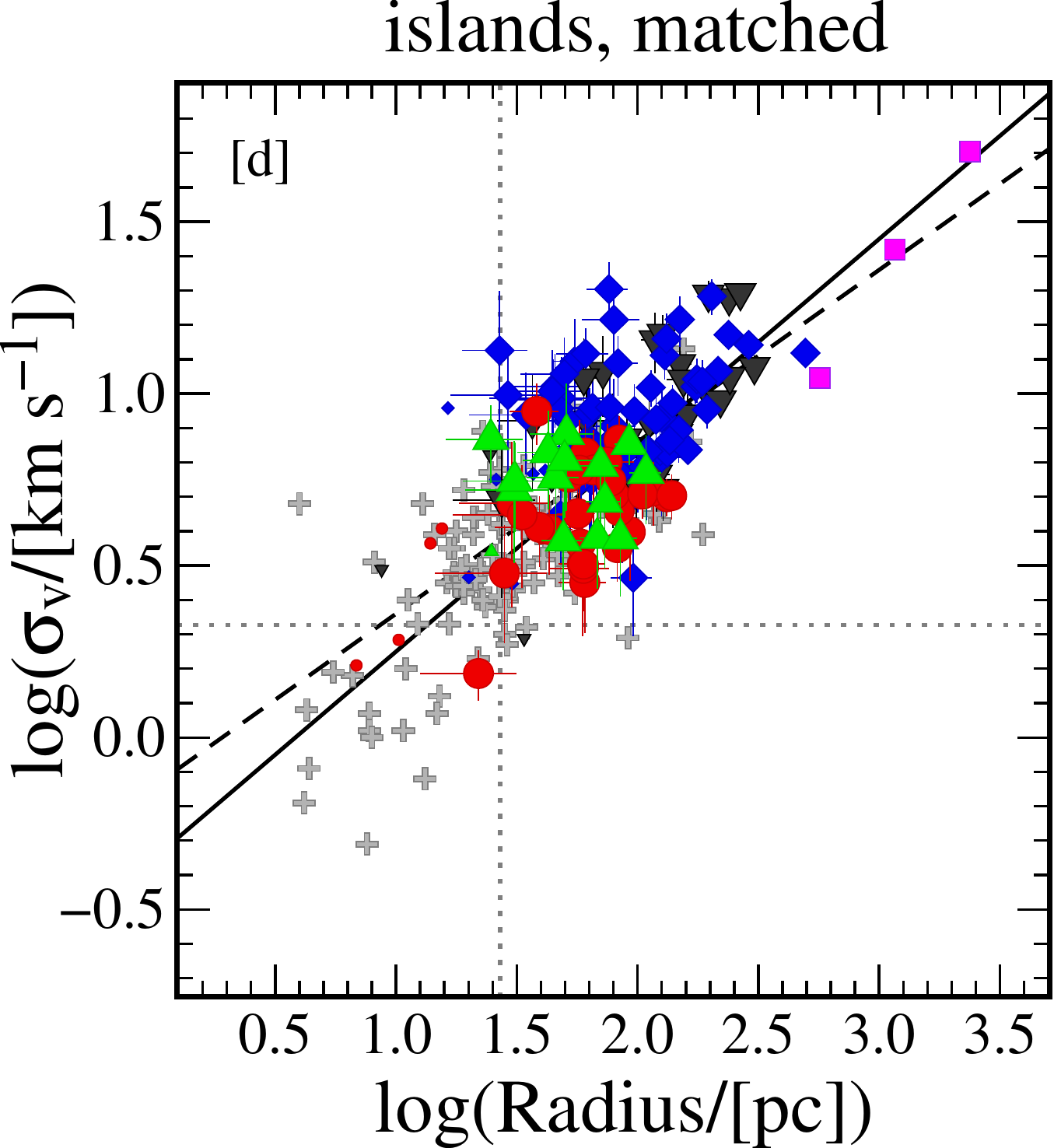}
\caption{\small A plot of radius versus velocity dispersion for
  objects identified within the CO datacubes for M51, M33, the
  LMC. Large symbols with error bars indicate resolved clouds where
  the relative uncertainty in both the size and linewidth measurements
  is less than 50\%; small symbols indicate resolved clouds with
  larger measurement uncertainties. We omit the error bars on the
  small symbols for clarity. In panel [a], we plot the relation for
  clouds identified in the cubes at their intrinsic resolution. The
  relation for clouds identified in the matched resolution cubes is
  shown in panel [b]. The relation for island structures are shown in
  panels [c] and [d] for the decompositions of the original and
  matched datacubes respectively. Islands with radii greater than
  $0.5$\,kpc (i.e. objects that are much larger than GMCs) are
  indicated by magenta squares in panels [c] and [d]. In all
  panels, the black dashed line indicates the relationship derived
  from the S87 inner Milky Way data, and the black solid line
  indicates the best-fitting relation for extragalactic GMCs
  determined by B08. The horizontal and vertical dotted lines indicate
  the radius and velocity dispersion corresponding to the spatial and
  spectral resolution of each survey. The sample of extragalactic GMCs
  analysed by B08 is indicated in each panel by grey crosses. }
\label{fig:rdv}
\end{center}
\end{figure*}

\begin{figure*}
\begin{center}
\hspace{-0.5cm}
\includegraphics[width=70mm,angle=0]{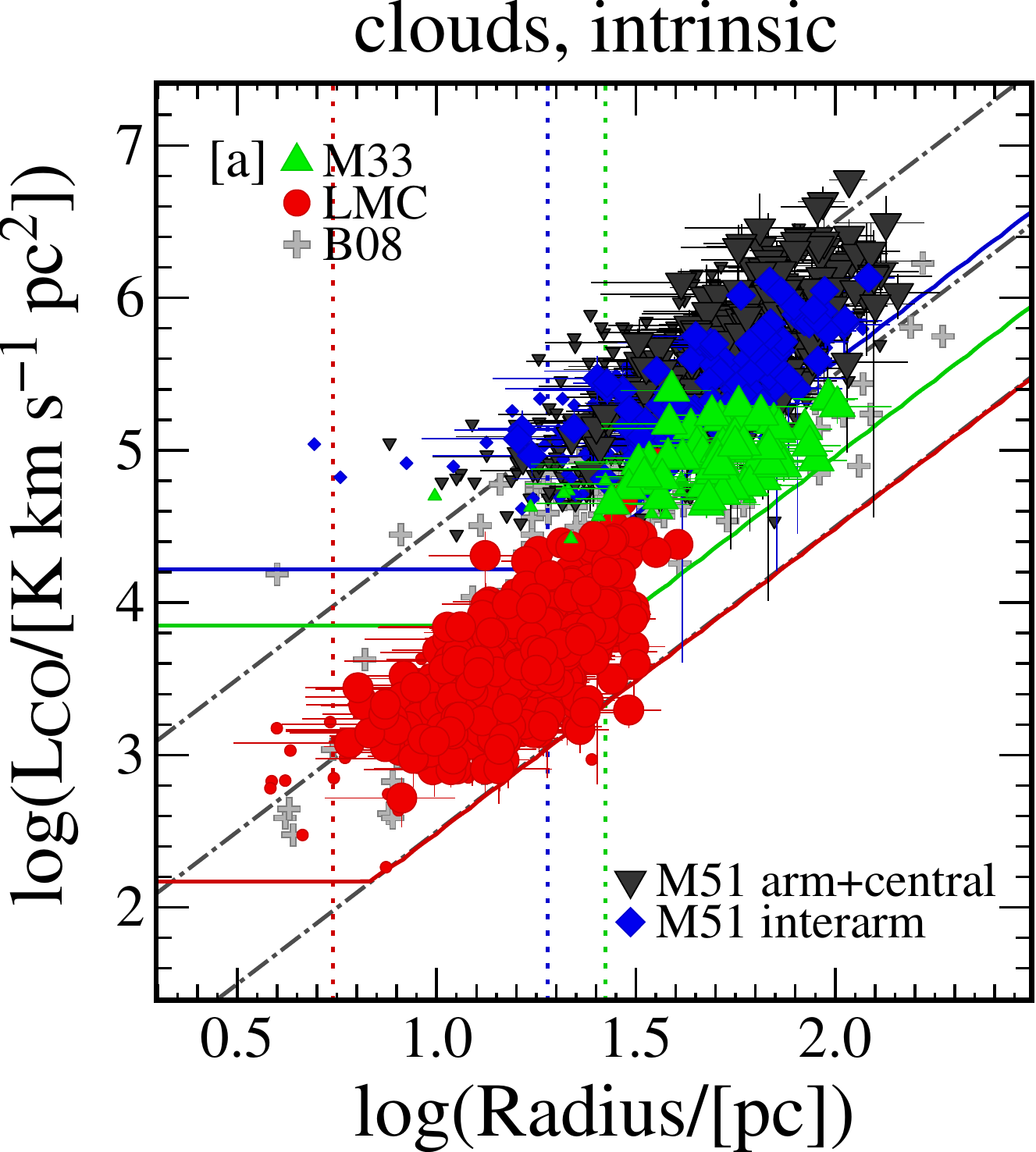}
\hspace{0.5cm}
\includegraphics[width=70mm,angle=0]{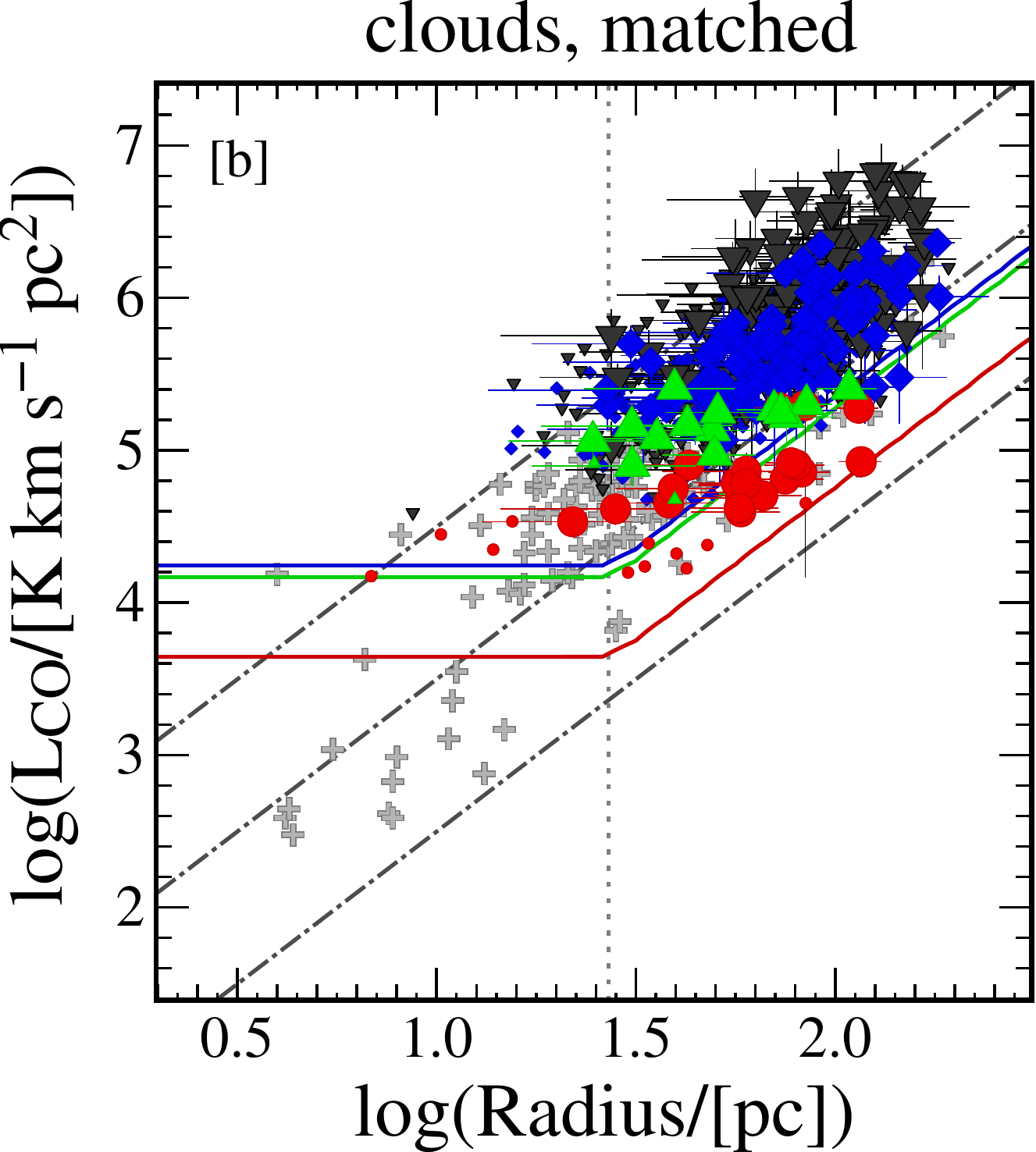}
\par \addvspace{0.5cm}
\hspace{-0.5cm}
\includegraphics[width=70mm,angle=0]{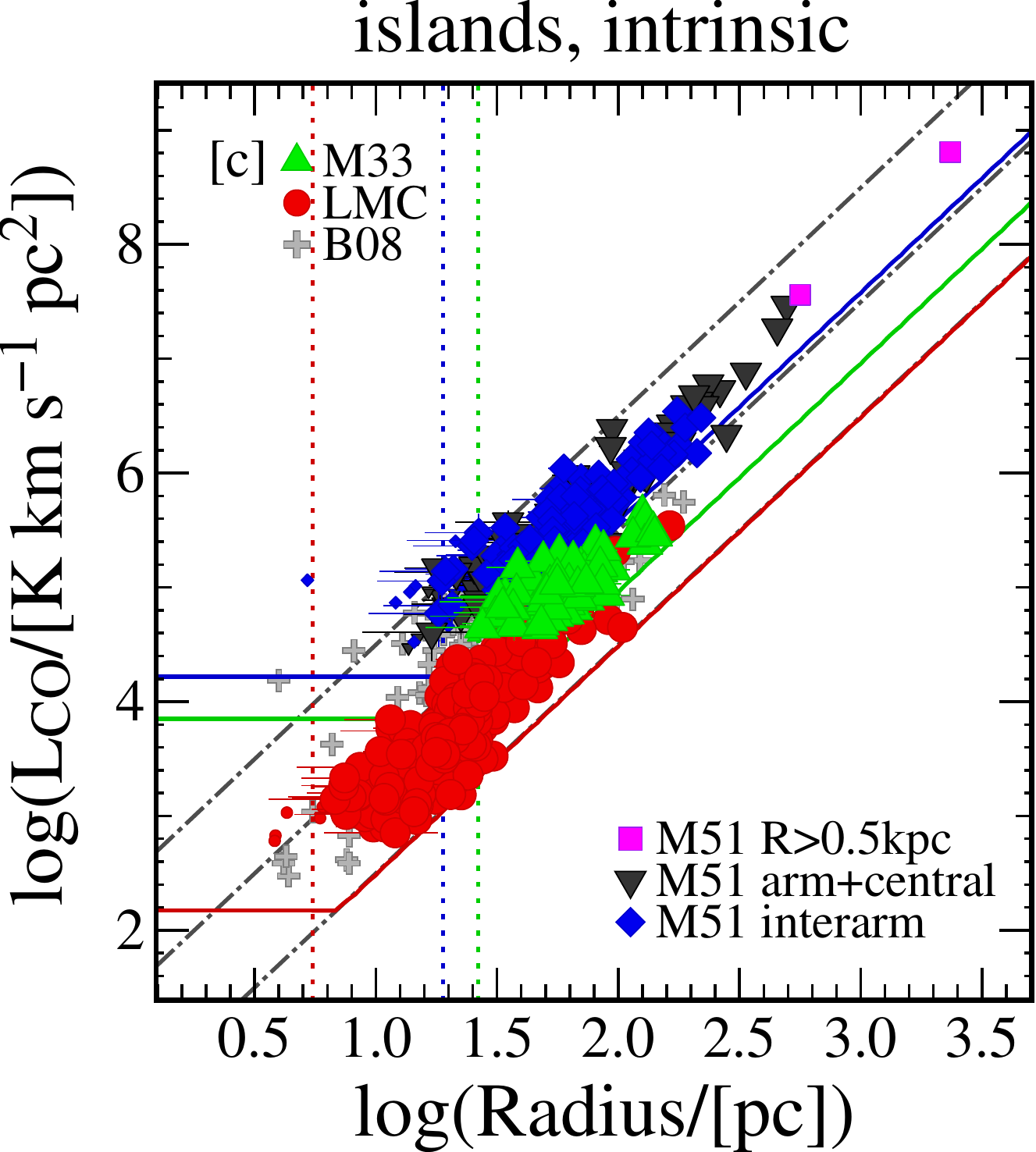}
\hspace{0.5cm}
\includegraphics[width=70mm,angle=0]{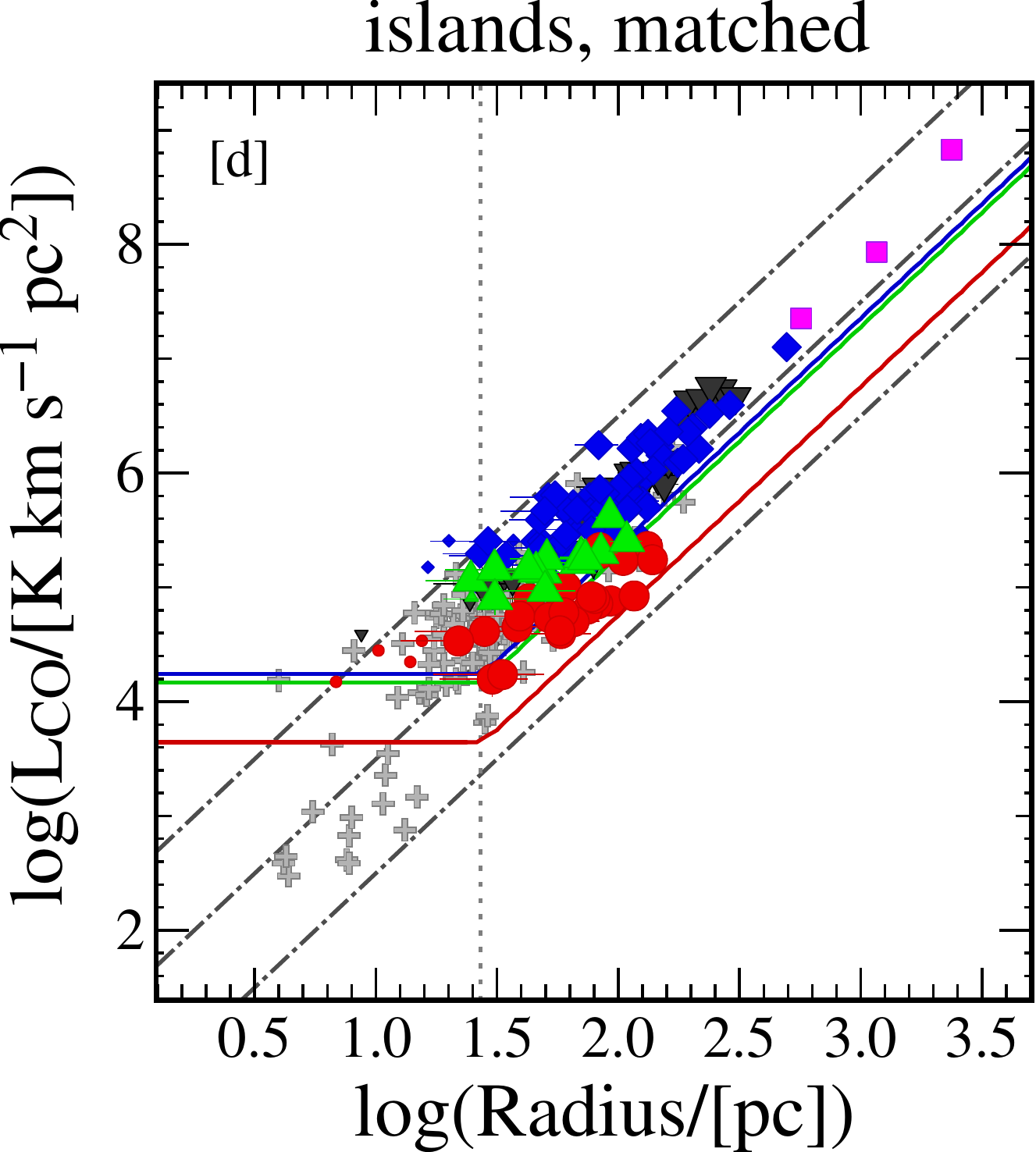}
\caption{\small A plot of radius versus CO luminosity for objects
  identified within the M51, M33, and LMC data cubes. The panels and
  plot symbols are the same as in Figure~\ref{fig:rdv}. The grey
  dot-dashed lines indicate constant values of CO surface brightness:
  $\ico = 1, 10, 100\,\kkms$, which correspond to \hh\ mass surface
  densities of $\Sigma_{\rm H_{2}} = 4, 44, 440$\,\mpcsq\ for $\xco =
  2.0 \times 10^{20}$\,\xcou. The solid coloured curves represent an
  estimate for the $3\sigma$ sensitivity limits of each survey (see
  also Section~\ref{sect:data}).}
\label{fig:rl}
\end{center}
\end{figure*}

\begin{figure*}
\begin{center}
\hspace{-0.5cm}
\includegraphics[width=70mm,angle=0]{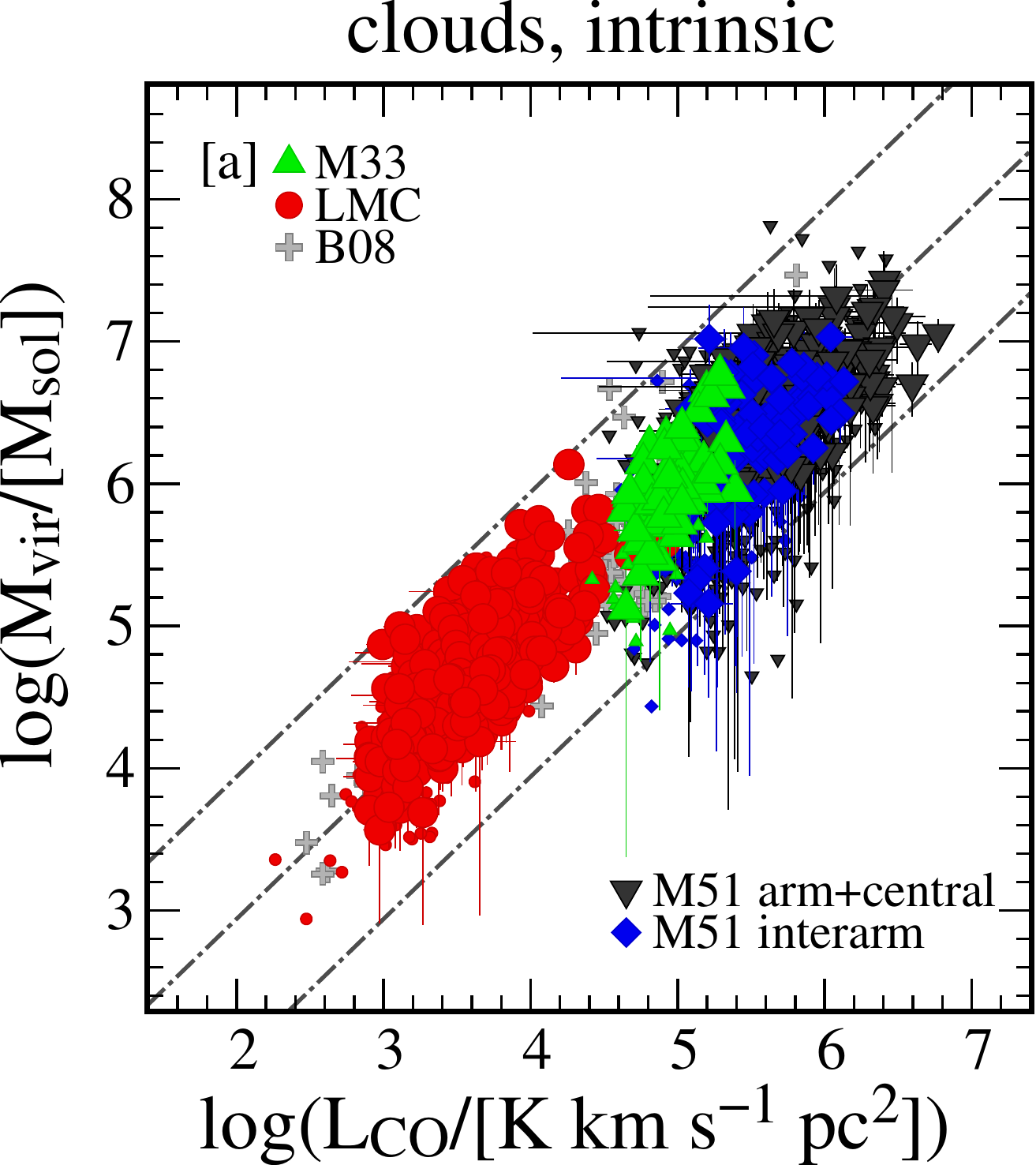}
\hspace{0.5cm}
\includegraphics[width=70mm,angle=0]{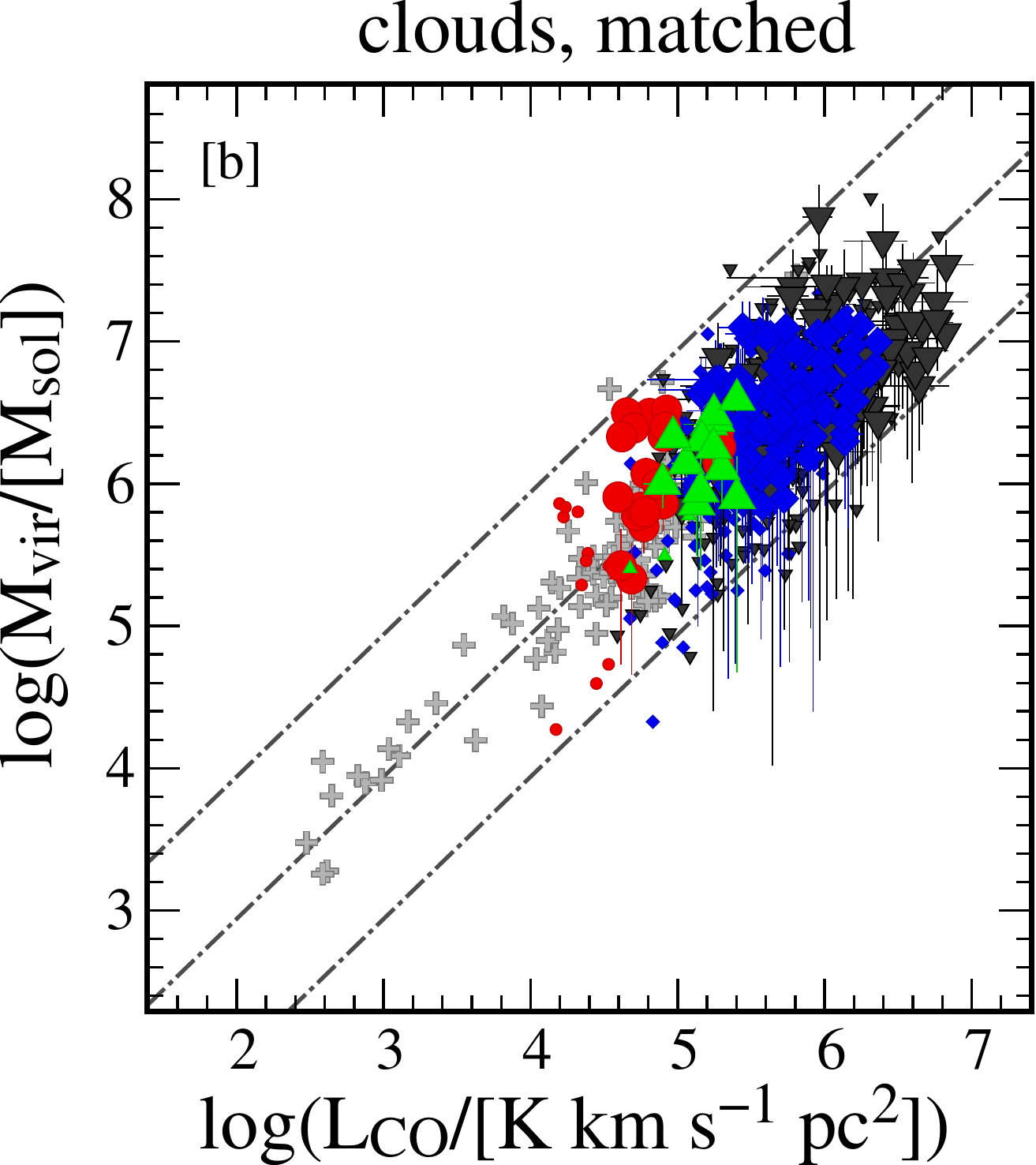}
\par \addvspace{0.5cm}
\hspace{-0.5cm}
\includegraphics[width=70mm,angle=0]{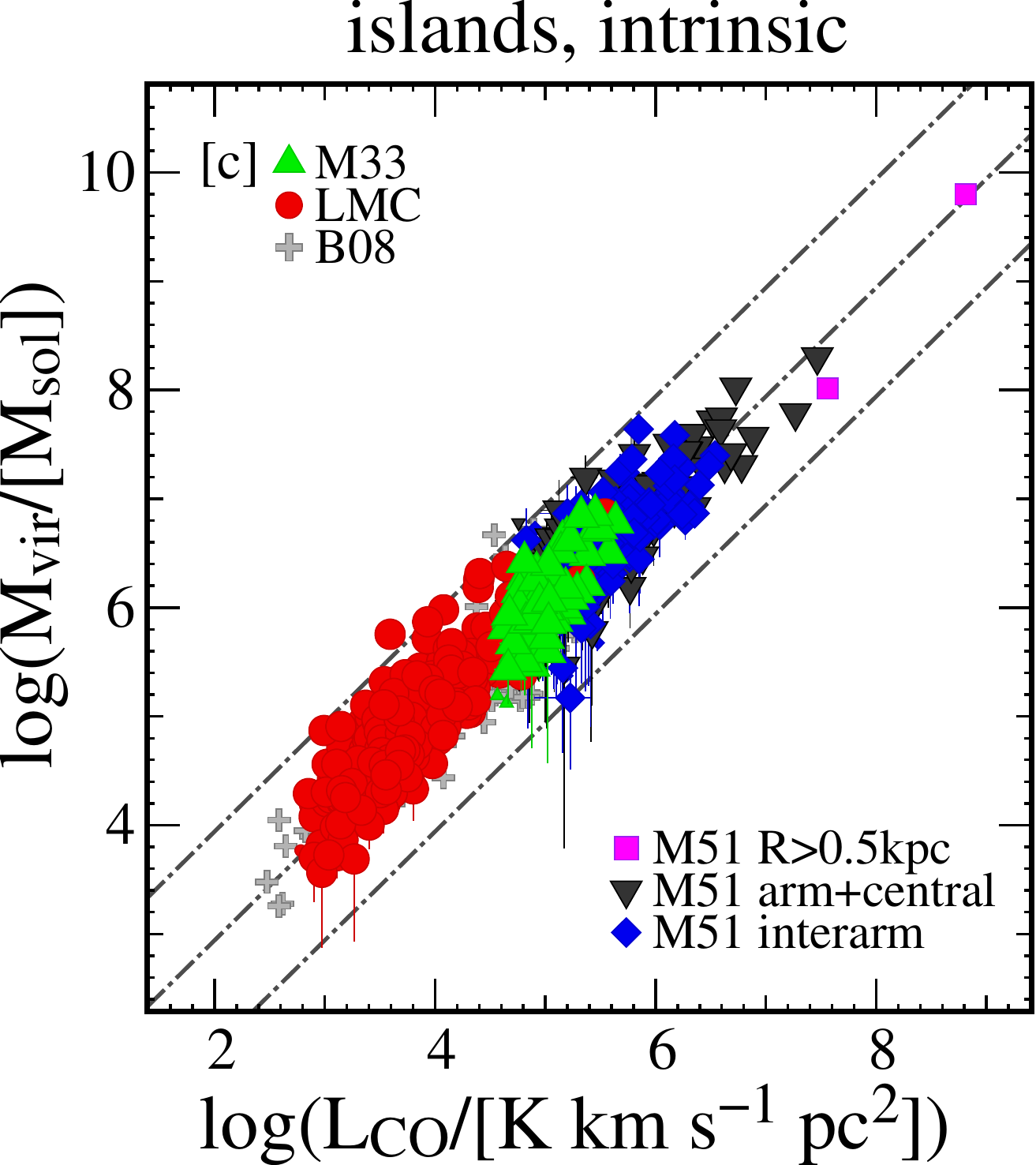}
\hspace{0.5cm}
\includegraphics[width=70mm,angle=0]{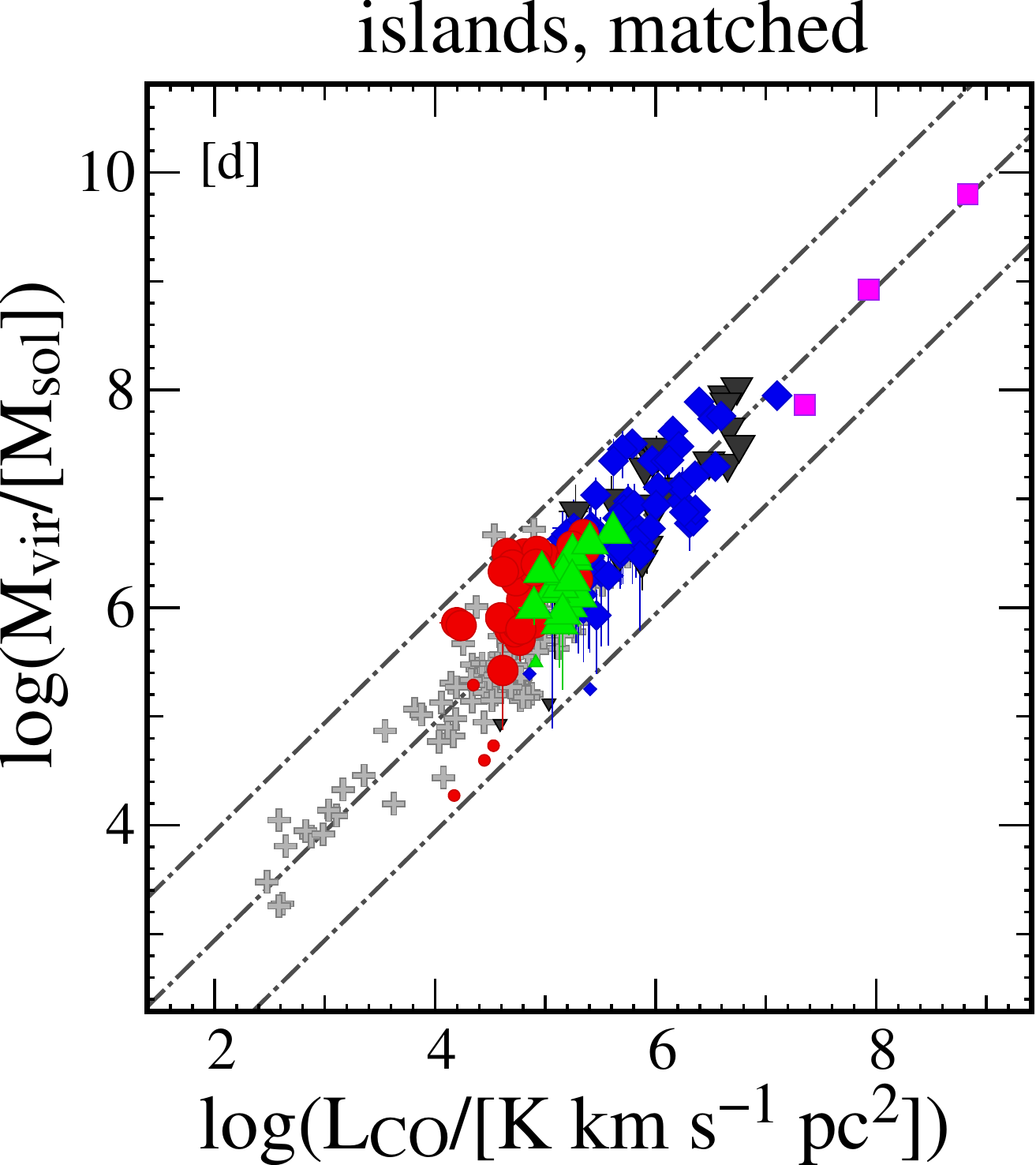}
\caption{\small A plot of virial mass versus CO luminosity for the
  objects identified in M51, M33, and the LMC.  The panels and plot
  symbols are the same as in Figure~\ref{fig:rdv}. The grey dashed
  lines indicate constant values of the CO-to-\hh\ conversion factor,
  $\xco = 0.4, 4.0, 40$\,\xcou. }
\label{fig:lm}
\end{center}
\end{figure*}

\noindent Since the very first studies of GMCs in the inner Milky Way
(e.g. S87), scaling relations between the physical properties of
molecular clouds have become a standard tool for assessing the
similarity of GMC populations \citep[e.g.][B08]{blitzetal07}. We plot
the relations between size and linewidth, size and luminosity, and
luminosity and virial mass for the objects identified in our M51, M33
and LMC cubes in Figures~\ref{fig:rdv} to~\ref{fig:lm}. In each
figure, we also indicate the extragalactic GMCs studied by B08 with
small grey crosses; note that we have not re-analysed these data and
simply adopt the cloud property measurements published by B08. The
relations for clouds identified in the intrinsic and matched
resolution data cubes are shown in panels [a] and [b]
respectively. Since the GMC identification procedure employed by S87
is most similar to our islands decomposition, we also plot the
relations for islands in panels [c] and [d] of each figure. \\

\noindent We assess the strength of scaling relations using the
Spearman rank correlation coefficient, $r_{sp}$. We regard $0.3 <
r_{sp} < 0.5$ to indicate a weak correlation, $0.5 < r_{sp} < 0.75$ to
indicate a moderate correlation, and $r_{sp} > 0.75$ to indicate a
strong correlation. For GMC samples where a correlation between size
and linewidth is evident, we estimated the best-fitting power-law
$\sigma_{v} = A R^{n}$ using the BCES bisector linear regression
method presented by \citet{akritasbershady96}. This method is designed
to take measurement errors in both the dependent and independent
variable, and the intrinsic scatter of a dataset into account. We use
the bisector method because our goal is to estimate the intrinsic
relation between the cloud properties
\citep[e.g.][]{babufeigelson96}. For the measurement errors, we adopt
the uncertainties derived by \textsc{CPROPS}. We have assumed that
measurement errors in the property measurements are uncorrelated,
although some pairs of parameters should have substantial
covariance. The resulting fits are tabulated in
Table~\ref{tbl:llfits}. To determine both the correlation strength and
the the best-fitting relations, we work with resolved clouds only. We
verified that our results are not driven by clouds with poorly
determined properties by repeating the calculations using a subsample
of resolved clouds where the relative uncertainty in the size and
linewidth measurements is less than 50\%.  \\

\noindent For the cloud decompositions (Figure~\ref{fig:rdv}[a]
and~[b]), there is no compelling evidence for a size-linewidth
correlation {\it within} any of the galaxies, and the different cloud
populations yield $r_{sp}$ between 0.07 and 0.37 (see
Table~\ref{tbl:spear}). For a composite sample containing all the
clouds in M33, M51 and the LMC, $r_{sp} = 0.62$ using the $R$ and
$\sigma_{\rm v}$ measurements determined from the original data
cubes. This good correlation is mostly a consequence of the
differences in the spatial and spectral resolutions of the LMC and M51
surveys, however, and disappears once the correlation is determined
using a composite cloud sample derived from the matched cubes, for
which $r_{sp} = 0.18$. \\

\begin{table}
\centering
\caption{\small Spearman rank correlation coefficients for size-linewidth correlations.}
\label{tbl:spear}
\par \addvspace{0.2cm}
\begin{threeparttable}
{\small
\begin{tabular}{@{}lcccccc}
\hline 
Galaxy/Region  & $r_{sp}$\tnote{a} & $r_{sp}$\tnote{b} & $r_{sp}$\tnote{c} & $r_{sp}$\tnote{d} & $r_{sp}$\tnote{e} & $r_{sp}$\tnote{f}  \\
\hline 
Composite\tnote{g}                  & 0.62        & 0.18        & 0.72       & 0.49   & 0.69  & 0.30 \\
M51                        & 0.16        & 0.16        & 0.37       & 0.44   & 0.26  & 0.18 \\ 
M51 arm+central            & 0.18        & 0.17        & 0.41       & 0.68   & 0.25  & 0.46 \\
M51 interarm               & 0.13        & 0.07        & 0.28       & 0.28   & 0.21  & 0.05 \\
LMC                        & 0.37        & 0.32        & 0.59       & 0.45   & 0.59  & 0.45\\
M33                        & 0.33        & 0.14        & 0.44       & 0.05   & 0.44  & 0.05\\ 
\hline
\end{tabular}}
{\small
\begin{tablenotes}
\item[a] Cloud decomposition, intrinsic resolution.
\item[b] Cloud decomposition, matched resolution.
\item[c] Island decomposition, intrinsic resolution, excluding structures with radius greater than 500pc.
\item[d] Island decomposition, matched resolution, , excluding structures with radius greater than 500pc.
\item[e] Island decomposition, intrinsic resolution, excluding structures with radius greater than 150pc.
\item[f] Island decomposition, matched resolution, , excluding structures with radius greater than 150pc.
\item[g] Our composite sample consists of all objects in M51, M33 and the LMC.
\end{tablenotes}}
\end{threeparttable}
\end{table}

\begin{table*}
\centering
\caption{\small Best-fitting parameters for the size-linewidth correlation, $\sigma_{\rm v} = A R^{n}$.}
\label{tbl:llfits}
\par \addvspace{0.2cm}
\begin{threeparttable}
{\small
\begin{tabular}{@{}lccccc}
\hline 
Galaxy/Region\tnote{a} & Resolution & Decomposition & Coefficient $A$ & Index $n$ & $\epsilon$\tnote{b} \\
\hline
Composite            & Intrinsic & Clouds   &  $0.06\pm0.02$ & $1.19\pm0.05$ & 0.25  \\
Composite            & Intrinsic & Islands  &  $0.11\pm0.02$ & $1.01\pm0.03$ & 0.24  \\
M51 arm+central\tnote{c}  & Matched & Islands    &  $0.46\pm0.37$ & $0.64\pm0.12$ & 0.15  \\
M51 arm+central\tnote{d}  & Matched & Islands    &  $0.34\pm0.56$ & $0.72\pm0.22$ & 0.17  \\
LMC                  & Intrinsic & Islands  &  $0.16\pm0.03$ & $0.84\pm0.05$ & 0.17  \\
\hline			
\end{tabular}}
{\footnotesize
\begin{tablenotes}
\item[a] We attempt to fit the size-linewidth relation for cloud samples where
  the Spearman rank correlation coefficient is greater than 0.5 (see Table~\ref{tbl:spear}).
\item[b] The final column lists the logarithmic scatter of the residuals about the best-fitting relationship. 
\item[c] Excludes structures with radius greater than 500pc. 
\item[d] Excludes structures with radius greater than 150pc. 
\end{tablenotes}}
\end{threeparttable}
\end{table*}

\noindent A stronger relationship between $R$ and $\sigma_{\rm v}$ is
apparent for the island decompositions (Figure~\ref{fig:rdv}[c]
and~[d]), with higher $r_{sp}$ values than those obtained using a
cloud decomposition for all three galaxies. The size-linewidth
relationships for the LMC and the M51 arm+central region yield
$r_{sp}$ values greater than 0.5 (see Table~\ref{tbl:spear}),
indicative of a moderate correlation. We caution, however, that the
correlation in the M51 arm+central region may be driven by the largest
islands with sizes greater than a few hundred parsecs. Although we
have excluded the largest CO-emitting structures in M51 (with
$R\gtrsim 0.5$\,kpc, represented by magenta squares in
Figure~\ref{fig:rdv}[c] and~[d]) from our correlation and regression
analysis, lowering the size threshold to 150\,pc tends to make the
correlation weaker and to steepen the best-fitting size-linewidth
relation (see Table~\ref{tbl:llfits}). It is remarkable how well the
large ($R \gtrsim 200$\,pc) island structures in M51 appear to follow
the classical S87 size-linewidth relation, since the physical
similarity of these structures to Galactic GMCs is remote. We would
expect the size-linewidth relationship for GMCs to break down on such
large scales, moreover, since existing interpretations for the
size-linewidth relation would predict no correlation for
non-virialized structures or if an object's size exceeds the
turbulence driving scale. We discuss how the size-linewidth
correlation depends on scale and decomposition approach in
Section~\ref{sect:llaw1_origin}.\\

\noindent Figure~\ref{fig:rl} presents the relationship between size
and luminosity. Once again, tighter correlations are detected for the
island decompositions than for the cloud decompositions. We note that
a good correlation between $R$ and $L_{\rm CO}$ is expected since
$L_{\rm CO} \approx \sigma_{\rm v}R^{2}\langle T \rangle$ (see
Equation~\ref{eqn:lcodef}), and the dynamic range of $R$, $\sigma_{\rm
  v}$ and $\langle T \rangle$ for each galaxy are limited by the
resolution and sensitivity of each survey. The robust-looking
correlations in Figure~\ref{fig:rl} should therefore not be regarded
as strong evidence that GMCs have constant mass surface density. In
particular, we note that GMCs in each panel typically lie close to the
surface brightness sensitivity limits of each survey. It is likely
that deeper observations would increase the number of low surface
brightness objects detected in each galaxy, and hence increase the
scatter in Figure~\ref{fig:rl}.\\

\noindent Despite these biases, Figure~\ref{fig:rl} indicates genuine
variations between the surface brightness of CO-emitting structures in
M51 compared to those in the low-mass galaxies. The size and
luminosity measurements from the matched cubes of the three galaxies
are clearly segregated: GMCs in the low-mass galaxies tend to be
smaller and fainter than clouds in M51. For the cloud decomposition of
the matched cubes, the median CO surface brightness of well-resolved
clouds ($R>50$\,pc) in M51 is $28$\,\kkms. In M51, the variation in
surface brightness measurements is also relatively large: the
brightest M51 cloud has a CO surface brightness of 347\,\kkms, more
than an order of magnitude above the population's median value, and
10\% of clouds have surface brightness greater than 100\,\kkms. For
well-resolved clouds in M33 and the LMC, by contrast, the median
(maximum) CO surface brightness is much lower: 10 (22) and 4
(11)\,\kkms respectively. While these estimates are biased by the
sensitivity of the input datasets, the fact that some M51 clouds
achieve CO luminosities more than an order of magnitude higher than
clouds of similar size in M33 and the LMC clouds is
meaningful. Assuming that the variation in $\xco$ between the three
galaxies is less than this variation in CO surface brightness, then
Figure~\ref{fig:rl} demonstrates that the molecular structures in M51
reach higher \hh\ surface densities than equivalent structures in the
low-mass galaxies. We discuss this result -- including the effect of
$\xco$ variations on the derived values of the mass surface density --
in Sections~\ref{sect:prev_results} and~\ref{sect:llaws_origin}.\\

\noindent The plot of CO luminosity versus virial mass in
Figure~\ref{fig:lm} shows that clouds and islands in all three
galaxies are distributed about the line corresponding to $\xco = 4.0
\times 10^{20}$\,\xcou\ but with considerable scatter, especially for
the cloud decompositions (Figure~\ref{fig:lm}[a] and~[b]). A slight
vertical offset between the M33, LMC and M51 cloud populations is
present in panel [b]: clouds in M33 and the LMC tend to lie above the
$\xco = 4.0 \times 10^{20}$\,\xcou\ line, while a larger proportion of
M51's cloud population lies on or below it. In previous works, the
normalization of the luminosity versus virial mass correlation has
been used to estimate the average value of \xco\ for a GMC population
\citep[e.g.][]{blitzetal07,fukuietal08}. We discuss the
  results of such a ``virial analysis'' in
  Section~\ref{sect:prev_results}, where we investigate possible
  variations in the \xco\ factor and their implication for our derived
  values of the cloud mass surface densities. Strictly, this method
requires independent evidence concerning the dynamical state of GMCs,
since it assumes that molecular clouds attain virial equilibrium on
average (i.e. $\langle \alpha \rangle = 1$). In other words, the
\xco\ values corresponding to the diagonal dashed lines in
Figure~\ref{fig:lm} depend on the average value of $\alpha$ that one
assumes for the cloud population: if GMCs tend to be globally
self-gravitating but not virialised, then $\langle \alpha \rangle \sim
2$ and the mean \xco\ value that should be inferred from a correlation
between luminosity versus virial mass is also smaller by a factor of
$\sim2$. Luminosity ($\sim \sigma_{\rm v}R^{2}\langle T_{\rm CO}
\rangle$) and virial mass ($\propto \sigma_{\rm v}^{2}R$) are
covariant quantities, so once again the physical significance of the
robust-looking correlations in Figure~\ref{fig:lm} should not be
over-interpreted. Nonetheless, the M33, M51 and the LMC data are not
located along widely separated tracks in Figure~\ref{fig:lm},
suggesting that the galaxy-wide averages of \xco\ and $\alpha$ do not
deviate from Galactic-like values ($\alpha = 1 - 2$, $\xco = 2.0 - 4.0
\times 10^{20}$\,\xcou) by more than a factor of a few. \\

%%%%%%%%%%%%%%%%%%%%%%%%%%%%%%
\section{Discussion}
%%%%%%%%%%%%%%%%%%%%%%%%%%%%%%
\label{sect:discussion}

%%%%%%%%%%%%%%%%%%%%%%%%%%%
\subsection{Comparison with Previous Results}
\label{sect:prev_results}
%%%%%%%%%%%%%%%%%%%%%%%%%%%

\noindent Variations between the properties of GMCs in different
galaxies are difficult to establish conclusively, and our analysis in
Section~\ref{sect:results} highlights the risk of conducting
comparisons on heterogeneous CO datasets. Nevertheless, we find
statistically significant differences between the GMC populations of
M51, M33 and the LMC after we account for observational effects
(i.e. by using the matched resolution cubes). Namely, GMCs in M51 are
intrinsically brighter, and they have larger velocity dispersions and
higher mass surface densities than GMCs with comparable size in the
two low-mass galaxies. In contrast to many previous studies, our two
main conclusions are therefore i) that Larson's scaling relations are
not an especially sensitive tool for comparing the physical properties
of GMC populations unless observational and methodological effects are
explicitly taken into account, and ii) that the physical properties of
GMCs are sensitive to their galactic environment. We discuss the
potential nature of this environmental dependence in the remainder of
this section, focussing on whether the trends that we observe are
better explained by blending (i.e. emission from clouds along the same
line-of-sight that overlap in velocity space), galaxy-to-galaxy
variations in the CO-to-\hh\ conversion factor, or whether they
indicate that external pressure plays a role in regulating the
physical properties of GMCs. We revisit the interpretation of Larson's
Laws in Section~\ref{sect:llaws_origin}.\\

\subsubsection{Emission from Overlapping Clouds}

\noindent In Section~\ref{sect:gmcproperties}, we found that clouds in
the arm+central region of M51 tend to have higher CO peak brightness
and larger velocity dispersions than clouds in the two low-mass
galaxies and M51's interarm region. One observational effect that
could contribute to this difference is blending of the CO emission
from discrete physical entities that overlap in $(x,y,v)$ space, and
which cannot be decomposed at our resolution. This effect is unlikely
to be dominant in regions where the CO emission is sparsely
distributed, and for this reason, blending would seem an improbable
explanation for the observed differences between the properties of
clouds in M33, the LMC and M51's interarm environment. On the other
hand, clouds may become crowded in M51's spiral arms. This would tend
to increase the size, brightness and velocity dispersion of the
structures that our decomposition algorithm identifies in the M51
spiral arm region. \\

\noindent While higher resolution observations -- especially in the
spectral domain -- are required to unambiguously assess the prevalence
of blending in M51's spiral arms, we suggest that blending is unable
to fully explain the trends that we observe for several
reasons. First, the scale height of the thin molecular disk in M51 is
only $\sim40$\,pc \citep{petyetal13}, which makes it unlikely that
several $\sim50$\,pc scale structures occuring along a single
line-of-sight through the galaxy would be a common
phenomenon. Furthermore, even though the typical linewidth of the
matched resolution clouds in the arm+central region of M51 is larger
than the typical linewidth of clouds in M33, the LMC and in M51's
interarm region, it is still a factor of $\sim3$ smaller than the
cloud-to-cloud velocity dispersion in M51 after we subtract a model of
galactic rotation from the cloud radial velocities ($\sim16$\,\kms,
see Section~\ref{sect:llaw1_origin} for a description of how we
subtract a galactic rotation model). This is consistent with the
appearance of the CO line profiles in the M51 arm region, which
typically exhibit a single peak or, more rarely, two peaks that are
well-separated along the velocity axis (which are then identified as
discrete clouds by our decomposition algorithm). At our resolution,
line profiles with multi-peaked velocity components are rare, even
though we might expect a significant number of such profiles if
emission in the spiral arms arose from distinct clouds with similar
radial velocities that overlapped along the line-of-sight.\\

\noindent A further piece of evidence that the clouds
  identified in M51's spiral arms are discrete objects, as opposed to
  blended emission from multiple overlapping clouds, is that we do not
  detect any variation in the scaling between the virial mass estimate
  and CO luminosity for clouds of similar size in the M51 arm and
  interarm regions. Assuming that blending is not a significant
  problem in the interarm region -- which is likely, since the
  observed interarm clouds are widely separated in $(x,y)$ space --
  this agreement suggests that we are identifying discrete clouds in
  both environments, since mass estimates derived from applying the
  virial theorem to unbound associations of molecular clouds tend to
  be significantly greater than the mass inferred from the CO
  luminosity \citep[e.g.][]{allenlequeux93,rand95}.\\

\noindent Finally, we note that if blending were solely responsible
for raising the brightness temperature and velocity dispersion in
M51's arm+central region, then we would expect the effect to be more
pronounced in the spiral arms than in the central zone, where the CO
emission is more sparsely distributed (see
Figure~\ref{fig:maps}[a]). Instead, we observe the opposite: the
median peak brightness and velocity dispersion of clouds in the
central zone is higher than in the arms by 1.1\,K and
0.8\,\kms\ respectively. While this does not imply that blended
emission from overlapping clouds is absent, it does suggest that there
are physical processes besides -- or in addition to -- blending that
determine the CO emission properties in these environments. In
conclusion, although we cannot definitively exclude the possibility
that blended emission from overlapping clouds contributes to the
higher peak brightness and velocity dispersion of the clouds in the
M51 arm+central region, it would seem insufficient to explain all the
trends that we describe in Section~\ref{sect:gmcproperties}.

\subsubsection{Variations in the CO-to-\hh\ factor}

\noindent A second potential explanation for the differences between the
GMC populations of M51, M33 and the LMC is that there is a systematic
difference in the way that \aco\ emission traces the underlying
\hh\ distribution. If this were true, then the underlying physical
properties of the molecular (i.e. \hh) clouds might be similar in all
three galaxies, despite the variations that we infer from our CO
observations. B08, for example, argued that the lower velocity
dispersions and CO luminosities of molecular clouds in the Small
Magellanic Cloud (SMC) were best understood in terms of selective
photodissociation of CO molecules. \citet{feldmannetal12}, on the
other hand, have argued that $\xco$ also increases at high \hh\ column
densities once the CO-emitting clumps within molecular clouds shadow
each other (i.e. when the CO filling factor within a velocity range
corresponding to the channel width reaches unity) and CO emission from
the cloud becomes globally optically thick. \\

\noindent In general, however, we do not expect large variations in
the value of \xco\ between M51, M33 and the LMC. In a companion paper
\citep{hughesetal13}, we argue that the absence of a truncation
and the width of the probability distribution functions of CO
integrated intensity and brightness provide evidence that the velocity
dispersion of the CO-emitting gas within the PAWS field is
sufficiently high that the saturation effect described by
\citet{feldmannetal12} does not yet apply. Selective CO
photodissociation should be an important effect at very low
metallicities, but it is not expected to cause large variations in
\xco\ for systems with metal abundances greater than $\sim0.3
Z_{\odot}$ \citep[e.g.][]{bolattoetal13}. The metallicity of M51's
inner disk is approximately solar
\citep[e.g.][]{moustakasetal10,bresolinetal04}, and several
independent analyses of dust and molecular line emission in M51
indicate that the \xco\ factor and dust-to-gas ratio are
consistent with local Milky Way values
\citep[e.g.][]{schinnereretal10,tanetal11,mentuchcooperetal12}. M33
and the LMC have a lower metallicity than M51 by a factor of $\sim2$,
but an empirical comparison between the \hh\ masses inferred from CO
and dust continuum emission also concludes that a Galactic value of
the \xco\ factor is applicable for GMCs in these two low-mass galaxies
\citep{leroyetal11}. Based on these studies, we would not expect the
\xco\ factor to vary by more than a factor of a few between all three
galaxies.  \\

\noindent We can assess whether small variations in the
  CO-to-\hh\ factor could nonetheless account for the differences in
  the mass surface density of GMCs that we infer using a virial
  analysis. The median values of the virial parameter in
  Table~\ref{tbl:cldprops} are 1.6, 3.1 and 2.9 for resolved clouds in
  M51, the LMC and M33 respectively. These values are obtained under
  the assumption that $\xco = 2 \times 10^{20}$\,\xcou. Alternatively,
  we can assume that the average dynamical state of GMCs is the same
  in all three galaxies, and that the differences in the virial
  parameter instead reflect variations in the true value of \xco. This
  is equivalent to attributing the small vertical offset between the
  GMC populations in Figure~\ref{fig:lm}[b] to variations in \xco,
  rather than inferring that the average dynamical state of GMCs
  varies between the different galaxies. If GMCs are typically just
  self-gravitating, then the median value of the virial parameter for
  a cloud population is $\langle \alpha \rangle = 2$
  \citep[e.g.][]{blitzetal07,leroyetal11}. Imposing this median value
  of $\alpha$ on all the cloud samples requires \xco\ values of 1.6,
  3.1 and $2.9\times 10^{20}$\,\xcou\ for M51, the LMC and M33
  respectively. \\

\noindent For resolved clouds identified in the matched resolution
cubes, the median mass surface densities that we infer for the LMC,
M33, and for M51's interarm and arm+central environments (assuming
$\xco = 2.0 \times 10^{20}$\,\xcou) are 22, 86, 122 and
167\,\mpcsq. The values that would be obtained using the
galaxy-dependent \xco\ values (i.e. derived under the assumption that
$\langle \alpha \rangle = 2$) are 34, 124, 98 and 134\,\mpcsq. We
repeated the KS tests on the $\Sigma_{\rm H_{2}}$ distributions that
we derive using the galaxy-dependent $\xco$ values and tabulate the
results in Table~\ref{tbl:kstest}. We find that there is still a
statistically significant difference $(p\leq 0.05$) between the mass
surface densities for the LMC clouds and all the other cloud
populations, and between clouds in the arm and interarm regions of
M51. On the other hand, there is no statistically significant
difference between the mass surface density distributions of clouds in
M33 and M51 using the galaxy-dependent
\xco\ values. Figure~\ref{fig:rl} shows that the clouds with CO
surface brightness greater than 10\,\kkms\ in M33 are mostly small
($R<50$\,pc). If we restrict our virial analysis to clouds that are
larger than this, then the median cloud mass surface density (obtained
using the galaxy-dependent \xco\ factor) for M33 reduces to
62\,\mpcsq, but the difference between the mass surface densities of
clouds with $R\geq50$\,pc in M33 and clouds in M51 remains
statistically insignificant.\\

\noindent In summary, it is possible that the \xco\ variations can
account for some of the differences between the CO-derived
properties of clouds in M33 and M51. The differences in CO-derived GMC
properties that we find for the LMC clouds and between the M51
environments, on the other hand, cannot be fully explained by
\xco\ variations that remain consistent with either a virial analysis
or independent analyses of dust emission and CO excitation. For these
galactic environments, the differences in the CO-derived properties
appear to reflect genuine variations in the physical properties of the
molecular (i.e. \hh) clouds, and not just the fidelity with which CO
emission traces the underlying \hh\ distribution. \\

\subsubsection{Variations in the Interstellar Pressure}

\noindent A third possible explanation for the observed differences
between the properties of GMCs in M33, the LMC and the inner disk of
M51 is that the typical density of GMCs is regulated by pressure
variations in the ambient ISM. Traditionally, a large discrepancy
between the high internal pressures of Milky Way GMCs and the much
lower kinetic (thermal plus turbulent) ISM pressure has been thought
to imply that GMCs are approximately in simple virial equilibrium
(i.e. with internal kinetic energy equal to half their gravitational
potential energy) and hence largely decoupled from the diffuse
interstellar gas that surrounds them \citep[e.g.][]{blitz93}. This
line of argumentation has been problematised, however, by renewed
attention to other potential sources of confining pressure for clouds,
such as ram pressure from inflowing material
\citep[e.g.][]{heitschetal09} and the (static) weight of surrounding
atomic gas \citep[e.g.][]{heyeretal01}, as well as a long history of
observations suggesting that pressure confinement is significant for
molecular clouds in certain galactic environments, such as the outer
Galaxy \citep[e.g.][]{heyeretal01} and at high latitude
\citep[e.g.][]{ketomyers86}. At the very least, we would expect the
internal pressure of molecular clouds to be comparable to the external
pressure, otherwise the clouds would be rapidly compressed and/or
destroyed.\\

\noindent The higher mass surface density of M51 clouds suggests that
they should also have higher internal pressures than clouds in the
low-mass galaxies (Section~\ref{sect:gmcproperties}). We can estimate
the internal pressure $P_{int}$ of a molecular cloud according to:
\begin{equation}
\frac{P_{int}}{k} = \rho_{g} \sigma_{\rm v}^{2} = 1176 \left (\frac{M}{M_{\odot}} \right )\left (\frac{R}{{\rm pc}} \right )^{-3}\left (\frac{\sigma_{\rm v}}{\kms} \right )^{2}\,\punit \, ,
\end{equation}
\noindent where $\rho_{g}$ is the \hh\ volume density. For the
resolved cloud populations identified in the matched resolution cubes
of M33, the LMC and M51, we find median internal pressures of $\langle
P_{int}/{k} \rangle \sim 1.9\times10^{5}$, and $3.0\times10^{4}$, and
$4.3\times10^{5}$\,\punit\ respectively. We note that clouds in the
spiral arms and central region of M51 tend to have higher internal
pressures ($\langle P_{int}/{k} \rangle \sim 5.3\times10^{5}$\,\punit)
than clouds in the interarm region ($\langle P_{int}/{k} \rangle \sim
2.6\times10^{5}$\,\punit).\\

\noindent The average kinetic pressure in the interstellar gas
  depends on the weight of the gas layer in the gravitational
  potential of the total mass (i.e. including gas, stars and dark
  matter) that lies within the gas layer. Assuming the contribution
  from dark matter is negligible, we can approximate the external
  pressure at the boundary of a molecular cloud using the expression
  for the hydrostatic pressure at the disk midplane derived by
  \citet{elmegreen89} for a two-component disk of gas and stars:
\begin{equation}
%\frac{P_{ext}}{k} = \frac{\sigma_{g,z}^{2}\Sigma_{g}}{\sqrt{2\pi}h_{g}} = \Sigma_{g} (G\Sigma_{%g} + ((G\Sigma_{g})^{2}+2G\rho_{*}\sigma_{g,z}^{2})^{1/2})\,\punit.
%\label{eqn:ko2009}
\label{eqn:elme89}
P_{ext} = \frac{\pi G}{2}\Sigma_{g} \left ( \Sigma_{g} + \frac{\sigma_{g}}{\sigma_{*}}\Sigma_{*} \right )
\end{equation}
In this expression, $\Sigma_{g}$ is the neutral (atomic + molecular)
gas surface density, $\Sigma_{*}$ is the stellar surface density, and
$\sigma_{g}$ and $\sigma_{*}$ are the velocity dispersions of the gas
and stars, respectively. This expression for the interstellar pressure
accounts for the gravity forces due to the stars and gas, as well as
the turbulent and thermal hydrodynamic pressure. It is obtained from
the definition of the midplane pressure $P_{midplane} =
\rho_{g}\sigma_{g}^{2}$, after substituting $\rho_{g} =
\Sigma_{g}/2h_{g}$ and $h_{g} = \frac{\sigma_{g}^{2}}{\pi G
  \Sigma_{total}}$, where $\rho_{g}$ is the gas density at the
midplane, $h_{g}$ is the scale height of the gas, and $\Sigma_{tot}
\approx (\Sigma_{g} + (\sigma_{g}/\sigma_{*})\Sigma_{*})$ is an
estimate for the total mass surface density within the gas layer.\\

\begin{figure*}
\begin{center}
\hspace{-0.5cm}
\includegraphics[width=70mm,angle=0]{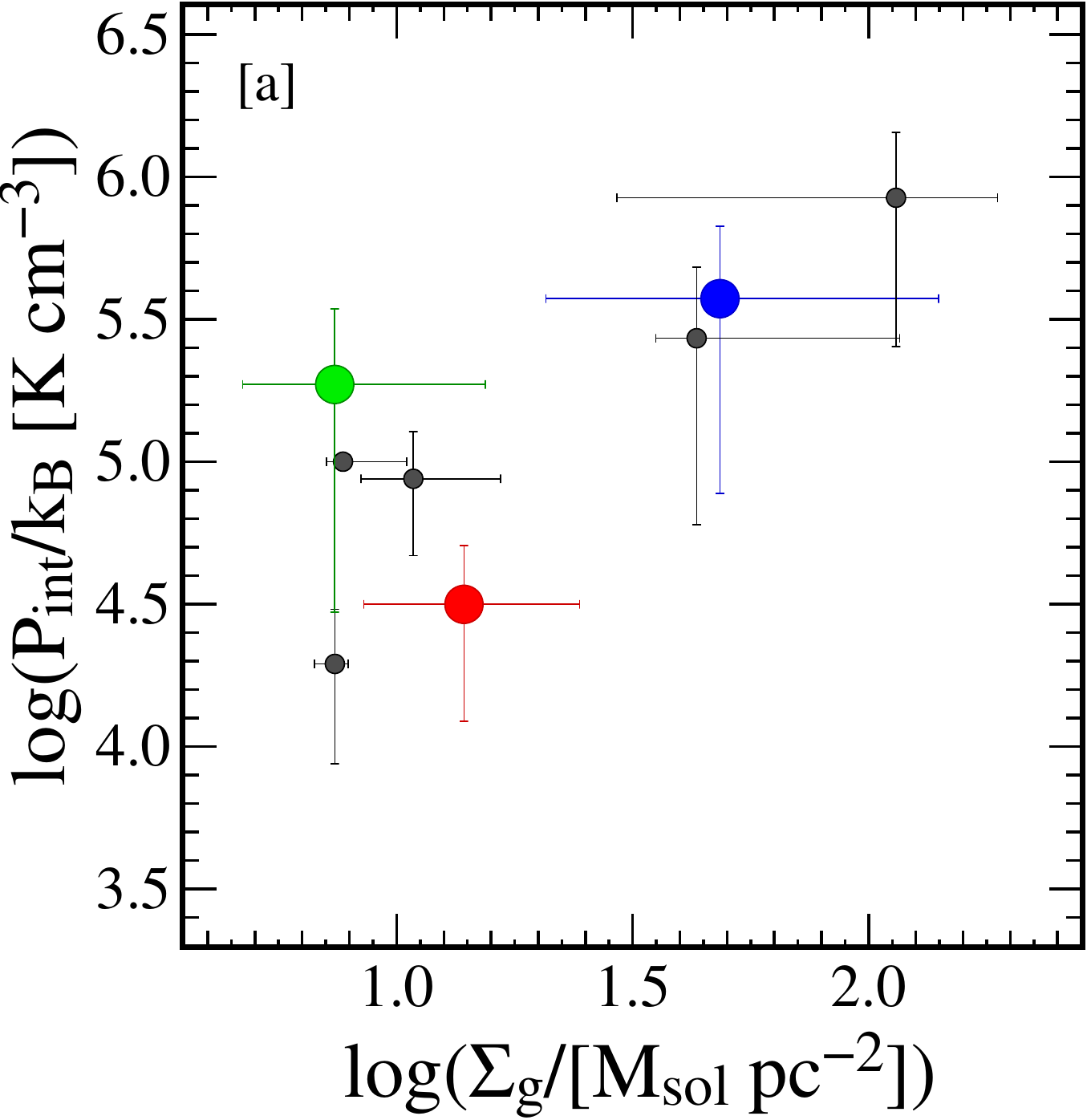}
\hspace{0.8mm}
\includegraphics[width=73mm,angle=0]{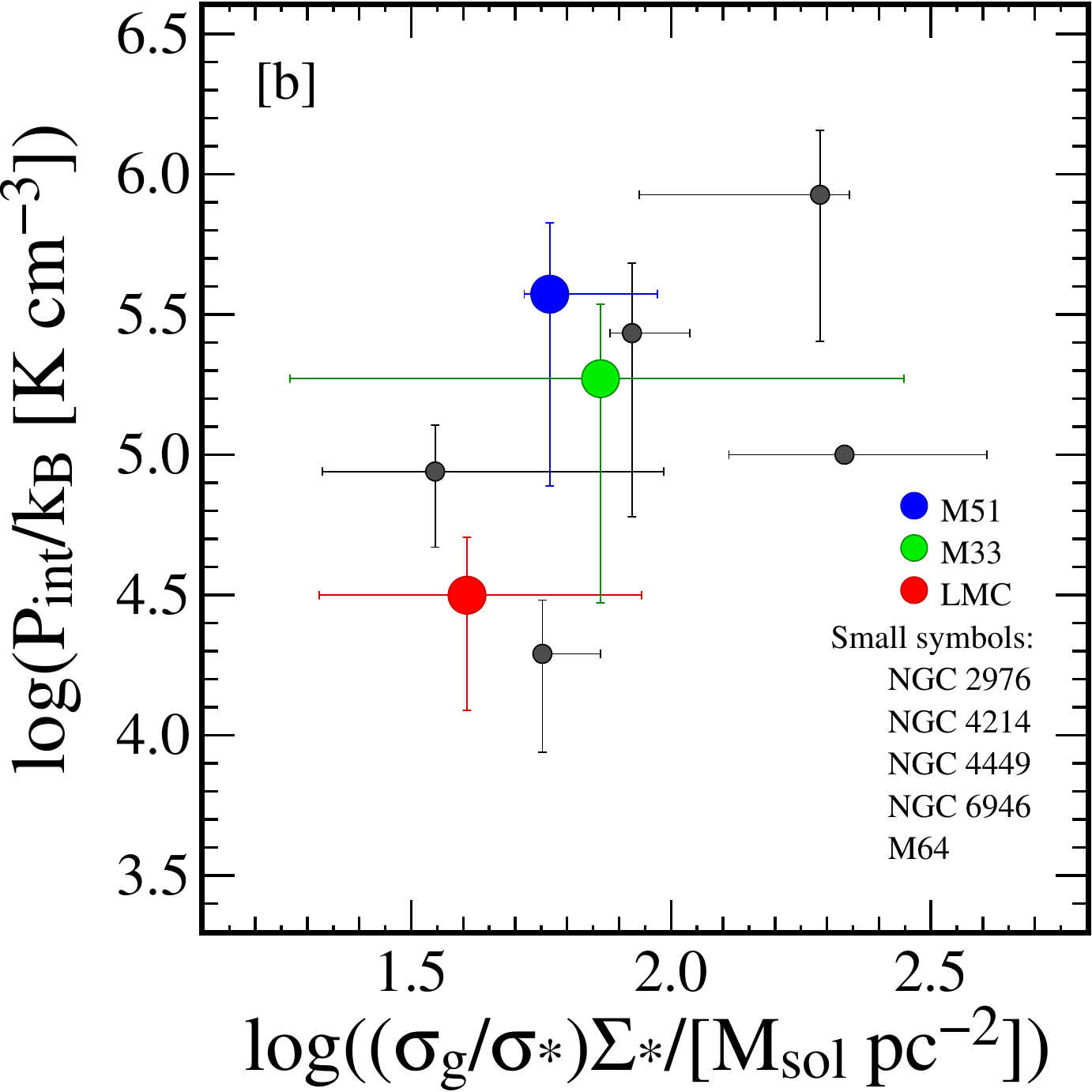}
\caption{\small The median internal pressure of the GMCs in a sample
  of galaxies where high spatial resolution CO surveys have been
  conducted versus an estimate of the gas (panel [a]) and stellar
  (panel [b]) components of the gravitational potential that acts upon
  the interstellar gas. Large coloured symbols represent M51, M33 and
  the LMC (this work), while small symbols are for extragalactic GMC
  populations studied by B08, \citet{rosolowskyblitz05} and
  \citet{donovanmeyeretal12}. The vertical error bars indicate the
  median absolute deviation of the internal pressure estimates for
  each GMC population. The horizontal bars indicate the range of
  abcissa values that we obtain across the region where GMCs are
  observed (see text).}
\label{fig:pcomponents}
\end{center}
\end{figure*}

\noindent Since our estimate for $P_{ext}$ involves a
  combination of quantities, we plot the internal pressure of the GMCs
  as a function of $\Sigma_{g}$ and
  $(\sigma_{g}/\sigma_{*})\Sigma_{*}$, i.e. the gas and stellar
  components of the gravitational potential, in
  Figure~\ref{fig:pcomponents}[a] and~[b] respectively. The origin and
  typical uncertainty associated with our observational estimates for
  $\Sigma_{g}$, $\Sigma_{*}$ and $\sigma_{*}$ are discussed below. We
  adopt a constant gas velocity dispersion $\sigma_{g} =
  10$\,\kms\ for all galaxies. The motivation for this choice is that
  a roughly constant gas velocity dispersion of $7 - 10$\,\kms\ has
  been reported across the disks of several nearby galaxies
  \citep[see][and references therein]{vanderkruitfreeman11}. By
  contrast, \citet{tamburroetal09} recently found that $\sigma_{g}$
  decreases linearly by $3 - 5$\,\kms\ per $\Delta R_{25}$ beyond the
  optical radius for a subsample of THINGS galaxies, as well as
  typical values of $\sigma_{g} \sim 15 - 20$\,\kms\ inside the
  optical radius. Assuming $\sigma_{g} = 20$\,\kms\ for all galaxies
  would shift the points in Figure~\ref{fig:pcomponents}[b] by
  $\sim0.3$\,dex towards higher values along the x-axis,
  i.e. increasing the contribution of stars to the gravitational
  potential acting on the gas layer. In addition to M51, M33, and the
  LMC, we include the GMC populations studied by B08 and the GMCs in
  M64 and NGC\,6946 identified by \citet{rosolowskyblitz05} and
  \citet{donovanmeyeretal12}, respectively. For these additional
  datasets, our estimates for $P_{int}/k$ are calculated using the
  published measurements of the GMC properties, i.e. we do not
  re-analyse the CO datacubes. The vertical error bars in each panel
  of Figure~\ref{fig:pcomponents} reflect the median absolute
  dispersion of the $P_{int}/k$ values, while the horizontal bars
  indicate the range of values that are observed across the
  field-of-view of each CO survey. For the B08 galaxies, the range on
  the x-axis applies to $R_{gal} < 0.4R_{25}$, where $R_{25}$ is the
  optical radius of the galaxy. We adopt this region based on the GMC
  positions published by B08. For M51, M33, the LMC, M64 and
  NGC\,6946, the range refers to the field-of-view of the original CO
  surveys. We note that our estimates of $\Sigma_{*}$, $\Sigma_{g}$
  and $\sigma_{*}$ are calculated using radial profiles of these
  quantities, rather than on a pixel-by-pixel basis. \\

\noindent Figure~\ref{fig:pcomponents}[a] and~[b] show that
  the median internal pressure of the GMCs increases with both the gas
  and stellar components of the gravitational potential, and that
  neither the gas nor the stellar term dominates our estimate for the
  external pressure. We note that a correlation with both quantities
  is to be expected if the internal pressure of molecular clouds is
  responding to the external pressure, since both stars and gas
  contribute to the total mass that determines gravitational force. By
  contrast, a good correlation with the stellar term would not be
  expected if the mass surface density of GMCs depends on a process
  such as the shielding of \hh\ molecules against the interstellar
  radiation field, which depends on the local gas column density
  only. The robust correlations in both panels of
  Figure~\ref{fig:pcomponents} further suggest that a relationship
  between our estimates for $P_{ext}$ and $P_{int}$ does not follow
  trivially from a correlation between $\Sigma_{g}$ and the average
  mass surface density of the GMCs. Although both quantities are
  measures of the gas surface density, \citet{leroyetal13} have
  recently shown that due to the clumpiness of the molecular ISM, the
  CO surface brightness measured on $\sim50$\,pc scales within
  galactic disks does not necessarily track measurements of the CO
  surface brightness on large ($\sim$\,kpc) scales or using radial
  profiles, since the latter is a combination of both the intrinsic CO
  surface brightness of small-scale structures and the filling factor
  of such structures within the $\sim$\,kpc scale region or
  annulus.\\

\noindent Our estimates for $\Sigma_{*}$ have several sources. For
M51, NGC\,6946 and the B08 targets, we use the $\Sigma_{*}$ radial
profile published by \citet{leroyetal08}, which uses an empirically
calibrated conversion from the 3.6\,\m\ intensity to the K-band flux,
and then a K-band mass-to-light ratio of 0.5 to convert from the
K-band intensity to stellar surface density \citep{belldejong01}. The
final calibration \citep[equation~C1 in ][]{leroyetal08} is:
\begin{equation}
\label{eqn:leroyc1}
\Sigma_{*} = 280 \, {\rm cos} i \, I_{3.6}
\end{equation}
\noindent where $I_{3.6}$ is the 3.6\,\m\ intensity in MJy sr$^{-1}$
and $i$ is the galaxy inclination. We apply the same method to the
{\it Spitzer} Local Volume Legacy Survey 3.6\,\m\ map of M64
\citep{daleetal09} to obtain a radial profile of $\Sigma_{*}$ in that
galaxy, assuming a central position of $\alpha_{2000} = 12:56:43.6$,
$\delta_{2000} = 21:40:59.3$ and an inclination $i=60\D$
\citep{garciaburilloetal03}. The largest uncertainty in our estimates
for $\Sigma_{*}$ is the mass-to-light ratio, which depends on the
metallicity, initial stellar mass function, and star formation history
of galaxies. \citet{belletal03} show that K-band stellar mass-to-light
ratios vary by 0.1\,dex for redder galaxies and 0.2\,dex for bluer
galaxies like M51. We note that the stellar surface density profiles
of NGC\,6946 and M51 rise sharply at small galactocentric radii
($R_{gal} \lesssim 1$\,kpc), which may reflect the presence of a
nuclear bulge. Since Equation~\ref{eqn:elme89} is invalid in such
regions, we only include galactocentric radii where the stellar
surface density follows a roughly exponential profile. For M33, we use
the stellar surface density profile at a lookback time of 0.6\,Gyr
published by \citet[][see their figure~4]{williamsetal09}. This
profile, which agrees within $\sim50$\% with the mass model of
\citet{corbelli03} inside our field-of-view ($R_{gal} < 5.5$\,kpc),
was constructed by modelling the star formation histories that best
reproduce the colour-magnitude diagrams (CMDs) obtained for four
fields in M33, imaged with the Advanced Camera for Surveys on the
Hubble Space Telescope. The uncertainties in $\Sigma_{*}$ quoted by
\citet{williamsetal09} are $\lesssim0.2$\,dex. To estimate
$\Sigma_{*}$ in the LMC, we use the stellar surface density map
published by \citet{yangetal07}, which was constructed using number
counts of red giant branch (RGB) and asymptotic giant branch (AGB)
stars in the Two Micron All Sky Survey Point Source Catalogue
\citep{skrutskieetal06}, and normalized for a total stellar mass of
$2\times10^{9}$\,\msol\ in the LMC \citep{kimetal98}.\\

\noindent To estimate $\Sigma_{g}$, we use radial profiles of
  \hi\ and CO integrated intensity, assuming $\xco = 2.0 \times
  10^{20}$\,\xcou, optically thin \hi\ emission and a helium
  contribution of 1.36 by mass to convert between measurements of
  integrated intensity and gas mass surface density. For M51,
  NGC\,6946 and the B08 galaxies, we use the gas radial profiles
  published by \citet{leroyetal08}. For the LMC, we use the radial
  profiles published by \citet{wongetal09}. For M33, we use the radial
  profiles published by \citet{gratieretal10}, re-calculating the
  molecular gas surface density using $\xco = 2.0 \times
  10^{20}$\,\xcou. For M64, we measured radial profiles of the atomic
  and molecular gas surface densities using the THINGS and BIMA-SONG
  integrated intensity maps of \hi\ and CO emission
  \citep{walteretal08,helferetal03}, adopting the same central
  position and inclination as for $\Sigma_{*}$.\\

\noindent With the exception of the LMC, our estimates for
$\sigma_{*}$ are based on measurements of the central stellar velocity
dispersion obtained by \citet{hoetal09} using the Palomar
spectroscopic survey of nearby galaxies \citep{hoetal95,hoetal97}. For
the LMC, we use the velocity dispersion of carbon stars measured by
\citet{vandermareletal02}. Many studies have indicated that the
exponential scale height of stellar disks is roughly constant with
galactocentric radius
\citep[e.g.][]{vanderkruitsearle81,degrijspeletier97,kregeletal02},
while $\sigma_{*}$ declines \citep[][and references
  therein]{vanderkruitfreeman11}. To estimate $\sigma_{*}$ across each
CO survey's field-of-view, we assume that $\sigma_{*}$ decreases
exponentially according to $\sigma_{*} = \sigma_{*,0} \exp \left
(-\frac{R_{gal}}{2R_{*}} \right )$
\citep[e.g.][]{bottema93,boissieretal03}, where $\sigma_{*,0}$ is the
central stellar velocity dispersion and $R_{*}$ is the exponential
scale length of the stellar disk $R_{*}$. We rely on literature values
for $R_{*}$ (see Table~\ref{tbl:pext_values} for exact references).\\

\begin{table*}
\centering
\caption{\small Adopted Parameters to Estimate the External Pressure on GMCs }
\label{tbl:pext_values}
\par \addvspace{0.2cm}
\begin{threeparttable}
{\scriptsize
\begin{tabular}{@{}lccccccc}
\hline 
Galaxy   & $\Sigma_{g}$\tnote{a} & $\Sigma_{*}$\tnote{a} & $\sigma_{g}$ & $\sigma_{*,0}$ & $R_{*}$ & $R_{lim}$ & References\tnote{b,c,d,e}\\
         & [\mpcsq]     & [\mpcsq]     & [\kms]         & [\kms]       &  [kpc]     & [kpc]    &    \\
\hline
M51\tnote{f}      &  [140,40] & [700,225]      & 10            &  96.0        & 2.8        & 4.2     &  1,5,7   \\
M33               &  [15,5]   & [590,10]       & 10            &  21.0        & 2.0        & 5.5     &  2,3,5,7   \\
LMC\tnote{g}      &  [8,16]  & [175,12]        & 10            &  20.0        & 1.4        & 3.5     &  6,7   \\
\hline		
NGC 2976          & [7,7]      & [250,90]            & 10            & 36.0         & 0.9        & 1.5     &  1,5,8    \\
NGC 4214          & [17,8]     & [465,50]            & 10            & 51.6         & 0.7        & 1.2     &  1,5,8   \\
NGC 4449          & [11,8]     & [680,125]           & 10            & 17.8         & 0.9        & 1.1     &  1,5,8    \\
M64\tnote{h}      & [170,5]    & [1880,270]          & 10            & 96.0         & 1.1        & 1.0     &  4,5,9    \\
NGC 6946\tnote{f} & [115,35]   & [500,190]           & 10            & 55.8         & 2.6        & 4.5     &  1,5,10    \\
\hline
\end{tabular}
}
{\scriptsize
\begin{tablenotes}
\item[a] {Values in square brackets indicate the maximum and
    minimum of the radial profile for galactocentric radii
    corresponding to the CO survey's field-of-view.}
\item[b]{References for gas radial profiles: 1. \citet{leroyetal08}; 2. \citet{gratieretal10}.  }
\item[c]{References for stellar radial profiles and stellar disk scale lengths:  1. \citet{leroyetal08}; 3. \citet{williamsetal09}; 4. \citet{reganetal01}.}
\item[d]{References for central stellar velocity dispersion: 5. \citet{hoetal09}; 6. \citet{vandermareletal02}.}
\item[e]{References for GMC properties: 7. this paper; 8. B08; 9. \citet{rosolowskyblitz05}; 10. \citet{donovanmeyeretal12}.}
\item[f]{External pressure estimate and radial profiles of gas and stellar surface density exclude the central $\sim1$\,kpc.}
\item[g]{Radial profile of $\Sigma_{*}$ determined by the author using
  the stellar mass surface density map of \citet{yangetal07}.}
\item[h]{Radial profiles of $\Sigma_{g}$ and $\Sigma_{*}$ determined
  by the author using the THINGS \hi\ survey data
  \citep{walteretal08}, BIMA-SONG CO survey data \citep{helferetal03}
  and {\it Spitzer} Local Volume Legacy Survey 3.6\,\m\ data for M64
  \citep{daleetal09}.}
\end{tablenotes}
}
\end{threeparttable}
\end{table*}

\begin{figure}
\begin{center}
\hspace{-0.5cm}
\includegraphics[width=70mm,angle=0]{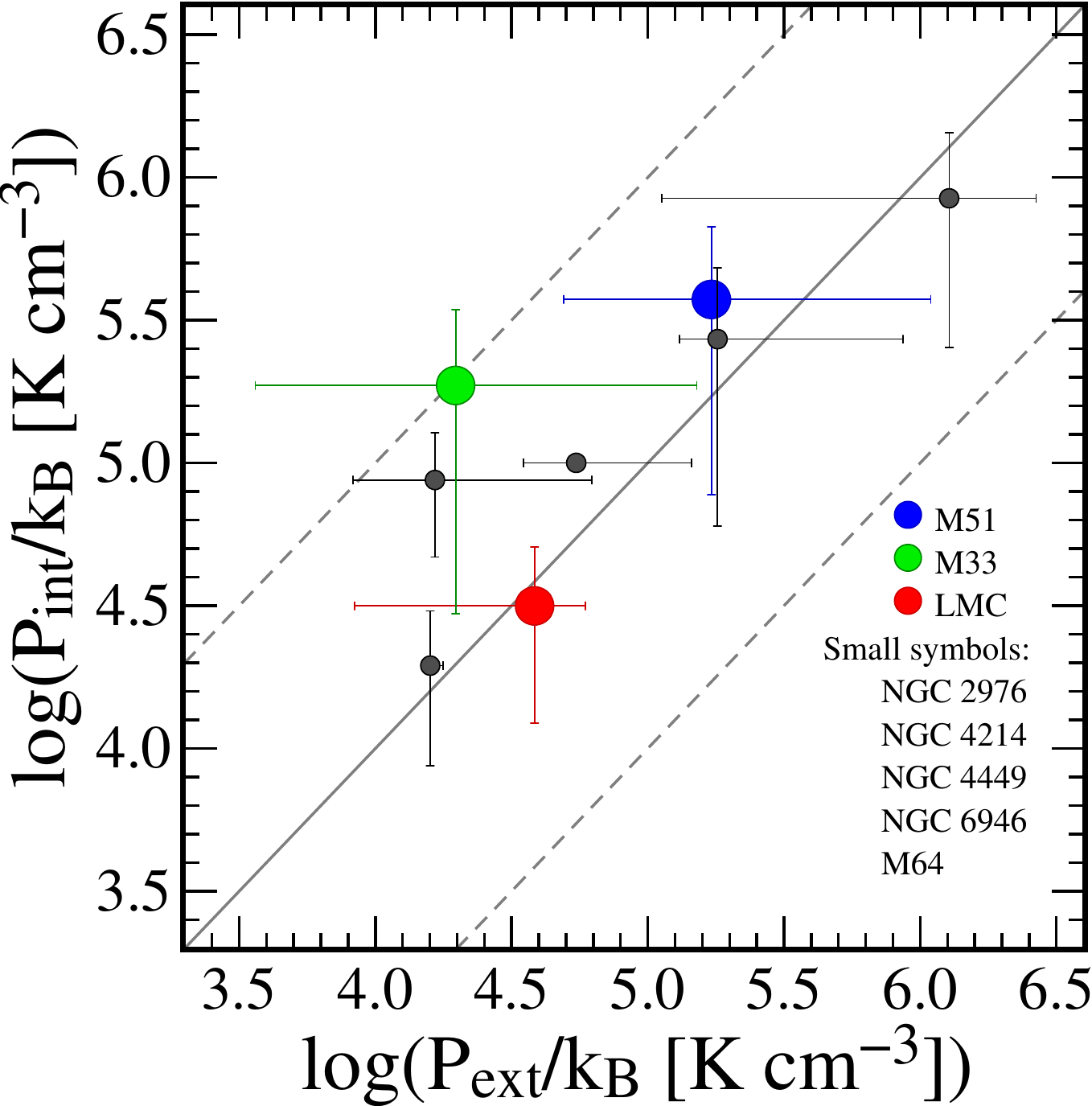}
\caption{\small The median internal pressure of the GMCs in a sample
  of galaxies where high spatial resolution CO surveys have been
  conducted versus an estimate of the external pressure within the
  region surveyed. Plot symbols are the same as in
  Figure~\ref{fig:pcomponents}. The internal pressures are estimated
  from GMC properties, the external pressures are estimated from the
  mass of stars and gas within the disk using an expression derived by
  \citet{elmegreen89} for the total hydrostatic pressure at the disk
  midplane. The solid diagonal line indicates equality; the dashed
  diagonal lines indicate where the internal and external pressures
  differ by an order of magnitude. As in Figure~\ref{fig:pcomponents},
  the vertical error bars indicate the median absolute deviation of
  the internal pressure estimates, while the horizontal bars indicate
  the range of external pressures in the region where GMCs are
  observed (see text).}
\label{fig:pint_pext}
\end{center}
\end{figure}

\noindent Finally in Figure~\ref{fig:pint_pext}, we plot the median
internal pressure of the GMC populations as a function of the external
pressure. As in Figure~\ref{fig:pcomponents}, the vertical error bars
correspond to the median absolute dispersion of the $P_{int}/k$
measurements of the GMCs in each galaxy, while the horizontal error
bars indicate a range of $P_{ext}/{k}$ values that characterise the
region of the galactic disk where the GMCs are located. It is clear
from Figure~\ref{fig:pint_pext} that there is a good correlation
between the internal and external pressures of GMCs, suggesting that
the variation in GMC mass surface densities that we observe between
M51, M33 and the LMC may arise because the external ISM pressure plays
a role in regulating the internal pressure (and hence velocity
dispersion and density) of molecular
clouds. Figure~\ref{fig:pint_pext} further suggests that GMCs are not
greatly overpressured with respect to their environment (i.e. $\langle
P_{int}/{k} \rangle \sim P_{ext}/{k}$).  Rather than simple virial
equilibrium between their gravitational and internal kinetic energies,
the implication is that GMCs may instead tend towards a
pressure-bounded equilibrium configuration. This result is satisfying
insofar as it suggests that the traditional dichotomy between strongly
gravitationally bound GMCs in the inner disk of the Milky Way and the
pressure-confined low-mass clouds in the outer Galaxy and at high
galactic latitude may be more apparent than real. If molecular
structures are bound by a combination of self-gravity and external
pressure, then self-gravity may appear dominant for samples that
preferentially include objects with high masses and densities, while
pressure confinement should appear more important for samples of
low-mass, low density objects. Moreover, the trend in
Figure~\ref{fig:pint_pext} would seem to confirm that the higher GMC
mass surface densities and line widths reported by studies of M64 and
M82 \citep{rosolowskyblitz05,ketoetal05} -- i.e. nearby systems where
the disk surface density is intermediate between conditions in Local
Group galaxies and true starbursts -- arise because GMCs exhibit a
continuum of properties \citep[as previously suggested
  by][]{rosolowsky06}, rather than an intrinsic bi-modality between
molecular gas properties in `normal' and `starburst' environments.\\

\noindent An important observable consequence of external pressure
regulating the properties of GMCs is that the scaling between a GMC's
size and linewidth -- i.e. the coefficient of the size-linewidth
relation -- should depend on the external pressure. We discuss whether
there is evidence for such variations in the GMC populations of M51,
M33 and the LMC in Section~\ref{sect:llaw1_origin}. Furthermore, if
the internal pressure of GMCs is comparable to the interstellar
pressure, then this shallow pressure gradient across GMC boundaries
means that the clouds' evolution should be more susceptible to
pressure fluctuations in the surrounding ISM than classical GMC models
have tended to assume. A detailed investigation of the importance of
dynamical pressure for the stability of gas and global patterns of
star formation in M51 is the subject of a companion paper \citep[][see
  also Jog et al. 2013]{meidtetal13}.

%%%%%%%%%%%%%%%%%%%%%%%%%%%
\subsection{The Origin of GMC Scaling Relations}
\label{sect:llaws_origin}
%%%%%%%%%%%%%%%%%%%%%%%%%%%

\noindent Empirical correlations between the size, line width and CO
luminosity of Galactic molecular clouds were initially reviewed by
\citet{larson81}. Although their interpretation remains controversial,
these scaling relations are regularly used to compare the
physical properties of molecular clouds in different galactic
environments. A key result of our analysis is that these scaling
relations -- as obtained from CO surveys of extragalactic GMC
populations -- are highly dependent on observational effects, such as
instrumental resolution and sensitivity, and on the techniques of GMC
identification and property measurement that are commonly applied to
CO spectral line cubes. In this Section, we discuss some caveats
regarding the physical significance of the empirical relations
observed for extragalactic GMCs, and whether they are sufficient to
demonstrate the universality of GMC properties.\\

%%%%%%%%%%%%%%%%%%%%%%%%%%%
\subsubsection{Larson's Third Law: GMCs have constant \hh\ surface densities}
\label{sect:llaw3_origin}
%%%%%%%%%%%%%%%%%%%%%%%%%%%

\noindent Larson's third ``law'' describes an inverse relationship
between the density of a molecular cloud and its size, implying that
molecular clouds have roughly constant molecular gas column
density. Several studies of 
extragalactic GMC populations via their CO emission (e.g. B08) have
reported that the average \hh\ surface density of extragalactic GMCs
is roughly constant within galaxies and, moreover, that it is in good
agreement with the value that is observed for GMCs in the inner Milky
Way, $\langle \Sigma_{\rm H_{2}} \rangle \sim 100$\,\mpcsq. Our results
in Section~\ref{sect:llaws}, by contrast, suggest that there are
subtle but genuine variations in the characteristic \hh\ surface
density for the GMC populations of M51, M33 and the LMC. \\

\noindent As noted by several previous authors
\citep[e.g.][]{kegel89,scalo90,ballesterosparedesmaclow02}, the
limited surface brightness sensitivity of extragalactic CO
observations is an evident source of bias for CO-based estimates of
GMC mass surface density, since sightlines with low to intermediate
\hh\ column densities will fall beneath the CO detection limit and
hence be excluded from the regions that are identified as
molecular. At high brightness, on the other hand, CO observations may
underestimate the true \hh\ column density if the CO-emitting regions
within shadow each other in velocity space
\citep[e.g.][]{feldmannetal12}. Coupled with the fact that widefield
extragalactic \aco\ observations are rarely designed to achieve
surface brightness sensitivities much deeper than $5\sigma$ for a
typical $\sim10^{5}$\,\msol\ cloud, these effects suggest that the
range of \hh\ column densities inferred from CO observations will
inevitably be quite restricted. Indeed, even though the minimum
CO-derived estimate of $\Sigma_{\rm H_{2}}$ for the PAWS GMCs still
likely reflects the survey's limiting CO surface brightness
sensitivity (see Figure~\ref{fig:rl}), the large velocity dispersion
of the CO-emitting gas within the PAWS field may explain why the
dispersion of the size-CO luminosity relation (and hence the width of
the inferred $\Sigma_{\rm H_{2}}$ distribution) is larger for M51 GMCs
than for the GMC populations of M33, the LMC and other Local Group
systems (cf figure~3 of B08).\\

\noindent Some further insight is provided by comparing the
size-luminosity relations obtained using different decompositions of
the PAWS data cube in Figure~\ref{fig:rl}. In particular, it is
evident that the CO surface brightness values obtained using a method
that preferentially identifies structures with a characteristic size
scale (i.e. the ``cloud-based'' decompositions in
Figure~\ref{fig:rl}[a] and~[b]) cover a wider range than the values
obtained when the boundaries of the identified structures are defined
using a fixed intensity threshold (i.e. the ``island-based''
decompositions in Figure~\ref{fig:rl}[c] and~[d]). Quantitatively, we
find that the scatter in the logarithm of the residuals about the
best-fitting size-luminosity relationships increases from
$\sim0.2$\,dex for islands in both the intrinsic and matched
resolution M51 datacubes to $\sim0.5$\,dex for the cloud structures
identified in the same cubes. \\

\noindent These decomposition-dependent results for the scatter in the
CO surface brightness (and hence $\Sigma_{H_{2}}$) values derived from
the PAWS data are qualitatively similar to the two cases considered by
\citet{lombardietal10} in their analysis of nearby Galactic clouds
using dust extinction to trace \hh\ column density: the average CO
surface brightness of molecular structures above a fixed brightness
threshold is approximately constant, while equivalent measurements
over a fixed size scale yield much larger variations in
$\Sigma_{H_{2}}$, both between and within the GMC populations of the
three galaxies that we investigate. \citet{lombardietal10} argue that
the former result arises because molecular clouds have an
approximately universal log-normal column density distribution. While
this hypothesis can be empirically verified using extinction data for
local clouds, the resolution and dynamic range of the extragalactic CO
data is insufficient to recover the detailed shape of the
\ico\ distribution for individual extragalactic GMCs. In our case, the
narrow range of CO surface brightness measurements in
Figure~\ref{fig:rl}[c] and~[d] arises because a large fraction of the
pixels within an island structure sample emission that is close to the
observational sensitivity limit. For the intrinsic resolution M51
cube, we find that for over half the islands (51\%), pixels with
integrated intensity values less than 5$\sigma$ make up more than half
the total number of pixels within the structure. This fraction
(i.e. where pixels with values $<5\sigma$ consitute the majority of
pixels within the structure) is similar in the LMC (46\% of islands)
and even greater in M33 (78\% of islands). \\

\noindent Our conclusion is that the appearance of constant CO surface
brightness among extragalactic GMCs is mostly an artifact due to the
combination of several conspiring effects: first, the
algebraically-imposed covariance of $L_{\rm CO}$ and $R$, which yields
a robust yet trivial correlation between these quantities; second, the
strategy of designing extragalactic CO surveys to detect a `typical'
Milky Way GMC at the $\sim5\sigma$ level, which limits a survey's CO
surface brightness sensitivity and may only reveal the high-mass,
CO-bright GMCs in low-mass galaxies (rather than the bulk of the
molecular cloud population); and third, the limited range of
environmental conditions that had been probed by extragalactic CO
observations with cloud-scale resolution prior to PAWS. \\

%%%%%%%%%%%%%%%%%%%%%%%%%%%
\subsubsection{Larson's First Law: The size-linewith relation}
\label{sect:llaw1_origin}
%%%%%%%%%%%%%%%%%%%%%%%%%%%

\noindent The possibility that the size-linewidth relation is an
observational artifact has received less attention in the literature,
with most debate focussing on whether it is a sign that GMCs attain
approximate virial balance between their gravitational and internal
kinetic energies, or a manifestation of interstellar turbulence
\citep[see e.g.][and references
  therein]{ballesterosparedes06}. Nevertheless, our analysis (see
Figure~\ref{fig:rdv}) suggests that some caution interpreting the
extragalactic GMC results is required. In particular, we find little
evidence for a correlation between the size and linewidth of cloud
structures {\it within} M51, M33 or the LMC. A combined sample of GMCs
from all three galaxies, on the other hand, yields a size-linewidth
relation similar to $\sigma_{\rm v} \propto R^{0.5}$ simply due to the
variations in spectral and spatial resolution of the input
datasets. For reasons noted in Section~\ref{sect:gmcproperties},
decomposition algorithms preferentially identify structures in
position and velocity space close to the resolution of a data
cube. Based on our analysis of the M51, M33 and LMC data, we would
therefore recommend that if measurements from high resolution datasets
yield the clouds with small radii and narrow linewidths, and low
resolution datasets populate the large $R$ and high $\sigma_{\rm v}$
end of a size-linewidth relation, then this bias should be explicitly
excluded before a physical explanation for the correlation is
invoked.\\

\noindent While cloud structures identified in the matched resolution
data cubes of all three galaxies fail to exhibit a strong correlation
between their size and linewidth, the vertical offset between the
cloud populations in M51 and the low-mass galaxies in
Figure~\ref{fig:rdv}[b] suggests there may be a genuine difference in
the physical state of their cloud populations. More precisely, clouds
in M51 -- and especially in the spiral arms and central regions --
have larger linewidths compared to clouds of an equivalent size in the
LMC or M33. Since regions with high velocity dispersion in M51's inner
disk are often associated with low levels of star formation activity
\citep{meidtetal13} and the star formation rates of M33 and
the LMC are high relative to their global CO luminosities
\citep[assuming a universal molecular gas depletion time of
  $\sim2$\,Gyr, e.g.][]{leroyetal08}, it seems unlikely that this
segregation is due to higher levels of internal turbulence generated
by star formation feedback. On the other hand, such an offset would be
expected if the dynamical state of the clouds is influenced by the
external pressure, as we suggested in
Section~\ref{sect:prev_results}. Following \citet[e.g.][]{elmegreen89}
(see also \citet{chieze87} and \citet{fieldetal11} for alternative
derivations), clouds that achieve equilibrium between self-gravity,
the external pressure and their internal kinetic energy should follow:
\begin{equation}
\sigma_{\rm v} \propto \left ( \frac{P_{ext}/k_{B}}{10^{4} \punit} \right )^{1/4} \left ( \frac{R}{\rm pc} \right )^{1/2}\,.
\label{eqn:rdv_pressure}
\end{equation}
\noindent According to our estimates in Section~\ref{sect:prev_results}, the
external pressure experienced by clouds in M51 is approximately an
order of magnitude higher than for clouds in M33 and the LMC. From
Equation~\ref{eqn:rdv_pressure}, we would then expect M51 clouds to
exhibit linewidths $\sim1.8$ times larger than clouds of similar size in
the low-mass galaxies, in good agreement with the vertical
offset in the size-linewidth plot that we observe (a factor of $\sim2$
at a fixed size scale).\\

\noindent A second important result from our analysis in
Section~\ref{sect:llaws} is that a tighter correlation between the
size and linewidth of molecular structures becomes apparent when we
identify GMCs as regions of connected CO emission
(i.e. islands). Arguably the most convincing example is for islands
identified in the LMC data cube with its intrinsic resolution
(Figure~\ref{fig:rdv}[c]), which follow $\sigma_{\rm v} =
(0.16\pm0.03) R^{0.84\pm0.05}$ over $\sim2$ orders of magnitude in
size. In the intrinsic resolution cube, island structures in the
spiral arms and central region of M51 exhibit a weak correlation that
is shallower than for LMC islands: $\sigma_{\rm v} = (1.1\pm0.4)
R^{0.5\pm0.1}$. On one hand, this could indicate a genuine difference
in the density structure of the molecular ISM between the two
environments. Numerical simulations by \citet{dobbsbonnell07b} show
that if the size-linewidth relation arises due to gas clumps being
brought together at the location of a shock (e.g a spiral shock, or
the interface of two colliding flows), then a steeper size-linewidth
relation is expected if molecular gas is more clumpy. The higher peak
brightness of the CO emission in M51 suggests that CO-emitting regions
may fill the beam more uniformly in M51 than in the LMC, which is at
least qualitatively consistent with the LMC hosting more clumpy
molecular material than M51, but higher resolution observations that
probe the internal density structure of GMCs in the LMC and M51 would
be required to validate this model. On the other hand, we caution that
the shallower correlation for islands in M51's spiral arms may be
partly driven by the $\sim10$ objects with $100 < R <
500$\,pc. Inspection of Figure~\ref{fig:rdv}[c] suggests that these
objects appear to follow a shallower relationship than the scattered
trend exhibited by the smaller islands. Excluding islands with $R >
150$\,pc from the fit yields $\sigma_{\rm v} = (0.6\pm0.3)
R^{0.6\pm0.1}$ for the M51 arm+central region, which has a slope that
is more similar to the LMC relation. \\

\noindent The possibility that the GMC linewidths that we measure
include gas motions unrelated to the cloud's intrinsic velocity
dispersion is underscored by the fact that structures with $R >
500$\,pc (the magenta squares in Figure~\ref{fig:rdv}[c] and~[d]) also
seem to follow a relation that is roughly consistent with the
canonical size-linewidth relationship derived for Galactic GMCs by
S87. This is somewhat surprising if the origin of the size-linewidth
relation is due to GMCs achieving dynamical equilibrium or turbulence
in the molecular ISM, since we would expect the size-linewidth
relationship to break down on scales corresponding to the largest
virialized structures or the spatial scale on which the turbulence is
driven, usually thought to be comparable to the scale height of the
gas disk ($\sim200$\,pc in M51). For structures on scales much larger
than a typical GMC, the linewidths that we measure are likely to be
broadened by systematic motions within the galactic disk, such as
galactic rotation or spiral arm streaming motions. To assess
  the importance of the former effect on the correlations in
  Figure~\ref{fig:rdv}[c] and~[d], we subtracted a model of the
  contribution of galactic rotation to the GMC linewidths from the M51
  datacube. In practice, we did this by generating a map of the
line-of-sight velocity that would be expected at each spatial position
within the PAWS and MAGMA fields from our preferred model of M51's and
LMC's rotation \citep{wongetal11,meidtetal13}. After identifying
islands of significant emission within the original data cubes, we
used the \textsc{MIRIAD} task {\tt specshift} to offset the individual
CO spectra belonging to an island along the spectral axis such that
the radial velocity corresponding to the model velocity field at each
$(x,y)$ position was shifted to the central channel of a new
`velocity-shifted' data cube. This is equivalent to
  subtracting the modeled rotation component from the observed
  velocity field map, although we manipulate the data in $(x,y,v)$
  space so that the result is a datacube that can be analysed using
  \textsc{CPROPS}. Next, we estimated the physical properties of each
``velocity-shifted'' island using the same method that we applied to
the original data cubes (see
Section~\ref{sect:cloudidentification}). The aim of this shifting
procedure was to suppress the contribution of galactic rotation to the
global velocity dispersion measurement that we obtain for each
island. \\

\begin{figure*}
\begin{center}
\hspace{-0.5cm}
\includegraphics[height=70mm,angle=0]{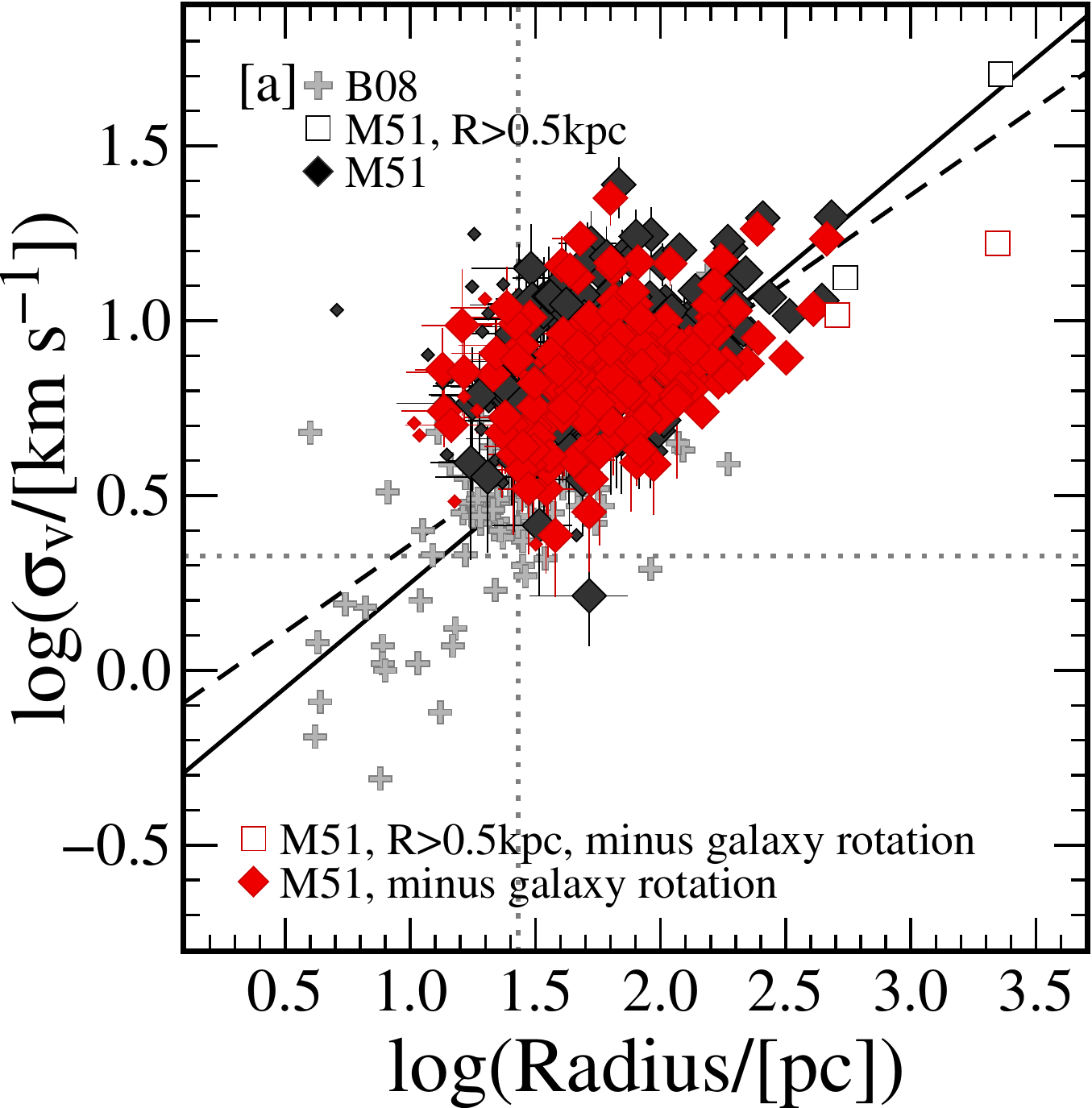}
\hspace{0.5cm}
\includegraphics[height=70mm,angle=0]{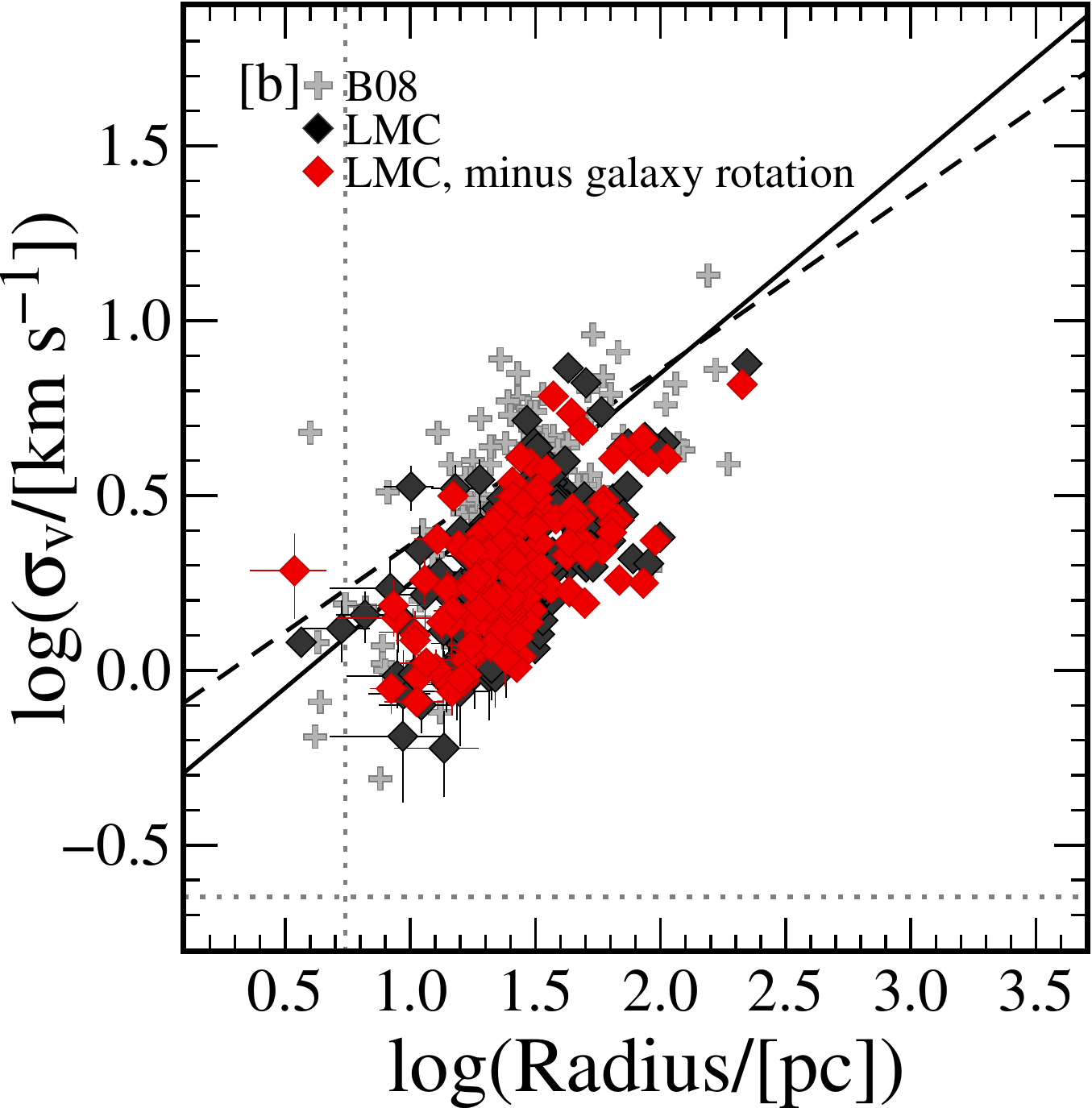}
\caption{\small A plot of radius versus velocity dispersion for
  islands identified within the [a] M51 and [b] LMC data cubes at
  their intrinsic resolution. The black points represent measurements
  obtained directly from the cube and are identical to those shown in
  Figure~\ref{fig:rdv}[c]. The red points represent measurements
  that were obtained after centring each line profile 
  to the radial velocity expected from a circular rotation model. The
  black dashed line indicates the relationship derived from the S87
  inner Milky Way data, and the black solid line indicates the
  best-fitting relation for extragalactic GMCs determined by
  B08. Extragalactic GMCs analysed by B08 are indicated with grey
  crosses.}
\label{fig:rdv_galsub}
\end{center}
\end{figure*}

\noindent In Figure~\ref{fig:rdv_galsub}, we plot the radius versus
the velocity dispersion for the ``velocity-shifted'' islands (red
diamonds) identified in M51 (panel [a]) and the LMC (panel [b]). In
both galaxies, there are fewer islands where {\textsc CPROPS} is able
to measure the velocity dispersion after the shifting procedure. In
M51, many of these objects correspond to islands in the original
sample with large velocity dispersions compared to their size, so the
outliers and upper envelope of the main distribution of red diamonds
in Figure~\ref{fig:rdv} are effectively removed. The result is to
bring the bulk of the ``velocity-shifted'' M51 data points slightly
closer to the measurements for the low-mass galaxies, and to the
size-linewidth relation for inner Milky Way GMCs (the dashed line in
Figure~\ref{fig:rdv_galsub}). Indeed, a BCES bisector fit to the
``velocity-shifted'' M51 islands yields $\sigma_{\rm v} =
(0.85\pm0.24) R^{0.51\pm0.06}$, which is indistinguishable from the
canonical S87 result. In the LMC, the data points appear less
scattered after the galactic rotation model is subtracted, but the
overall distribution is not shifted towards significantly lower
velocity dispersions, an effect that {\it is} seen for large ($R
\gtrsim 300$\,pc) islands in M51. The best-fitting size-linewidth
relation for the velocity-shifted LMC islands is identical to the
relation obtained from islands identified in the original data
cube. In summary, we find that contamination of GMC linewidths by
systematic motions associated with galactic rotation is significant
for the largest CO-emitting structures in M51, but it does not appear
to strongly determine the correlation between size and linewidth for
structures with spatial scales corresponding to the characteristic
size of GMCs (i.e. 10 to 100\,pc).  \\

\noindent Once again, our conclusion is that further investigation is
required to establish whether extragalactic GMC populations follow the
same size-linewidth relation as GMCs in the inner Milky Way, and that
particular care must be taken to eliminate the effects of resolution,
survey design -- since the observing configuration of most
extragalactic CO surveys is selected to optimise sensitivity to
structures with sizes and linewidths similar to Galactic GMCs -- and
analysis methods. Alternative explanations for the physical origin of
the size-linewidth relation -- besides simple virial equilibrium and
interstellar turbulence -- also merit further consideration. Here we
have suggested that the larger velocity dispersion of M51 clouds
relative to clouds with similar size in the LMC and M33 may be due to
the higher pressure at the cloud surface in M51's inner disk. By
contrast, the best explanation for the more robust size-linewidth
relations that we recover when we identify GMCs as `islands' of CO
emission may be that an islands decomposition yields regions where
external processes -- such as a converging flow, or a spiral shock --
are bringing pre-existing smaller molecular structures into the same
spatial location. These `cloud associations' are likely to be globally
unbound, with CO linewidths that reflect the macroscopic motions of
their constituent gas clumps, rather than interstellar turbulence {\it
  per se} \citep[e.g.][]{dobbsbonnell07b}. In this case, a scaling
between the size and linewidth might simply reflect the inhomogeneous
density distribution and velocity fluctuations in the neutral ISM:
over small scales, the constituent clumps encounter relatively
homogenous material with similar density and local velocity resulting
in a low clump-clump velocity dispersion. Over larger scales, however,
the clumps interact with material with a larger range of densities and
peculiar velocities, and hence experience different
decelerations. This produces a higher velocity dispersion between
widely-separated clumps. In other words, the tighter correlation
between size and linewidth when we identify GMCs as regions of
connected CO emission may simply reflect more adequate sampling of the
density and velocity structure of the ISM by the dense clumps that
constitute the CO islands \citep[e.g.][]{bonnelletal06}.

%%%%%%%%%%%%%%%%%%%%%%%%%%%%%%
\section{Conclusions}
%%%%%%%%%%%%%%%%%%%%%%%%%%%%%%
\label{sect:conclusions}

\noindent In this paper, we compared the properties of GMCs identified
in the PAWS survey of M51's inner disk to the GMC populations of M33
and the Large Magellanic Cloud (LMC). In contrast to previous
comparative studies of extragalactic GMC populations, our datasets
contain a statistically significant sample of clouds for each galaxy,
have sufficient resolution to resolve individual GMCs and include
single-dish measurements to recover total flux information. We
explicitly homogenize the resolution, sensitivity and gridding scheme
of the CO datasets to suppress these important sources of bias on the
derived GMC properties. Our key results are:\\

\noindent 1. We find genuine differences in the physical properties of
GMCs in M51, M33 and the LMC: on average, GMCs in M51 have higher peak
CO brightness, CO surface brightness, and velocity dispersion than
GMCs of equivalent size in the low-mass galaxies, consistent with the
dynamical state of clouds being influenced by the ambient interstellar
pressure. The observed differences are especially pronouced when we
restrict our M51 GMC sample to objects in the spiral arm and central
region. For this comparison, we took care to homogenize the CO
datasets and we note that this procedure was essential to minimize
observational bias.  \\

\noindent 2. The presence of a correlation between size and linewidth
depends sensitively on how we define clouds. If we apply an aggressive
decomposition algorithm to the CO data cubes, we find no compelling
evidence for a correlation between size and line width for the GMC
populations of M51, M33 or the LMC. A strong correlation similar to
the canonical size-linewidth relationship for GMCs in the inner Milky
Way is apparent when we identify GMCs as regions of contiguous CO
emission. We propose that these structures are more like cloud
associations, i.e. regions where an external process has caused
smaller pre-existing molecular gas structures to converge.  In
addition to simple virial equilibrium or a classical turbulent
cascade, the observed size-linewidth relation may also be a reflection
of more adequate sampling of the inhomeogeneous density and velocity
structure of the interstellar medium by the clouds that belong to
these associations. In general, more observational effort to identify
the processes that contribute to the global linewidths of molecular
clouds in different galactic environments would be highly desirable,
and would significantly improve our understanding of the physical
origin of the size-linewidth relation.\\

\noindent 3. Within M51, M33 and the LMC, CO islands exhibit a
relatively narrow range of surface brightness measurements. We argue
that the appearance of uniform surface brightness for these structures
may be imposed by using a intensity threshold close to the survey's
sensitivity limit to define the island boundary, since for many
islands (between 46 and 78\% depending on the galaxy), the majority of
pixels sample emission that is less than the 5$\sigma$ sensitivity
limit.\\

\noindent 4. The dynamic range of surface brightness measurements
increases for a more aggressive cloud decomposition of the CO
datacubes, which tends to retain the high brightness substructure
within the islands, and discard the low brightness emission
surrounding the high intensity peaks. There appear to be genuine
variations in the average surface brightness -- and, we infer, average
mass surface density -- of the GMC populations in M51, M33 and the
LMC. Combining our analysis with literature measurements of resolved
extragalactic GMC properties suggests that the average GMC mass
surface density varies with the characteristic interstellar pressure
of the galactic environment where the GMCs are located.\\

\noindent Our results highlight the difficulties and limitations of
decomposing molecular gas into clouds. This type of analysis becomes
especially problematic for molecule-rich environments -- like the
inner disk of M51 -- where the emission is both bright and extended
over spatial scales many times greater than the observational
resolution. In light of the sensitive dependence of GMC property
measurements and scaling relations on instrumental resolution,
observational sensitivity and decomposition approach, methods that can
analyse hierarchical structure
\citep[e.g.][]{rosolowskyetal08,kauffmannetal10,shettyetal12} would
seem to offer a promising approach to studying the physical properties
of the molecular ISM, since they can quantify the importance of
e.g. self-gravity or external pressure as a function of spatial scale
without imposing a cloud-like model for the molecular gas {\it a
  priori}. An analysis of the PAWS data using hierarchical
decomposition methods will be presented in a forthcoming paper (Leroy
et al., in preparation).\\

\noindent Finally, we note that the tendency for molecular clouds in
low-mass galaxies and the outer Milky Way to be smaller and fainter
than molecular structures in the inner Milky Way -- and, conversely,
for GMCs in molecule-rich, high-pressure environments to be denser and
more massive than local clouds -- was already suggested by several
previous observational studies
\citep[e.g.][]{ketomyers86,heyeretal01,okaetal01,rosolowskyblitz05,gratieretal10,wongetal11}.
Firmly establishing a dependence between GMC properties and galactic
environment has proven difficult, however, in part because assembling
large samples of extragalactic GMCs (especially in CO-faint dwarf
galaxies) remains technically challenging with the current generation
of millimetre telescopes. Future observations with ALMA that
efficiently survey the CO emission across a significant fraction of
nearby galactic disks will be invaluable for increasing the number of
extragalactic GMC samples and the range of environments where GMC
properties can be studied. Such observations will be crucial for
establishing the physical mechanisms that are responsible for
variations in GMC properties and regulate their ability to form
stars. \\

\acknowledgments

\noindent We thank the anonymous referee for their helpful comments
and suggestions, which significantly improved the quality of this
paper. We thank the IRAM staff for their support during the
observations with the Plateau de Bure interferometer and the 30m
telescope.  DC and AH acknowledge funding from the Deutsche
Forschungsgemeinschaft (DFG) via grant SCHI 536/5-1 and SCHI 536/7-1
as part of the priority program SPP 1573 'ISM-SPP: Physics of the
Interstellar Medium'.  CLD acknowledges funding from the European
Research Council for the FP7 ERC starting grant project LOCALSTAR.
TAT acknowledges support from NASA grant \#NNX10AD01G.  During this
work, J.~Pety was partially funded by the grant ANR-09-BLAN-0231-01
from the French {\it Agence Nationale de la Recherche} as part of the
SCHISM project (\url{http://schism.ens.fr/}).  ES, AH and DC thank
NRAO for their support and hospitality during their visits in
Charlottesville.  ES thanks the Aspen Center for Physics and the NSF
Grant No. 1066293 for hospitality during the development and writing
of this paper.

\clearpage

\bibliographystyle{apj}

\end{document}